\documentclass[twoside]{article}
\usepackage[accepted]{aistats2024}
\usepackage[round]{natbib}

\usepackage{amsmath}
\usepackage{subcaption}
 
\usepackage{tikz}
\usetikzlibrary{arrows}
\usepackage[inline]{enumitem}
\usepackage{comment}

\usepackage{hyperref}
\usepackage{dsfont}

\usepackage[ruled,linesnumbered]{algorithm2e}
\SetKwInOut{Input}{input}
\SetKwInOut{Output}{output}
\SetKw{Throw}{throw}

\RequirePackage{amssymb}

\newtheorem{example}{Example} 
\newtheorem{theorem}{Theorem}
\newtheorem{lemma}[theorem]{Lemma} 
\newtheorem{proposition}[theorem]{Proposition}

\newtheorem{definition}[theorem]{Definition}

\DeclareMathOperator{\Var}{Var}
\DeclareMathOperator{\Cov}{Cov}
\DeclareMathOperator{\An}{An}

\DeclareMathOperator{\Pa}{Pa}
\DeclareMathOperator{\pa}{pa}
\DeclareMathOperator{\De}{De}
\DeclareMathOperator{\PossDe}{PossDe}
\DeclareMathOperator{\PossPa}{PossPa}
\DeclareMathOperator{\PossAn}{PossAn}

\DeclareMathOperator{\Adj}{Adj}

\DeclareMathOperator{\Forb}{Forb}
\newcommand{\fb}[2][X,Y]{\Forb(\mb{#1},#2)}
\newcommand{\gpbd}[2][G]{\mathcal{#1}_{\mb{#2}}^{pbd}}
\newcommand{\dpbd}[2][G]{\mathcal{D}_{\mb{#2}}^{pbd}}

\DeclareMathOperator{\Adjust}{Adjust}
\newcommand{\adjustb}[2][X,Y,Z]{\Adjust(\mb{#1},#2)}

\DeclareMathOperator{\Mediators}{Med}
\newcommand{\mediatb}[2][X,Y]{\Mediators(\mb{#1},#2)}
\newcommand{\mediat}[2][X,Y]{\Mediators({#1},#2)}

\DeclareMathOperator{\PossMediators}{PossMed}
\newcommand{\possmediatb}[2][X,Y]{\PossMediators(\mb{#1},#2)}
\newcommand{\possmediat}[2][X,Y]{\PossMediators({#1},#2)}

\DeclareMathOperator{\Opt}{O}
\newcommand{\opt}[2][X,Y]{\Opt(\mb{#1},#2)}
\newcommand{\optb}[2][X,Y]{\Opt(\mb{#1},#2)}

\DeclareMathOperator{\E}{\mathbb{E}}
\newcommand\given[1][]{\:#1\vert\:}

\newcommand{\mb}[1]{\mathbf{#1}}
\newcommand{\dsepp}{\perp_{d}}
\newcommand{\g}[1][G]{\mathcal{#1}}
\newcommand*\diff{\mathop{}\!\mathrm{d}}
\newcommand{\ind}{\perp\!\!\!\!\perp}

\newenvironment{proofof}[1][]{\begin{trivlist}
\item[\hskip \labelsep {\bfseries Proof of #1.}]}{\hfill\BlackBox\end{trivlist}}

\newcommand{\BlackBox}{\rule{1.5ex}{1.5ex}}
\ifdefined\proof
    \renewenvironment{proof}{\par\noindent{\bf Proof\ }}{\hfill\BlackBox\\[2mm]}
\else
    \newenvironment{proof}{\par\noindent{\bf Proof\ }}{\hfill\BlackBox\\[2mm]}
\fi

\newcommand{\bulletcirc}{
  \setlength{\unitlength}{1mm}
  \begin{picture}(5,1)(0,0)
    \put(0.2,0){$\bullet$}
    \put(1.1, 1){\line(1,0){2.4}}
    \put(4, 1){\circle{1}}
  \end{picture}
}

\newcommand{\circbullet}{
  \setlength{\unitlength}{1mm}
  \begin{picture}(5,1)(0,0)
    \put(1,1){\circle{1}}
    \put(1.5,1){\line(1,0){2.4}}
    \put(2.9,0){$\bullet$}
  \end{picture}
}

\newcommand{\bulletarrow}{
  \setlength{\unitlength}{1mm}
  \begin{picture}(5,1)(0,0)
    \put(0.2,0){$\bullet$}
    \put(1,0){$\rightarrow$}
  \end{picture}
}

\newcommand{\arrowbullet}{
  \setlength{\unitlength}{1mm}
  \begin{picture}(5,1)(0,0)
    \put(0.2,0){$\leftarrow$}
    \put(3,0){$\bullet$}
  \end{picture}
}

\newcommand{\circarrow}{
  \setlength{\unitlength}{1mm}
  \begin{picture}(5,1)(0,0)
    \put(1,1){\circle{1}}
    \put(1.2,0){$\rightarrow$}
  \end{picture}
}

\newcommand{\circcirc}{
  \setlength{\unitlength}{1mm}
  \begin{picture}(5,1)(0,0)
    \put(1,1){\circle{1}}
    \put(1.5,1){\line(1,0){2}}
    \put(4,1){\circle{1}}
  \end{picture}
}

\begin{document}

\twocolumn[
\aistatstitle{Conditional Adjustment in a Markov Equivalence Class}
\aistatsauthor{ Sara LaPlante \And Emilija Perkovi\'c }
\aistatsaddress{ University of Washington \And University of Washington } ]


\begin{abstract}
We consider the problem of identifying a conditional causal effect through covariate adjustment. We focus on the setting where the causal graph is known up to one of two types of graphs: a maximally oriented partially directed acyclic graph (MPDAG) or a partial ancestral graph (PAG). Both MPDAGs and PAGs represent equivalence classes of possible underlying causal models. After defining adjustment sets in this setting, we provide a necessary and sufficient graphical criterion -- the \textit{conditional adjustment criterion} -- for finding these sets under conditioning on variables unaffected by treatment. We further provide explicit sets from the graph that satisfy the conditional adjustment criterion, and therefore, can be used as adjustment sets for conditional causal effect identification.
\end{abstract}


\section{INTRODUCTION}
\label{sec:intro}

Many scientific disciplines have an interest in identifying and estimating causal effects for specific subgroups of a population. For instance, researchers may want to know if a medical treatment is beneficial for people with heart disease or if the treatment will harm older patients \citep{brand2010benefits, world2010medical}. Such causal effects are referred to as \textit{conditional causal effects} or \textit{heterogeneous causal effects}. The identification of these conditional causal effects from observational data is the subject of this work. 

Much of the literature on estimating conditional causal effects from observational data focuses on the conditional average treatment effect (CATE; \citealp{athey2016recursive, wager2018estimation, kunzel2019metalearners, nie2021quasi, kennedy2022minimax}). The CATE is represented as a contrast of means for a response $Y$ under different do-interventions (see Section \ref{sec:prelim} for definition) of a treatment $X$ when conditioning on a set of covariate values $\mb{z}$. These means take the form $\E[Y | do(X=x), \mb{Z} = \mb{z}]$.

Some results on CATE estimation assume that the conditioning set $\mb{Z}$ is rich enough to capture all relevant common causes of $X$ and $Y$ -- meaning that $X$ and $Y$ are \textit{unconfounded} given $\mb{Z}$. This implies 
\begin{align}
    \E[Y | do(X=x), \mb{Z} = \mb{z}] =\E[Y | X=x, \mb{Z} =\mb{z}], \label{eq:cate-claim-1}
\end{align}
 which allows the CATE to be estimated as a difference of means from observational data. 

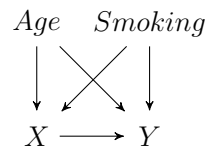
\begin{figure}
        \centering
        \begin{tikzpicture}[>=stealth',shorten >=1pt,auto,node distance=2cm,main node/.style={minimum size=0.8cm,font=\sffamily\Large\bfseries},scale=0.75,transform shape]
        \node[main node]   (X)                        {$X$};
        \node[main node]   (V1)  at (0,2)             {$Age$};
        \node[main node]   (V2) [right of       = V1] {$Smoking$};
        \node[main node]   (Y)  [right of = X] {$Y$};
        \draw[->]   (X)  edge (Y);
        \draw[->]   (V2) edge (X);
        \draw[->]   (V2) edge (Y);
        \draw[->]   (V1) edge (X);
        \draw[->]   (V1) edge (Y);
        \end{tikzpicture}
    \caption{A causal DAG used in Section \ref{sec:intro}.}
    \label{fig:intro}
\end{figure}

However, this assumption does not hold in all applications. Consider, for example, the setting depicted in the causal directed acyclic graph (DAG) of Figure \ref{fig:intro}, where we want to compute a causal effect of $X$ on $Y$ given some set $\mb{Z}$. In this setting, age and smoking status are common causes of $X$ and $Y$, and therefore, $X$ and $Y$ are confounded unless we condition on both age and smoking status ($\mb{Z} = \{Age, Smoking\}$). But we may want to know the causal effect of $X$ on $Y$ conditional on age alone ($\mb{Z} =\{Age\}$). 


To allow for estimation of the CATE in such cases, various recent works \citep{abrevaya2015estimating, fan2022estimation,chernozhukov2023simple, smucler2020efficient} have proposed estimation methods that rely on knowing an additional set of covariates $\mb{S}$ that -- together with $\mb{Z}$ -- leads to $X$ and $Y$ being unconfounded. We refer to this set of variables as a \textit{conditional adjustment set} (Definition \ref{def:cas}). For such a set $\mb{S}$, 
\begin{align}
    \E[Y| d&o(X=x), \mb{Z} = \mb{z}] \label{eq:cate-claim-2} \\ 
            &= \E_{\mb{S}} \Big[ \E [Y| X=x, \mb{Z} , \mb{S}] \ \Big| \ \mb{Z} = \mb{z} \Big]. \nonumber
\end{align}
In the example above, if $\mb{Z} = \{Smoking\}$, then $\mb{S} = \{Age\}$.


Of course, not all conditional causal effect research focuses on estimation through the functional in Equation \eqref{eq:cate-claim-2}. Notably, other work has explored identifiability without limiting focus to a particular functional. For example, \cite{shpitser2008complete} and \cite{jaber2019identification, jaber2022causal} focus on the conditions under which the interventional distribution $f(\mb{y}|do(\mb{x}),\mb{z})$ is identifiable given a causal graph. Though these results broaden the options for identification, estimators based on these results would have to rely on functionals that may prove difficult to estimate, such as $\frac{f(\mb{y},\mb{z}|do(\mb{x}))}{f(\mb{z}|do(\mb{x}))}$ \citep{shpitser2008complete, jaber2019identification, jaber2022causal}. Our work addresses this by focusing on identification of the same interventional distribution given a causal graph -- but through the use of conditional adjustment sets, which may lead to more desirable estimators. To the best of our knowledge, this area of research is largely unexplored.

Our main contribution is the \textit{conditional adjustment criterion} (Definitions \ref{def:cac} and \ref{def:cac-pag}), a graphical criterion that we show is necessary and sufficient for identifying a conditional adjustment set (Theorems \ref{thm:cac-cas} and \ref{thm:cac-cas-pag}). We additionally provide explicit sets that satisfy this criterion when any such set exists. We note, however, that these results are restricted to a setting where the conditioning set $\mb{Z}$ consists of variables known to be unaffected by treatment. While this restricted setting produces limitations (see the second example in the discussion, Section \ref{sec:discussion}), our results are broadly applicable to a variety of research questions. For example, the restriction is met when the conditioning set includes exclusively pre-treatment variables.


In considering the problem of identifying a conditional adjustment set, we assume that the underlying causal system can be represented by a causal DAG. When we collect observational data on all variables in the system, we can attempt to learn this causal DAG by relying on the constraints present in the data \citep{spirtes1999algorithm, chickering2002learning, zhang2008completeness, hauser2012characterization, mooij2020joint, squires2022causal}. However, this task is often impossible from observational data alone, regardless of the available sample size. And further, we cannot always observe every variable. 

Thus, our work focuses on causal models that represent \textit{Markov equivalence classes} of graphs that can be learned from observational data: a maximally oriented partially directed acyclic graph (MPDAG; \citealp{meek1995causal}) and a maximally oriented partial ancestral graph (PAG; \citealp{richardson2002ancestral}). An MPDAG represents a restriction of the Markov equivalence class of DAGs that can be learned from observational data and background knowledge when all variables are observed \citep{andersson1997characterization, meek1995causal, chickering2002learning}. A PAG represents a Markov equivalence class of maximal ancestral graphs (MAGs; \citealp{richardson2002ancestral}), which can be learned from observational data and which allows for unobserved variables \citep{spirtes2000causation, zhang2008completeness, ali2009markov}. A MAG, in turn, can be seen as a marginalization of a DAG containing only the observed variables \citep{richardson2002ancestral}. See Section \ref{sec:prelim} and Supp.\ \ref{supp:defs} for further definitions.


The structure of this paper is as follows: Section \ref{sec:prelim} provides preliminary definitions, with the remaining definitions given in Supp.\ \ref{supp:defs}. Section \ref{sec:mpdags} contains all results for the MPDAG setting. In particular, we introduce our conditional adjustment criterion in Section \ref{sec:cond-adj}; Section \ref{sec:exs} illustrates applications of our criterion with examples; Section \ref{sec:constructing} provides several methods for constructing conditional adjustment sets; and Section \ref{sec:comparison} includes a discussion of the similarities of our conditional adjustment criterion with both the adjustment criterion of \cite{perkovic2017interpreting} and the $\mb{Z}$-dependent dynamic adjustment criterion of \cite{smucler2020efficient}. We present some analogous results for PAGs in Section \ref{sec:pags}, and we discuss some limitations of our results and areas for future work in Section \ref{sec:discussion}.


\section{PRELIMINARIES}
\label{sec:prelim}

We use capital letters (e.g.\ $X$) to denote nodes in a graph as well as random variables that these nodes represent. Similarly, bold capital letters (e.g.\ $\mb{X}$) are used to denote node sets and random vectors.

\textbf{Nodes, Edges, and Subgraphs.}
A graph $\g=(\mb{V},\mb{E})$ consists of a set of nodes (variables) $\mb{V}=\left\lbrace V_{1},\dots,V_{p}\right\rbrace, p \ge 1$, and a set of edges $\mb{E}$.  Edges can be \textit{directed} ($\rightarrow$),  \textit{bi-directed}  ($\leftrightarrow$), \textit{undirected} ($\circcirc$ or $-$), or \textit{partially directed} ($\circarrow$). We use $\bullet$ as a stand in for any of the allowed edge marks. An edge is \textit{into} (\textit{out of}) a node $X$ if the edge has an arrowhead (tail) at $X$. An \textit{induced subgraph} $\g_{\mb{V'}} =(\mb{V'}, \mb{E'})$ of $\g$ consists of $\mb{V'} \subseteq \mb{V}$ and $\mb{E'} \subseteq \mb{E}$ where $\mb{E'}$ are all edges in $\mb{E}$ between nodes in $\mb{V'}$. 

\textbf{Directed and Partially Directed Graphs.}
A \textit{directed graph} contains only directed edges ($\to$). A \textit{partially directed graph} may contain undirected edges ($-$) and directed edges ($\to$).

\textbf{Mixed and Partially Directed Mixed Graphs.}
A \textit{mixed graph} may contain directed and bi-directed edges. The \textit{partially directed mixed graphs} we consider can contain any of the following edge types: $\circcirc$, $\circarrow$, $\to$, and $\leftrightarrow$. Hence, an edge $\bulletarrow$ in a partially directed graph can only refer to edge $\to$, whereas in a partially directed mixed graph, $\bulletarrow$ can represent $\to$, $\leftrightarrow$, or $\circarrow$.

\textbf{Paths and Cycles.}
For disjoint node sets $\mb{X}$ and $\mb{Y}$, a path from $\mb{X}$ to $\mb{Y}$ is a sequence of distinct nodes $\langle X, \dots,Y \rangle$ from some $X \in \mb{X}$ to some $Y \in \mb{Y}$ for which every pair of successive nodes is adjacent. A path consisting of undirected edges ($-$ or $\circcirc$) is an \textit{undirected path}. A \textit{directed path} from $X$ to $Y$ is a path of the form $X \to \dots \to Y$. A directed path from $X$ to $Y$ and the edge $Y \to X$ form a \textit{directed cycle}. A directed path from $X$ to $Y$ and the edge $X \to Y$ form an \textit{almost directed cycle}. A path $\langle V_1, \dots, V_k \rangle$, $k>1$, in a graph $\g$ is a \textit{possibly directed path} if no edge $V_i \arrowbullet V_j, 1 \le i < j \le k$, is in $\g$ (\citealp{perkovic2017interpreting}, \citealp{zhang2008causal}).

A path from $\mb{X}$ to $\mb{Y}$ is \textit{proper} (w.r.t. $\mb{X}$) if only its first node is in $\mb{X}$. A path from $X$ to $Y$ is a \textit{back-door path} if does not begin with a visible edge out of $X$ (see definition of \textit{visible} below; \citealp{pearl2009causality}, \citealp{maathuis2015generalized}). For a path $p = \langle X_1, X_2, \dots, X_k \rangle$ and $i,j,k$ such that $1 \le i < j \le k$, we define the \textit{subpath} of $p$ from $X_i$ to $X_j$ as the path $p(X_i,X_j) = \langle X_i,X_{i+1},\dots,X_j \rangle$.

\textbf{Colliders, Shields, and Definite Status Paths.}
If a path $p$ contains $X_i \bulletarrow X_j \arrowbullet X_k$ as a subpath, then $X_j$ is a \textit{collider} on $p$. A path $\langle X_{i},X_{j},X_{k} \rangle$ is an \textit{unshielded triple} if $X_{i}$ and $X_{k}$ are not adjacent. A path is \textit{unshielded} if all successive triples on the path are unshielded. A node $X_{j}$ is a \textit{definite non-collider} on a path $p$ if the edge $X_i \leftarrow X_j$ or $X_j \rightarrow X_k$ is on $p$, or if $\langle X_{i}, X_j,X_k \rangle$ is an undirected subpath of $p$ and $X_i$ is not adjacent to $X_k$. A node is of \textit{definite status} on a path if it is a collider, a definite non-collider, or an endpoint on the path. A path $p$ is of definite status if every node on $p$ is of definite status.

\textbf{Blocking, D-separation, and M-separation.}
Let $\mb{X}$, $\mb{Y}$, and $\mb{Z}$ be pairwise disjoint node sets in a directed or partially directed graph $\g$. A definite-status path $p$ from $\mb{X}$ to $\mb{Y}$ is \textit{d-connecting} given $\mb{Z}$ if every definite non-collider on $p$ is not in $\mb{Z}$ and every collider on $p$ has a descendant in $\mb{Z}$. Otherwise, $\mb{Z}$ \textit{blocks} $p$. If $\mb{Z}$ blocks all definite status paths between $\mb{X}$ and $\mb{Y}$ in $\g$, then $\mb{X}$ is \textit{d-separated} from $\mb{Y}$ given $\mb{Z}$ in $\g$ and we write $(\mb{X} \dsepp \mb{Y} | \mb{Z})_{\g}$ \citep{pearl2009causality}.

If $\g$ is a mixed or partially directed mixed graph, the analogous terms to d-connection and d-separation are called \textit{m-connection} and \textit{m-separation} \citep{richardson2002ancestral}. If a path is not m-connecting in such a graph $\g$ we will also call it blocked. We will also use the same notation $\dsepp$ to denote m-separation in a mixed or partially directed mixed graph $\g$.

\textbf{Ancestral Relationships.}
If $X \to Y$, then $X$ is a \textit{parent} of $Y$. If $X-Y$, $X \circcirc Y$, $X \circarrow Y$, or $X \to Y$, then $X$ is a \textit{possible parent} of $Y$. If there is a directed path from $X$ to $Y$, such as  $X \to M_1 \to \dots \to M_k$, $M_k = Y$, $k \ge 1$, then $X$ is an \textit{ancestor} of $Y$, $Y$ is a \textit{descendant} of $X$, and $M_1, \dots, M_k$ are \textit{mediators} for $X$ and $Y$.  We use the convention that if $Y$ is a descendant of $X$, then $Y$ is also a mediator for $X$ and $Y$. If there is a possibly directed path from $X$ to $Y$, then $X$ is a \textit{possible ancestor} of $Y$, $Y$ is a \textit{possible descendant} of $X$, and any node on this path that is not $X$ is a \textit{possible mediator} of $X$ and $Y$. We use the convention that if $Y$ is a possible descendant of $X$, then $Y$ is also a possible mediator for $X$ and $Y$. We also use the convention that every node is an ancestor, descendant, possible ancestor, and possible descendant of itself. The sets of parents, possible parents, ancestors, descendants, possible ancestors, and possible descendants of $X$ in $\g$ are denoted by $\Pa(X,\g)$, $\PossPa(X,\g)$, $\An(X,\g)$, $\De(X,\g)$, $\PossAn(X,\g)$, and $\PossDe(X,\g)$, respectively. Similarly, we denote the sets of mediators and possible mediators for $X$ and $Y$ in $\g$ by $\mediat{\g}$ and $\possmediat{\g}$. 

We let $\An(\mb{X},\g) = \cup_{X \in \mb{X}} \An(X,\g)$, with analogous definitions for $\De(\mb{X},\g)$, $\PossAn(\mb{X},\g)$, and $\PossDe(\mb{X},\g)$. For disjoint node sets $\mb{X}$ and $\mb{Y}$, we let $\mediatb{\g}$ be the union of all mediators of $X \in \mb{X}$ and $Y \in \mb{Y}$ that lie on a proper causal path from $\mb{X}$ to $\mb{Y}$, with an analogous definition for $\possmediatb{\g}$. Unconventionally, we define $\Pa(\mb{X},\g) =(\cup_{X \in \mb{X}} \Pa(X,\g)) \setminus \mb{X}$. We denote that $X$ is adjacent to $Y$ in $\g$ by $X \in \Adj(Y,\g)$.

\textbf{DAGs and PDAGs.}
A directed graph without directed cycles is a \textit{directed acyclic graph} (DAG). A \textit{partially directed acyclic graph} (PDAG) is a partially directed graph without directed cycles.

\textbf{MAGs.}
A mixed graph without directed or almost directed cycles is called \textit{ancestral}. Note that we do not consider ancestral graphs that represent selection bias (see \citealp{zhang2008causal}, for details). A \textit{maximal ancestral graph} (MAG) is an ancestral graph $\g[M] = (\mb{V,E})$ where every pair of non-adjacent nodes $X$ and $Y$ in $\g[M]$ can be m-separated by a set $\mb{Z} \subseteq \mb{V} \setminus \{X,Y\}$. A DAG $\g[D] = (\mb{V,E})$ with unobserved variables $\mb{U} \subseteq \mb{V}$ can be uniquely \textit{represented by} a MAG $\g[M] = (\mb{V} \setminus \mb{U}, \mb{E'})$, which preserves the ancestry and m-separations among the observed variables \citep{richardson2002ancestral}.

\textbf{MPDAGs and Markov Equivalence.}
All DAGs over a node set $\mb{V}$ with the same adjacencies and unshielded colliders can be uniquely \textit{represented by} a \textit{completed PDAG} (CPDAG). These DAGs form a Markov equivalence class with the same set of d-separations. A \textit{maximally oriented PDAG} (MPDAG) is formed by taking a CPDAG, adding background knowledge (by directing undirected edges), and completing \cite{meek1995causal}'s orientation rules. We say a DAG is \textit{represented by} an MPDAG $\g$ if it has the same nodes, adjacencies, and directed edges as $\g$. The set of such DAGs -- denoted by $[\g]$ -- forms a restriction of the Markov equivalence class so that all DAGs in $[\g]$ have same set of d-separations. Note that if $\g$ has the edge $A - B$, then $[\g]$ contains at least one DAG with $A \to B$ and one DAG with $A \leftarrow B$ \citep{meek1995causal}. Further, note that all DAGs and CPDAGs are MPDAGs.

\textbf{PAGs and Markov Equivalence.}
All MAGs that encode the same set of m-separations form a Markov equivalence class, which can be uniquely \textit{represented by} a \textit{partial ancestral graph} (PAG; \citealp{richardson2002ancestral, ali2009markov}). $[\g]$ denotes all MAGs represented by a PAG $\g$. We say a DAG $\g[D]$ is \textit{represented by} a PAG $\g$ if there is a MAG $\g[M] \in [\g]$ such that $\g[D]$ is represented by $\g[M]$.

We do not consider PAGs that represent selection bias (see \citealp{zhang2008completeness}). Further, we only consider maximally informative PAGs \citep{zhang2008completeness}. That is, if a PAG $\g$ has the edge $A \bulletcirc B$, then $[\g]$ contains a MAG with $A \bulletarrow B$ and a MAG with $A \gets B$. (We preclude MAGs with $A - B$ by assuming no selection bias.) Any arrowhead or tail edge mark in a PAG $\g$ corresponds to that same arrowhead or tail edge mark in every MAG in $[\g]$. The edge orientations in every PAG we consider are completed with respect to orientation rules $R1-R4$ and $R8-R10$ of \cite{zhang2008completeness}.

\textbf{Visible and Invisible Edges.}
Given a MAG or PAG $\g$, a directed edge $X \rightarrow Y$ is \textit{visible} in $\g$ if there is a node $V \notin \Adj(Y, \g)$ such that $\g$ contains either $V \bulletarrow X$ or $V \bulletarrow V_1 \leftrightarrow \dots \leftrightarrow V_k \leftrightarrow X$, where $k \ge 1$ and $V_1, \dots, V_k \in \Pa(Y, \g) \setminus \{V,X,Y\}$ \citep{zhang2006causal}. A directed edge that is not visible in a MAG or PAG is said to be \textit{invisible}. 

\textbf{Markov Compatibility and Positivity.}
An \textit{observational density} $f(\mb{v})$ is \textit{Markov compatible} with a DAG $\g[D] = (\mb{V},\mb{E})$ if $f(\mb{v})= \prod_{V_i \in \mb{V}}f(v_i|\pa(v_i,\g[D]))$. If $f(\mb{v})$ is Markov compatible with a DAG $\g[D]$, then it is Markov compatible with every DAG that is Markov equivalent to $\g[D]$ \citep{pearl2009causality}. Hence, we say that a density is \textit{Markov compatible} with an MPDAG, MAG, or PAG $\g$ if it is Markov compatible with a DAG represented by $\g$. Throughout, we assume positivity. That is, we only consider distributions that satisfy $f(\mb{v})>0$ for all valid values of $\mb{V}$ \citep{kivva2023identifiability}.

\textbf{Probabilistic Implications of Graph Separation.}
Let $\mb{X}$, $\mb{Y}$, and $\mb{Z}$ be pairwise disjoint node sets in a DAG, MPDAG, MAG, or PAG $\g$. If $\mb{X}$ and $\mb{Y}$ are d-separated or m-separated given $\mb{Z}$ in $\g$, then $\mb{X}$ and $\mb{Y}$ are conditionally independent given $\mb{Z}$ in any \textit{observational density} that is Markov compatible with $\g$ \citep{lauritzen1990independence, zhang2008causal, henckel2022graphical}. 


\textbf{Causal Graphs.}
Let $\g$ be a graph with nodes $V_i$ and $V_j$. When $\g$ is an MPDAG, it is a \textit{causal MPDAG} if every edge $V_i \to V_j$ represents a direct causal effect of $V_i$ on $V_j$ and if every edge $V_i - V_j$ represents a direct causal effect of unknown direction (either $V_i$ affects $V_j$ or $V_j$ affects $V_i$). Note that all DAGs are MPDAGs.

When $\g$ is a MAG or PAG, it is a \textit{causal MAG} or \textit{causal PAG}, respectively, if every edge $V_i \to V_j$ represents the presence of a causal path from $V_i$ to $V_j$; every edge $V_i \arrowbullet V_j$ represents the absence of a causal path from $V_i$ to $V_j$; and every edge $V_i \circcirc V_j$ represents the presence of a causal path of unknown direction or a common cause in the underlying causal DAG.

\textbf{Causal and Non-causal Paths.}
Note that any directed or possibly directed path in a causal graph is \textit{causal} or \textit{possibly causal}, respectively. However, since we focus on causal graphs, we will use this causal terminology for paths in any of our graphs. We will say a path is \textit{non-causal} if it is not possibly causal.

\textbf{Consistency.}
Let $f(\mb{v})$ be an observational density over $\mb{V}$. The notation $do(\mb{X} = \mb{x})$, or $do(\mb{x})$ for short, represents an outside intervention that sets $\mb{X} \subseteq \mb{V}$ to fixed values $\mb{x}$. An \textit{interventional density} $f(\mb{v}|do(\mb{x}))$ is a density resulting from such an intervention.

Let $\mb{F^*}$ denote the set of all interventional densities $f(\mb{v}|do(\mb{x}))$ such that $\mb{X} \subseteq \mb{V}$ (including $\mb{X} = \emptyset$). A causal DAG $\g[D] = (\mb{V,E})$ is a \textit{causal Bayesian network compatible with} $\mb{F^*}$ if and only if for all $f(\mb{v} | do(\mb{x})) \in \mb{F^*}$, the following \textit{truncated factorization} holds:
\begin{align}
    f(\mb{v} | do(\mb{x})) = \prod_{V_i \in \mb{V} \setminus \mb{X}} f(v_i|\pa(v_i,\g[D])) \mathds{1}(\mb{X} = \mb{x}) \label{eq:trunc-fact}
\end{align}
\citep{pearl2009causality, bareinboim2012local}. We say an interventional density is \textit{consistent} with a causal DAG $\g[D]$ if it belongs to a set of interventional densities $\mb{F^*}$ such that $\g[D]$ is compatible with $\mb{F^*}$. Note that any observational density that is Markov compatible with $\g[D]$ is consistent with $\g[D]$. We say an interventional density is \textit{consistent} with a causal MPDAG, MAG, or PAG $\g$ if it is consistent with each DAG represented by $\g$ -- were the DAG to be causal.


\textbf{Identifiability.}
Let $\mb{X}$, $\mb{Y}$, and $\mb{Z}$ be pairwise disjoint node sets in a causal MPDAG or PAG $\g = (\mb{V,E})$, and let $\mb{F^*_i} = \{ f_i(\mb{v}|do(\mb{x'})) : \mb{X'} \subseteq \mb{V} \}$ be a set with which a DAG $\g[D]_i$ represented by $\g$ is compatible -- were $\g[D]_i$ to be causal. We say the conditional causal effect of $\mb{X}$ on $\mb{Y}$ given $\mb{Z}$ is \textit{identifiable} in $\g$ if for any $\mb{F^*_1}, \mb{F^*_2}$ where $f_1(\mb{v}) = f_2(\mb{v})$, we have $f_1(\mb{y}|do(\mb{x}), \mb{z}) = f_2(\mb{y}|do(\mb{x}), \mb{z})$ \citep{pearl2009causality}.

\textbf{Forbidden Set.}
Let $\mb{X}$ and $\mb{Y}$ be disjoint node sets in an MPDAG or PAG $\g$. Then the \textit{forbidden set} relative to $(\mb{X,Y})$ in $\g$ is
\begin{align}
    \text{Fo}&\text{rb}(\mb{X},\mb{Y},\g)
    =&\left\lbrace
    \begin{array}{@{}l@{}l@{}}
        \text{nodes in}\, \PossDe(W,\g) \text{, where} \notag\\
        \,\,\,\,\,\,\,\,\,\,W \in \PossMediators(\mb{X}, \mb{Y}, \g)
    \end{array}
    \right\rbrace .
\end{align}


\section{RESULTS - MPDAGS}
\label{sec:mpdags}

In this section, we present our results on identifying a conditional causal effect via our conditional adjustment criterion in the setting of an MPDAG (Definition \ref{def:cac}). Examples of how to use our criterion and explicit conditional adjustment sets based on our criterion follow these results. We remark here that our criterion shares similarities with the adjustment criterion for total effect identification of \cite{perkovic2017interpreting} and with the $\mb{Z}$-dependent dynamic adjustment criterion of \cite{smucler2020efficient}, but we save these results and reflections for Section \ref{sec:comparison}. 

Note that the results of this section hold when a fully oriented DAG is known, since all DAGs are MPDAGs. Throughout, our goal is to identify the conditional causal effect of treatments $\mb{X}$ on responses $\mb{Y}$ conditional on covariates $\mb{Z}$ and given a known graph $\g$.


\subsection{Conditional Adjustment Criterion}
\label{sec:cond-adj}

We include our definition of a conditional adjustment set below (Definition \ref{def:cas}). Note that, while this section focuses on MPDAGs, we write Definition \ref{def:cas} broadly for further use in Section \ref{sec:pags}. Our goal in this section is to find an equivalent graphical characterization of a conditional adjustment set. Theorem \ref{thm:cac-cas} establishes that Definition \ref{def:cac} provides such a graphical characterization, which we call the conditional adjustment criterion, under the assumption that the conditioning set does not contain variables affected by treatment ($\mb{Z} \cap \PossDe(\mb{X}, \g) = \emptyset$).

\begin{definition}
\label{def:cas}
{\normalfont (\textbf{Conditional Adjustment Set for MPDAGs, PAGs})}
    Let $\mb{X}$, $\mb{Y}$, $\mb{Z}$, and $\mb{S}$ be pairwise disjoint node sets in a causal MPDAG or PAG $\g$. Then $\mb{S}$ is a conditional adjustment set relative to $(\mb{X},\mb{Y},\mb{Z})$ in $\g$ if for any density $f$ consistent with $\g$ 
    \begin{align}
        f(\mb{y}|do(\mb{x}), \mb{z}) = 
        \begin{cases} 
            f(\mb{y} | \mb{x}, \mb{z}) & \mb{S}=\emptyset \\
            \int f(\mb{y} | \mb{x}, \mb{z}, \mb{s}) f(\mb{s}|\mb{z}) \diff \mb{s} & \mb{S}\neq\emptyset.
        \end{cases}
    \end{align}
\end{definition}

\begin{definition}
\label{def:cac}
{\normalfont (\textbf{Conditional Adjustment Criterion for MPDAGs})}
    Let $\mb{X}$, $\mb{Y}$, $\mb{Z}$, and $\mb{S}$ be pairwise disjoint node sets in an MPDAG $\g$, where $\mb{Z} \cap \PossDe(\mb{X},\g) = \emptyset$ and where every proper possibly causal path from $\mb{X}$ to $\mb{Y}$ in $\g$ starts with a directed edge. Then $\mb{S}$ satisfies the conditional adjustment criterion relative to $(\mb{X},\mb{Y},\mb{Z})$ in $\g$ if 
    \begin{enumerate}[label=\emph{(\alph*)}]
        \item $\mb{S} \cap \fb{\g} = \emptyset$, and \label{def:cac-a}
        \item $\mb{S} \cup \mb{Z}$ blocks all proper non-causal definite status paths from $\mb{X}$ to $\mb{Y}$ in $\g$. \label{def:cac-b}
    \end{enumerate}
\end{definition}

\begin{theorem}
\label{thm:cac-cas}
{\normalfont (\textbf{Completeness, Soundness of Conditional Adjustment Criterion for MPDAGs})}
    Let $\mb{X,Y}, \mb{Z}$, and $\mb{S}$ be pairwise disjoint node sets in a causal MPDAG $\g$, where $\mb{Z} \cap \PossDe(\mb{X},\g) = \emptyset$. Then $\mb{S}$ is a conditional adjustment set relative to $(\mb{X,Y,Z})$ in $\g$ (Definition \ref{def:cas}) if and only if $\mb{S}$ satisfies the conditional adjustment criterion relative to $(\mb{X,Y,Z})$ in $\g$ (Definition \ref{def:cac}).
\end{theorem}

\begin{proofof}[Theorem \ref{thm:cac-cas}]
    First note the following facts. 
    \begin{enumerate}[label=(\roman*)]
        \item Every proper possibly causal path from $\mb{X}$ to $\mb{Y}$ in $\g$ starts with a directed edge. \label{pf:cac-cas-1}
        \item $\mb{Z} \cap \De(\mb{X},\g[D]) = \emptyset$ in every DAG $\g[D]$ in $[\g]$. \label{pf:cac-cas-2} 
        \item $\mb{Z} \cap \fb{\g} = \emptyset$. \label{pf:cac-cas-3} 
    \end{enumerate}

    We have that \ref{pf:cac-cas-1} holds in either direction -- by definition ($\Leftarrow$) or by Proposition \ref{prop:id-necessary} (Supp.\ \ref{supp:necessary}) ($\Rightarrow$). Then Lemmas \ref{lem:equiv-z-mpdag} and \ref{lem:henckel-e6} (Supp.\ \ref{supp:existing}) imply \ref{pf:cac-cas-2} and \ref{pf:cac-cas-3}, respectively, given $\mb{Z} \cap \PossDe(\mb{X},\g) = \emptyset$ and \ref{pf:cac-cas-1}.
    
    Now consider the following statements.
    \begin{enumerate}[label=(\alph*)]
        \item $\mb{S}$ is a conditional adjustment set relative to $(\mb{X},\mb{Y},\mb{Z})$ in $\g$. \label{pf:cac-cas-a}

        \item $\mb{S}$ is a conditional adjustment set relative to $(\mb{X},\mb{Y},\mb{Z})$ in each DAG in $[\g]$ -- were the DAG to be causal.\label{pf:cac-cas-b}
        
        \item $\mb{S}$ satisfies the conditional adjustment criterion relative to $(\mb{X,Y,Z})$ in each DAG in $[\g]$. \label{pf:cac-cas-c}
        
        \item $\mb{S}$ satisfies the conditional adjustment criterion relative to $(\mb{X,Y,Z})$ in $\g$. \label{pf:cac-cas-d}
    \end{enumerate}

    By definition, \ref{pf:cac-cas-a} $\Leftrightarrow$ \ref{pf:cac-cas-b}. Then \ref{pf:cac-cas-b} $\Leftrightarrow$ \ref{pf:cac-cas-c} by Theorems \ref{thm:completeness} and \ref{thm:soundness} (Supp.\ \ref{supp:adjustment}) and the fact that the conditional adjustment criterion does not require a causal DAG. Lastly, by the facts above and by applying Lemmas \ref{lem:equiv-forb} and \ref{lem:equiv-block} (Supp.\ \ref{supp:existing}) in turn, \ref{pf:cac-cas-c} $\Leftrightarrow$ \ref{pf:cac-cas-d}. 
\end{proofof}


\subsection{Examples}
\label{sec:exs}

To illustrate the usefulness of the results above, we provide examples below where we aim to find $f(\mb{y}|do(\mb{x}),\mb{z})$ when $\mb{Z} \cap \PossDe(\mb{X},\g) = \emptyset$. Theorem \ref{thm:cac-cas} allows us to use the conditional adjustment criterion to (a) check whether a set can be used for conditional adjustment (Examples \ref{ex:cas-empty}-\ref{ex:cas-descendants}) or (b) determine if no such set exists (Example \ref{ex:nocas-noamen}).

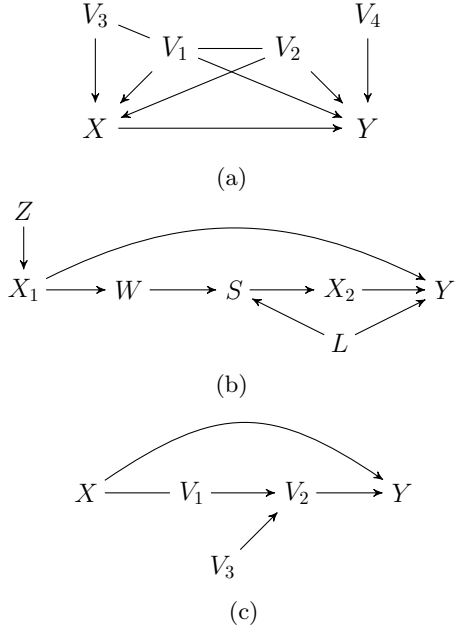
\begin{figure}
    \centering
    \begin{subfigure}{.45\textwidth}
        \vspace{1cm}
        \centering
        \begin{tikzpicture}[>=stealth',shorten >=1pt,auto,node distance=2cm,main node/.style={minimum size=0.8cm,font=\sffamily\Large\bfseries},scale=0.75,transform shape]
        \node[main node]   (X)                        {$X$};
        \node[main node]   (V1) [above right of = X]  {$V_{1}$};
        \node[main node]   (V2) [right of       = V1] {$V_2$};
        \node[main node]   (V3) [above of       = X]  {$V_3$};
        \node[main node]   (Y)  [below right of = V2] {$Y$};
        \node[main node]   (V4) [above of       = Y]  {$V_4$};
        \draw[->]   (V3) edge (X);
        \draw[-]    (V3) edge (V1);
        \draw[->]   (X)  edge (Y);
        \draw[-]    (V1) edge (V2);
        \draw[->]   (V2) edge (X);
        \draw[->]   (V2) edge (Y);
        \draw[->]   (V1) edge (X);
        \draw[->]   (V1) edge (Y);
         \draw[->]   (V4) edge (Y);
        \end{tikzpicture}
        \caption{}
        \label{fig:examples-a}
    \end{subfigure}
    \begin{subfigure}{.45\textwidth}
        \centering
     \begin{tikzpicture}[>=stealth',shorten >=1pt,auto,node distance=2cm,main node/.style={minimum size=0.5cm,font=\sffamily\Large\bfseries},scale=0.7,transform shape]
            \node[main node]         (X1)                   {$X_1$};
            \node[main node]         (Z)   at (0,1.5)        {$Z$};
            \node[main node]         (W)  [right of= X1]    {$W$};
            \node[main node]       	 (S)  [right of= W]     {$S$};
            \node[main node]         (X2) [right of= S]     {$X_2$};
            \node[main node]         (Y)  [right of= X2]    {$Y$};
            \node[main node]         (U)  at (6,-1)		    {$L$};
            \draw[->] (Z)   edge    (X1);
            \draw[->] (X1)  edge    (W);
            \draw[->] (W)   edge    (S);
            \draw[->] (S)   edge    (X2);
            \draw[->] (X2)  edge    (Y);
            \draw[->] (U)   edge    (S);
            \draw[->] (U)   edge    (Y);
            \draw[->] (X1) ..  controls (3,1.5) and (5,1.5)   ..     (Y);
            \end{tikzpicture}
        \caption{}
        \label{fig:examples-b}
    \end{subfigure}
    \begin{subfigure}{.5\textwidth}
        \centering
        \begin{tikzpicture}[>=stealth',shorten >=1pt,auto,node distance=2cm,main node/.style={minimum size=0.5cm,font=\sffamily\Large\bfseries},scale=.7,transform shape]
            \node[main node]    (X1)                            {$X$};
            \node[main node]    (V1)    [right of      = X1]    {$V_1$};
            \node[main node]    (V2)    [right of      = V1]    {$V_2$};
            \node[main node]    (V3)    [below left of = V2]    {$V_3$};
            \node[main node]    (Y1)    [right of      = V2]    {$Y$};
            \draw[-]    (X1)  edge    (V1);
            \draw[->]   (V1)  edge    (V2);
            \draw[->]   (V2)  edge    (Y1);
            \draw[->]   (V3)  edge    (V2);
            \draw[->] (X1) ..  controls (2.5,1.7) and (3.5,1.7) ..   (Y1);
        \end{tikzpicture}
        \caption{}
        \label{fig:examples-c}
    \end{subfigure}
    \caption{\small Causal MPDAGs used in Examples \ref{ex:cas-empty}-\ref{ex:nocas-noamen}.}
\end{figure}


\begin{example} 
\label{ex:cas-empty}
{\normalfont (\textbf{Empty Conditional Adjustment Set.})}
    Let $\g$ be the causal MPDAG in Figure \ref{fig:examples-a} \footnote{Compare to Figure 5(a) of \cite{perkovic2020identifying}.}, and let $\mb{X} = \{X\}$, $\mb{Y}=\{Y\}$, and $\mb{Z}=\{V_1, V_2\}$. Note that $\mb{Z} \cap \PossDe(\mb{X}, \g) = \emptyset$ and that every possibly causal path from $\mb{X}$ to $\mb{Y}$ in $\g$ starts with a directed edge.

    Let $\mb{S} = \emptyset$. Note that $\mb{S} \cap (\mb{X} \cup \mb{Y} \cup \mb{Z}) = \emptyset$, $\mb{S} \cap \fb{\g} = \emptyset$, and $\mb{S} \cup \mb{Z}$ blocks all non-causal definite status paths from $\mb{X}$ to $\mb{Y}$. Thus, $\mb{S}$ satisfies the conditional adjustment criterion relative to $(\mb{X},\mb{Y},\mb{Z})$ in $\g$, and by Theorem \ref{thm:cac-cas}, $f(\mb{y}|do(\mb{x}), \mb{z}) = f(y | x, v_1, v_2)$.
\end{example}


\begin{example}
\label{ex:cas-nonempty}
{\normalfont (\textbf{Only Nonempty Conditional Adjustment Sets.})}
    Again let $\g$ be the causal MPDAG in Figure \ref{fig:examples-a}, where $\mb{X} = \{X\}$ and $\mb{Y}=\{Y\}$. But now let $\mb{Z}=\{V_1\}$. We still have that $\mb{Z} \cap \PossDe(\mb{X}, \g) = \emptyset$ and that every possibly causal path from $\mb{X}$ to $\mb{Y}$ in $\g$ starts with a directed edge.

    Note that if we let $\mb{S} = \emptyset$, $\mb{S} \cup \mb{Z}$ does not block the path $X \gets V_2 \to Y$, which is a proper non-causal definite status path from $\mb{X}$ to $\mb{Y}$. Thus, the empty set is not a conditional adjustment set relative to $(\mb{X,Y,Z})$ in $\g$.

    Consider, instead, the set $\mb{S} = \{V_2\}$. Note that $\mb{S} \cap (\mb{X} \cup \mb{Y} \cup \mb{Z}) = \emptyset$, $\mb{S} \cap \fb{\g} = \emptyset$, and $\mb{S} \cup \mb{Z}$ blocks all non-causal definite status paths from $\mb{X}$ to $\mb{Y}$. Thus, $\mb{S}$ satisfies the conditional adjustment criterion relative to $(\mb{X,Y,Z})$ in $\g$, and by Theorem \ref{thm:cac-cas}, $f(\mb{y}|do(\mb{x}), \mb{z}) = \int f(y | x, v_1, v_2) f(v_2|v_1) \diff v_2$.
\end{example}


\begin{example}
\label{ex:cas-descendants}
{\normalfont (\textbf{Conditional Adjustment Set Contains Descendants of $\mb{X}$.})}
    Let $\g$ be the causal DAG (and therefore, MPDAG) in Figure \ref{fig:examples-b} \footnote{Compare to Figure 6(a) of \cite{perkovic2018complete}.}, where we assume $L$ is a variable that cannot be measured. Define $\mb{X} = \{X_1,X_2\}$, $\mb{Y}=\{Y\}$, and $\mb{Z} = \{Z\}$. Note that $\mb{Z} \cap \De(\mb{X}, \g) = \emptyset$.
    
    Consider the set $\mb{S} = \{S,W\}$. Note that $\mb{S} \cap (\mb{X} \cup \mb{Y} \cup \mb{Z}) = \emptyset$, $\mb{S} \cap \fb{\g} = \emptyset$, and $\mb{S}$ blocks all proper non-causal paths from $\mb{X}$ to $\mb{Y}$ in $\g$. Hence, $\mb{S}$ satisfies the conditional adjustment criterion relative to $(\mb{X,Y,Z})$ in $\g$, and by Theorem \ref{thm:cac-cas}, $f(\mb{y}|do(\mb{x}), \mb{z}) = \int f(y | x_1,x_2,z,s,w) f(s,w|z) \diff s \diff w$.
\end{example}


\begin{example}
\label{ex:nocas-noamen}
{\normalfont (\textbf{No Conditional Adjustment Set, Effect Non-identifiable}.)}
    Let $\g$ be the causal MPDAG in Figure \ref{fig:examples-c}, and let $\mb{X} = \{X\}$, $\mb{Y}=\{Y\}$, and $\mb{Z}=\{V_3\}$. Note that $\mb{Z} \cap \PossDe(\mb{X}, \g) = \emptyset$. However, $X - V_1 \to V_2 \to Y$ is a proper possibly causal path from $\mb{X}$ to $\mb{Y}$ in $\g$ that starts with an undirected edge. Thus, by Theorem \ref{thm:cac-cas}, there can be no conditional adjustment set relative to $(\mb{X,Y,Z})$ in $\g$. In fact, by Proposition \ref{prop:id-necessary} (Supp.\ \ref{supp:necessary}), $f(\mb{y}| do(\mb{x}), \mb{z})$ is not identifiable in $\g$ using any method. 
\end{example}


\subsection{Constructing Adjustment Sets}
\label{sec:constructing}

The conditional adjustment criterion provides a way to check if a set can be used for conditional adjustment given an MPDAG $\g$, but it does not provide a way to construct a conditional adjustment set -- a task that may be difficult when $\g$ is large. The results in this section provide such a roadmap under certain assumptions. The proofs can be found in Supp.\ \ref{supp:constructing}.

\begin{lemma}
\label{lem:parent-set}
    Let $\mb{X} =\{X\}$, $\mb{Y}$, and $\mb{Z}$ be pairwise disjoint node sets in a causal MPDAG $\g$, where $\mb{Z} \cap \PossDe(X, \g) = \emptyset$ and where every possibly causal path from $X$ to $\mb{Y}$ in $\g$ starts with a directed edge. If $\mb{Y} \cap \Pa(X, \g) = \emptyset$, then the following is a conditional adjustment set relative to $(\mb{X,Y,Z})$ in $\g$:
    \begin{align}
        \Pa(X, \g) \setminus \mb{Z}. \label{eq:parent-set}
    \end{align}
\end{lemma}

\begin{theorem}
\label{thm:adjust-and-o} 
    Let $\mb{X}$, $\mb{Y}$, and $\mb{Z}$ be pairwise disjoint node sets in a causal MPDAG $\g$, where $\mb{Z} \cap \PossDe(\mb{X}, \g) = \emptyset$ and where every proper possibly causal path from $\mb{X}$ to $\mb{Y}$ in $\g$ starts with a directed edge. 
    \begin{enumerate}[label = (\alph*)]
        \item If there is any conditional adjustment set relative to $(\mb{X,Y,Z})$ in $\g$, then the following set is one:
        \begin{align}
            \Adjust(&\mb{X},\mb{Y},\mb{Z},\g) \label{eq:adjust-mpdag}\\
                &= \big[ \PossAn(\mb{X \cup Y},\g) \cup \An(\mb{Z}, \g) \big]  \nonumber \\
                &  \hspace{.25in} \setminus \big[ \fb{\g} \cup \mb{X} \cup \mb{Y} \cup \mb{Z} \big]. \nonumber
        \end{align}
        
        \item Suppose $\mb{Y} \subseteq \PossDe(\mb{X}, \g)$. If there is any conditional adjustment set relative to $(\mb{X,Y,Z})$ in $\g$, then the following set is one:
        \begin{align}\label{eq:o-set}
            \optb{\g} &= \emph{Pa}\Big(\possmediatb{\g},\g \Big)  \\
            &\hspace{.25in} \setminus \Big[ \fb{\g} \cup \mb{X} \cup \mb{Y} \cup \mb{Z} \Big]. \nonumber
        \end{align}
    \end{enumerate}
\end{theorem}

\begin{example}
\label{ex:construct}
    Consider again the causal MPDAG $\g$ in Figure \ref{fig:examples-a}, where $\mb{X} = \{X\}$, $\mb{Y} = \{Y\}$, and $\mb{Z} = \{V_1\}$. Note that the conditions of Lemma \ref{lem:parent-set} and Theorem \ref{thm:adjust-and-o} are met, so we can construct three valid conditional adjustment sets using Equations \eqref{eq:parent-set}, \eqref{eq:adjust-mpdag}, and \eqref{eq:o-set}.
    \begingroup
    \allowdisplaybreaks
    \begin{align}
        \Pa(X, \g) \setminus \mb{Z} &= \{V_1,V_2,V_3\} \setminus \{V_1\} \notag\\
        &= \{V_2,V_3\}. \nonumber\\
        \Adjust(\mb{X,Y,Z},\g) &= \{X,Y, V_1,V_2,V_3,V_4\} \setminus \{X,Y,V_1\} \nonumber\\
        &= \{V_2,V_3,V_4\}. \nonumber\\
        \opt{\g} &= \{X,V_1,V_2,V_4\} \setminus \{X,Y,V_1\} \nonumber\\
        &= \{V_2,V_4\}. \nonumber
    \end{align}
    \endgroup
\end{example}


\subsection{Comparison of Contexts} 
\label{sec:comparison}

In this section, we point out a bridge between our conditional adjustment results and prior literature on unconditional adjustment and adjustment under \textit{dynamic treatment}. We begin by presenting Lemma \ref{lem:comparison}, which provides an equivalence between our criterion and the criterion of \cite{perkovic2017interpreting} used for unconditional adjustment given an MPDAG. Note that this lemma is used to prove Theorem \ref{thm:cac-cas} (see Figure \ref{fig:proof-map} in Supp.\ \ref{supp:adjustment}). See Supp.\ \ref{supp:adjustment} for the lemma's proof.


\begin{lemma}
\label{lem:comparison}
    Let $\mb{X}$, $\mb{Y}$, $\mb{Z}$, and $\mb{S}$ be pairwise disjoint node sets in an MPDAG $\g$, where $\mb{Z} \cap \PossDe(\mb{X}, \g) = \emptyset$. Then we have the following.
    \begin{enumerate}[label=\emph{(\alph*)}]
        \item \textbf{Comparison of Adjustment Criteria:}\\ 
        $\mb{S}$ satisfies the conditional adjustment criterion relative to $(\mb{X,Y,Z})$ in $\g$ (Definition \ref{def:cac}) if and only if $\mb{S} \cup \mb{Z}$ satisfies the adjustment criterion relative to $(\mb{X}, \mb{Y})$ in $\g$ (Definition \ref{def:ac}, Supp.\ \ref{supp:defs}). \label{lem:comparison-a}
        
        \item \textbf{Comparison of Adjustment Sets:}\\
        $\mb{S}$ is a conditional adjustment set relative to $(\mb{X,Y,Z})$ in $\g$ (Definition \ref{def:cas}) if and only if $\mb{S} \cup \mb{Z}$ is an adjustment set relative to $(\mb{X}, \mb{Y})$ in $\g$ (Definition \ref{def:as}, Supp.\ \ref{supp:defs}). \label{lem:comparison-b}
    \end{enumerate}
\end{lemma}

Next we turn to the work of \cite{smucler2020efficient}, where the authors consider causal effect estimation under a dynamic treatment. For this purpose, \cite{smucler2020efficient} define a dynamic adjustment set, which they then relate to the set used by \cite{maathuis2015generalized} for unconditional adjustment (Definition \ref{def:as}, Supp.\ \ref{supp:defs}). Lemma \ref{lem:comparison} allows us to connect this dynamic adjustment to our work.

Before making this connection, we briefly describe the context of these authors' work. Unlike a do-intervention that sets $\mb{X}$ to fixed values $\mb{x}$, a dynamic intervention sets $\mb{X}$ to values $\mb{x}$ with probability $\pi(\mb{x} | \mb{Z} = \mb{z})$. However, a do-intervention can be seen as a special case of a dynamic intervention where $\pi(\mb{x} |\mb{Z} = \mb{z}) = \mathds{1}(\mb{X} =\mb{x})$. Dynamic interventions are often of interest in personalized medicine \citep{robins1993analytic, murphy2001marginal, chakraborty2013statistical}.

\cite{smucler2020efficient} refer to a causal effect under a dynamic intervention, whose assignment probability depends on $\mb{Z}$, as a $\boldsymbol{\mathit{Z}}$-\textit{dependent dynamic causal effect} (also called a \textit{single stage dynamic treatment effect} in \cite{chakraborty2013statistical}). They consider these causal effects in the setting where $\mb{X}$ and $\mb{Y}$ are nodes, the given graph $\g$ is a DAG, and the following assumption holds: $\mb{Z} \cap \De(X,\g) = \emptyset$. They then define a $\boldsymbol{\mathit{Z}}$-\textit{dependent dynamic adjustment set} as a set $\mb{S}$ that satisfies
\begin{align*} 
    f(y|\pi(x | \mb{z})) = 
    \begin{cases} 
          \pi(x|\mb{z})f(y | x, \mb{z}) & \mb{S}=\emptyset, \\
          \pi(x|\mb{z}) \int f(y | x, \mb{z}, \mb{s}) f(\mb{s}|\mb{z}) \diff \mb{s} & \mb{S}\neq\emptyset.
    \end{cases}
    \end{align*}

To compare these sets to our conditional adjustment sets, we reference Proposition 1 of \cite{smucler2020efficient}. This result states that, under their assumptions, $\mb{S}\cup \mb{Z}$ is a $\mb{Z}$-dependent dynamic adjustment set if and only if $\mb{S} \cup \mb{Z}$ is an adjustment set relative to $(X,Y)$ in $\g$ (Definition \ref{def:as}, Supp.\ \ref{supp:defs}). It follows from Lemma \ref{lem:comparison} that $\mb{S} \cup \mb{Z}$ is a $\mb{Z}$-dependent dynamic adjustment set if and only if $\mb{S}$ is a conditional adjustment set relative to $(X,Y,\mb{Z})$ in $\g$ -- when $\g$ is a DAG such that $\mb{Z} \cap \De(X,\g) = \emptyset$. Thus, our results can be seen as generalizations of \cite{smucler2020efficient} for $|\mb{X}| >1$ and, therefore, can be used for $\mb{Z}$-dependent dynamic causal effect identification.


\section{RESULTS - PAGS}
\label{sec:pags}
We now extend our results on conditional adjustment to the setting of a PAG.


\subsection{Conditional Adjustment Criterion}
\label{sec:cond-adj-pag}

We first introduce our conditional adjustment criterion for PAGs (Definition \ref{def:cac-pag}). Note that the difference between this criterion and the analogous criterion for MPDAGs is the use of a \textit{visible} as opposed to a \textit{directed} edge. Visibility is a stronger condition introduced by \cite{zhang2008causal} (see Supp.\ \ref{supp:defs} for definition).

Following this, Lemma \ref{lem:comparison-pag} provides an equivalence between our criterion and the criterion of \cite{perkovic2018complete} used for unconditional adjustment given a PAG. Theorem \ref{thm:cac-cas-pag} is our main result in this section. It establishes that, under restrictions on $\mb{Z}$, the conditional adjustment criterion is an equivalent graphical characterization of a conditional adjustment set in causal PAGs. Proofs of these results are given in Supp.\ \ref{supp:comparison-pag}.

\begin{definition}
\label{def:cac-pag}
{\normalfont (\textbf{Conditional Adjustment Criterion for PAGs})} 
    Let $\mb{X}$, $\mb{Y}$, $\mb{Z}$, and $\mb{S}$ be pairwise disjoint node sets in a PAG $\g$, where $\mb{Z} \cap \PossDe(\mb{X},\g) = \emptyset$ and where every proper possibly causal path from $\mb{X}$ to $\mb{Y}$ in $\g$ starts with a \textbf{visible edge} out of $\mb{X}$. Then $\mb{S}$ satisfies the conditional adjustment criterion relative to $(\mb{X},\mb{Y},\mb{Z})$ in $\g$ if 
    \begin{enumerate}[label=\emph{(\alph*)}]
        \item \label{def:cac-pag-a} $\mb{S} \cap \fb{\g} = \emptyset$, and
        \item $\mb{S} \cup \mb{Z}$ blocks all proper non-causal definite status paths from $\mb{X}$ to $\mb{Y}$ in $\g$.
    \end{enumerate}
\end{definition}



\begin{lemma}
\label{lem:comparison-pag}
    Let $\mb{X}$, $\mb{Y}$, $\mb{Z}$, and $\mb{S}$ be pairwise disjoint node sets in a PAG $\g$, where $\mb{Z} \cap \PossDe(\mb{X}, \g) = \emptyset$. Then we have the following.
    \begin{enumerate}[label=\emph{(\alph*)}]
        \item \textbf{Comparison of Adjustment Criteria:}\\
        $\mb{S}$ satisfies the conditional adjustment criterion relative to $(\mb{X,Y,Z})$ in $\g$ (Definition \ref{def:cac-pag}) if and only if $\mb{S} \cup \mb{Z}$ satisfies the adjustment criterion relative to $(\mb{X}, \mb{Y})$ in $\g$ (Definition \ref{def:ac}, Supp.\ \ref{supp:defs}). \label{lem:comparison-pag-a}
        \item \textbf{Comparison of Adjustment Sets:}\\
        $\mb{S}$ is a conditional adjustment set relative to $(\mb{X,Y,Z})$ in $\g$ (Definition \ref{def:cas}) if and only if $\mb{S} \cup \mb{Z}$ is an adjustment set relative to $(\mb{X}, \mb{Y})$ in $\g$ (Definition \ref{def:as}, Supp.\ \ref{supp:defs}). \label{lem:comparison-pag-b}
    \end{enumerate}
\end{lemma}

\begin{proofof}[Lemma \ref{lem:comparison-pag}] 
    \ref{lem:comparison-pag-a} Follows from the fact that $\fb{\g} \subseteq \PossDe(\mb{X}, \g)$. 

    \ref{lem:comparison-pag-b} We start by noting the following fact. Since $\mb{Z} \cap \PossDe(\mb{X}, \g) = \emptyset$, then $\mb{Z} \cap \De(\mb{X},\g[D]) = \emptyset$ in every DAG represented by $\g$ (Lemma \ref{lem:equiv-z-pag}, Supp.\ \ref{supp:comparison-pag}).
    Then consider the following statements.
    \begin{enumerate}[label=(\alph*)]
    	\item $\mb{S}$ is a conditional adjustment set relative to $(\mb{X},\mb{Y},\mb{Z})$ in $\g$. \label{pf:cas-as-pag-a}
    
     	\item $\mb{S}$ is a conditional adjustment set relative to $(\mb{X},\mb{Y},\mb{Z})$ in each DAG represented by $\g$ -- were the DAG to be causal.\label{pf:cas-as-pag-b}
    
      	\item $\mb{S} \cup \mb{Z}$ is an adjustment set relative to $(\mb{X},\mb{Y})$ in each DAG represented by $\g$ -- were the DAG to be causal. \label{pf:cas-as-pag-c}
    
      	\item $\mb{S} \cup \mb{Z}$ is an adjustment set relative to $(\mb{X},\mb{Y})$ in $\g$. \label{pf:cas-as-pag-d}
    \end{enumerate}

    By definition, \ref{pf:cas-as-pag-a} $\Leftrightarrow$ \ref{pf:cas-as-pag-b}. Then by Lemma \ref{lem:comparison}\ref{lem:comparison-b} and the fact above, we have \ref{pf:cas-as-pag-b} $\Leftrightarrow$ \ref{pf:cas-as-pag-c}. The statement \ref{pf:cas-as-pag-c} $\Leftrightarrow$ \ref{pf:cas-as-pag-d} follows again by definition.
\end{proofof}


\begin{theorem}
\label{thm:cac-cas-pag}
{\normalfont (\textbf{Completeness, Soundness of Conditional Adjustment Criterion for PAGs})}
    Let $\mb{X,Y}, \mb{Z}$, and $\mb{S}$ be pairwise disjoint node sets in a causal PAG $\g$, where $\mb{Z} \cap \PossDe(\mb{X},\g) = \emptyset$. Then $\mb{S}$ is a conditional adjustment set relative to $(\mb{X,Y,Z})$ in $\g$ (Definition \ref{def:cas}) if and only if $\mb{S}$ satisfies the conditional adjustment criterion relative to $(\mb{X,Y,Z})$ in $\g$ (Definition \ref{def:cac-pag}).
\end{theorem}


\subsection{Constructing Adjustment Sets}
\label{sec:constructing-pag}

We now provide a method for constructing conditional adjustment sets given a causal PAG (Theorem \ref{thm:adjust-and-o-pag}). We illustrate this result in Example \ref{ex:construct-pag}. The proof of Theorem \ref{thm:adjust-and-o-pag} can be found in Supp.\ \ref{supp:constructing-pag}.

\begin{theorem}
\label{thm:adjust-and-o-pag}  
    Let $\mb{X}$, $\mb{Y}$, and $\mb{Z}$ be pairwise disjoint node sets in a causal PAG $\g$, where $\mb{Z} \cap \PossDe(\mb{X}, \g) = \emptyset$ and where every proper possibly causal path from $\mb{X}$ to $\mb{Y}$ in $\g$ starts with a visible edge out of $\mb{X}$. If there is any conditional adjustment set relative to $(\mb{X,Y,Z})$ in $\g$, then the following set is one:
    \begin{align}\label{eq:adjust-pag}
        \Adjust&(\mb{X},\mb{Y},\mb{Z},\g) \\
            &= \big[ \PossAn(\mb{X \cup Y}, \g) \cup \PossAn(\mb{Z}, \g) \big] \nonumber\\
            & \hspace{.25in} \setminus \Big[ \fb{\g} \cup \mb{X} \cup \mb{Y} \cup \mb{Z} \Big]. \nonumber
    \end{align}
\end{theorem}

\begin{figure}
        \vspace{1cm}
        \centering
        \begin{tikzpicture}[>=stealth',shorten >=1pt,auto,node distance=2cm,main node/.style={minimum size=0.8cm,font=\sffamily\Large\bfseries},scale=0.75,transform shape]
        \node[main node]   (X)                        {$X$};
        \node[main node]   (V1) [above right of = X]  {$V_{1}$};
        \node[main node]   (V2) [right of       = V1] {$V_2$};
        \node[main node]   (V3) [above of       = X]  {$V_3$};
        \node[main node]   (V5) [left of    = V3]     {$V_5$};
        \node[main node]   (Y)  [below right of = V2] {$Y$};
        \node[main node]   (V4) [above of       = Y]  {$V_4$};
        \draw[<->]   (V3) edge (X);
        \draw[o->]   (V5) edge (V3);
        \draw[<-o]   (V3) edge (V1);
        \draw[->]    (X)  edge (Y);
        \draw[o-o]   (V1) edge (V2);
        \draw[o->]   (V2) edge (X);
        \draw[->]    (V2) edge (Y);
        \draw[o->]   (V1) edge (X);
        \draw[->]    (V1) edge (Y);
        \draw[o->]   (V4) edge (Y);
        \end{tikzpicture}
        \caption{A causal PAG used in Example \ref{ex:construct-pag}.}
    \label{fig:example-pag}
\end{figure}
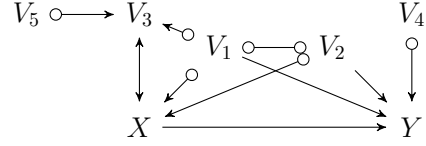

\begin{example}
\label{ex:construct-pag}
    Let $\g$ be the causal PAG in Figure \ref{fig:example-pag}, and let $\mb{X} = \{X\}$, $\mb{Y}=\{Y\}$, and $\mb{Z}=\{V_1\}$. Note that $\mb{Z} \cap \PossDe(\mb{X}, \g) = \emptyset$. Furthermore, the only possibly causal path from $\mb{X}$ to $\mb{Y}$ is the edge $X \to Y$, which is visible due to the presence of $V_3 \leftrightarrow X$, where $V_3 \notin \Adj(Y,\g)$. If there is any conditional adjustment set relative to $(\mb{X,Y,Z})$ in $\g$, then the conditions of Theorem \ref{thm:adjust-and-o-pag} are met. We consider the set from Equation \eqref{eq:adjust-pag}.
    \begin{align}
        \Adjust(\mb{X},\mb{Y},\mb{Z},\g) 
            &= \{X,Y,V_1,V_2,V_4\} \setminus \{X,Y,V_1\} \nonumber\\
            &= \{V_2,V_4\}. \nonumber
    \end{align}
    To see that this is a conditional adjustment set relative to $(\mb{X,Y,Z})$ in $\g$, we note that it fulfills the requirements of Definition \ref{def:cac-pag}. That is, $\adjustb{\g} \cap \fb{\g} = \emptyset$ and $\adjustb{\g} \cup \mb{Z} =\{V_1,V_2, V_4\}$ blocks all proper non-causal definite status paths from $\mb{X}$ to $\mb{Y}$ in $\g$.
\end{example}


\section{DISCUSSION}
\label{sec:discussion}

This paper defines a conditional adjustment set that can be used to identify a causal effect in a setting where a causal MPDAG or PAG is known (Definition \ref{def:cas}). We give necessary and sufficient graphical conditions for identifying such a set when $\mb{Z} \cap \PossDe(\mb{X},\g) =\emptyset$ (Theorems \ref{thm:cac-cas} and \ref{thm:cac-cas-pag}). Further, we provide multiple methods for constructing these sets (Sections \ref{sec:constructing} and \ref{sec:constructing-pag}). While our results can be used to identify a broad class of conditional causal effects, we discuss some limitations below. 

One such limitation is that there are conditional causal effects that can be identified but cannot be identified using conditional adjustment sets. As an example, consider the causal DAG (and therefore, MPDAG) $\g$ in Figure \ref{fig:discussion}, and let $\mb{X} = \{X_1, X_2\}$, $\mb{Y}=\{Y\}$, and $\mb{Z}=\{V_2\}$. Note that the conditional causal effect of $\mb{X}$ on $\mb{Y}$ given $\mb{Z}$ is identifiable using do calculus rules (\citealp{pearl2009causality}, see Equations \eqref{eq:rule1-do}-\eqref{eq:rule3-do} in Supp.\ \ref{supp:existing}):
\begingroup
\allowdisplaybreaks
\begin{align}
    f&(\mb{y}|do(\mb{x}),\mb{z}) \nonumber\\
            &= \int_{v_1} f(y,v_1|do(\mb{x}),v_2) \diff v_1 \nonumber \\
            &= \int_{v_1} f(y|do(\mb{x}),v_1,v_2) f(v_1|do(\mb{x}),v_2) \diff v_1 \nonumber \\
            &= \int_{v_1} f(y|do(\mb{x}),v_1,v_2) f(v_1|do(\mb{x})) \diff v_1 \label{eq:discussion-1-a} \\
            &= \int_{v_1} f(y|do(x_2),v_1,v_2) f(v_1|do(x_1)) \diff v_1 \label{eq:discussion-1-b} \\
            &= \int_{v_1} f(y|x_2,v_1,v_2) f(v_1|x_1) \diff v_1. \label{eq:discussion-1-c}
\end{align}
The first two equalities follow from basic probability rules. Equation \eqref{eq:discussion-1-a} follows from Rule 1 of the do calculus, since $V_1 \dsepp V_2 | X_1,X_2$ in $\g_{\overline{\{X_1,X_2\}}}$. Equation \eqref{eq:discussion-1-b} follows from Rule 3 of the do calculus, since $Y \dsepp X_1 | V_1,V_2,X_2$ in $\g_{\overline{X_2}}$ and $V_1 \dsepp X_2 | X_1$ in $\g_{\overline{\{X_1,X_2\}}}$. Equation \eqref{eq:discussion-1-c} follows from Rule 2 of the do calculus, since $Y \dsepp X_2 | V_1,V_2$ in $\g_{\underline{X_2}}$ and $V_1 \dsepp X_1$ in $\g_{\underline{X_1}}$.

However, we can show that there is no conditional adjustment set relative to $(\mb{X},\mb{Y},\mb{Z})$ in $\g$ that could have been used to identify the effect above. To see this, note that since $\mb{Z} \cap \PossDe(\mb{X}, \g) = \emptyset$, we can use Theorem \ref{thm:cac-cas} to state the following. A set $\mb{S}$ must satisfy the conditional adjustment criterion relative to $(\mb{X},\mb{Y},\mb{Z})$ in $\g$ (Definition \ref{def:cac}) in order to be a conditional adjustment set. Definition \ref{def:cac} requires that $\mb{S}$ block the path $X_2 \gets V_1 \to Y$, since it is a proper non-causal definite status path from $\mb{X}$ to $\mb{Y}$. It follows that $\mb{S}$ must contain $V_1 \in \fb{\g}$, but this contradicts Definition \ref{def:cac}'s requirement that $\mb{S} \cap \fb{\g} = \emptyset$. 

Adding to the limitation above, there are conditional causal effects that can be identified using conditional adjustment sets but where these conditional adjustment sets cannot be identified using our criterion. This can occur when $\mb{Z} \cap \PossDe(\mb{X},\g) \neq \emptyset$, since our graphical criterion requires this restriction but our conditional adjustment set definition does not. As an example, consider again the causal DAG $\g$ given in Figure \ref{fig:examples-b}, and let $\mb{X} =\{X_1,X_2\}$, $\mb{Y}=\{Y\}$, and $\mb{Z}=\{Z, W\}$. Since $\mb{Z} \cap \PossDe(\mb{X},\g) \neq \emptyset$, no set satisfies the conditional adjustment criterion. However, using do calculus rules \citep{pearl2009causality}, we can show that $\mb{S}=\{S\}$ is a conditional adjustment set relative to $(\mb{X,Y,Z})$ in $\g$:
\begin{align}
    f(\mb{y}|do(\mb{x}),\mb{z}) 
            &= \int_\mb{s} f(\mb{y},\mb{s}|do(\mb{x}),\mb{z}) \diff \mb{s} \nonumber \\
            &= \int_\mb{s} f(\mb{y}|do(\mb{x}),\mb{z},\mb{s}) f(\mb{s}|do(\mb{x}),\mb{z}) \diff \mb{s} \nonumber \\
            &= \int_\mb{s} f(\mb{y}|\mb{x},\mb{z},\mb{s}) f(\mb{s}|\mb{z}) \diff \mb{s}. \nonumber
\end{align}
The first and second equality follow from basic probability rules. The third follows by Rules 2 and 3 of the do calculus, since $\mb{Y} \dsepp \mb{X} \given \mb{Z} \cup \mb{S}$ in $\g_{\underline{\mb{X}}}$ and $\mb{S} \dsepp \mb{X} \given \mb{Z}$ in $\g_{\overline{\mb{X}(\mb{Z})}}$. Future work could address identification in this setting by expanding our graphical criterion to allow for arbitrary conditioning.

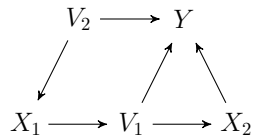
\begin{figure}
    \centering
    \begin{tikzpicture}[>=stealth',shorten >=1pt,auto,node distance=2cm,main node/.style={minimum size=0.8cm,font=\sffamily\Large\bfseries},scale=.7,transform shape]
    \node[main node]   (X1)                  {$X_1$};
    \node[main node]   (V1) [right of = X1]  {$V_{1}$};
    \node[main node]   (Y)  at (3,2)         {$Y$};
    \node[main node]   (X2) [right of = V1]  {$X_{2}$};
    \node[main node]   (V2) at (1,2)         {$V_{2}$};
    \draw[->]   (X1) edge (V1);
    \draw[->]   (V1) edge (Y);
    \draw[->]   (V1) edge (X2);
    \draw[->]   (X2) edge (Y);
    \draw[->]   (V2) edge (X1);
    \draw[->]   (V2) edge (Y);
    \end{tikzpicture}
    \caption{A causal DAG used in Section \ref{sec:discussion}.}
    \label{fig:discussion}
\end{figure}


\subsubsection*{Acknowledgements}
This material is based upon work supported by the National Science Foundation under Grant No. 2210210.


\bibliographystyle{abbrvnat}
\bibliography{main}


\section*{Checklist}

\begin{enumerate}
    \item For all models and algorithms presented, check if you include:
    \begin{enumerate}
        \item A clear description of the mathematical setting, assumptions, algorithm, and/or model. [Yes/No/Not Applicable] \textbf{Not Applicable}
        \item An analysis of the properties and complexity (time, space, sample size) of any algorithm. [Yes/No/Not Applicable]  \textbf{Not Applicable}
        \item (Optional) Anonymized source code, with specification of all dependencies, including external libraries. [Yes/No/Not Applicable] \textbf{Not Applicable}
    \end{enumerate}
    \item For any theoretical claim, check if you include:
    \begin{enumerate}
        \item Statements of the full set of assumptions of all theoretical results. [Yes/No/Not Applicable] \textbf{Yes}
        \item Complete proofs of all theoretical results. [Yes/No/Not Applicable] \textbf{Yes}
        \item Clear explanations of any assumptions. [Yes/No/Not Applicable]      \textbf{Yes}
    \end{enumerate}
    \item For all figures and tables that present empirical results, check if you include:
    \begin{enumerate}
        \item The code, data, and instructions needed to reproduce the main experimental results (either in the Supplemental material or as a URL). [Yes/No/Not Applicable] \textbf{Not Applicable}
        \item All the training details (e.g., data splits, hyperparameters, how they were chosen). [Yes/No/Not Applicable] \textbf{Not Applicable}
            \item A clear definition of the specific measure or statistics and error bars (e.g., with respect to the random seed after running experiments multiple times). [Yes/No/Not Applicable] \textbf{Not Applicable}
            \item A description of the computing infrastructure used. (e.g., type of GPUs, internal cluster, or cloud provider). [Yes/No/Not Applicable] \textbf{Not Applicable}
    \end{enumerate}
    \item If you are using existing assets (e.g., code, data, models) or curating/releasing new assets, check if you include:
    \begin{enumerate}
        \item Citations of the creator If your work uses existing assets. [Yes/No/Not Applicable] \textbf{Not Applicable}
        \item The license information of the assets, if applicable. [Yes/No/Not Applicable] \textbf{Not Applicable}
        \item New assets either in the Supplemental material or as a URL, if applicable. [Yes/No/Not Applicable] \textbf{Not Applicable}
        \item Information about consent from data providers/curators. [Yes/No/Not Applicable] \textbf{Not Applicable}
        \item Discussion of sensible content if applicable, e.g., personally identifiable information or offensive content. [Yes/No/Not Applicable] \textbf{Not Applicable}
    \end{enumerate}
    \item If you used crowdsourcing or conducted research with human subjects, check if you include:
    \begin{enumerate}
        \item The full text of instructions given to participants and screenshots. [Yes/No/Not Applicable] \textbf{Not Applicable}
        \item Descriptions of potential participant risks, with links to Institutional Review Board (IRB) approvals if applicable. [Yes/No/Not Applicable] \textbf{Not Applicable}
        \item The estimated hourly wage paid to participants and the total amount spent on participant compensation. [Yes/No/Not Applicable] \textbf{Not Applicable}
    \end{enumerate}
\end{enumerate}

%
%

\appendix
\onecolumn

\pagenumbering{arabic}
\setcounter{page}{1}
\renewcommand*{\thepage}{\arabic{page}}

\title{Supplement to:\\ Conditional Adjustment in a Markov Equivalence Class}
\date{}
\maketitle


\section{FURTHER PRELIMINARIES AND DEFINITIONS}
\label{supp:defs}


\subsection{Preliminaries}
\textbf{Path Construction.}
A \textit{subsequence} of a path $p$ is a path obtained by deleting non-endpoint nodes from $p$ without changing the order of the remaining nodes. Let $p = \langle X_1, X_2, \dots, X_k \rangle$ and $i,j,k$ such that $1 \le i < j \le k$. We denote the concatenation of paths by the symbol $\oplus$, so that $p = p(X_1, X_i) \oplus p(X_i, X_k)$. We use the notation $(-p)(X_j,X_i)$ to denote the path $\langle X_j, X_{j-1}, \dots, X_i \rangle$.


\subsection{Definitions}

\begin{definition}
\label{def:as}
{\normalfont (\textbf{Adjustment Set for MPDAGs (PAGs)}; \citealp{perkovic2017interpreting, perkovic2018complete, perkovic2015complete}; cf. \citealp{maathuis2015generalized})}
    Let $\mb{X}$, $\mb{Y}$, and $\mb{S}$ be pairwise disjoint node sets in a causal MPDAG (PAG) $\g$. Then $\mb{S}$ is an adjustment set relative to $(\mb{X},\mb{Y})$ in $\g$ if for any density $f$ consistent with $\g$
    \[ f(\mb{y}|do(\mb{x}))  = 
    \begin{cases} 
          f(\mb{y} | \mb{x}) & \mb{S}=\emptyset \\
          \int f(\mb{y} | \mb{x}, \mb{s}) f(\mb{s}) \diff \mb{s}  & \mb{S}\neq\emptyset.
    \end{cases}
    \]
\end{definition}

\begin{definition}
\label{def:ac}
{\normalfont (\textbf{Adjustment Criterion for MPDAGs (PAGs)}; \citealp{perkovic2017interpreting, perkovic2018complete})}
    Let $\mb{X}$, $\mb{Y}$, and $\mb{S}$ be pairwise disjoint node sets in an MPDAG (PAG) $\g$, where every proper possibly causal path from $\mb{X}$ to $\mb{Y}$ in $\g$ starts with a directed (visible) edge out of $\mb{X}$. Then $\mb{S}$ satisfies the adjustment criterion relative to $(\mb{X},\mb{Y})$ in $\g$ if 
    \begin{enumerate}[label=\emph{(\alph*)}]
        \item $\mb{S} \cap \fb{\g} = \emptyset$, and
        \item $\mb{S}$ blocks all proper non-causal definite status paths from $\mb{X}$ to $\mb{Y}$ in $\g$. \label{def:ac-b}
    \end{enumerate}
\end{definition}

\begin{definition}
\label{def:gbc}
{\normalfont (\textbf{Generalized Back-Door Criterion for DAGs}; cf. \citealp{maathuis2015generalized})}
    Let $\mb{X}$, $\mb{Y}$, and $\mb{S}$ be pairwise disjoint node sets in a DAG $\g[D]$. Then $\mb{S}$ satisfies the generalized back-door criterion relative to $(\mb{X},\mb{Y})$ in $\g[D]$ if 
    \begin{enumerate}[label=\emph{(\alph*)}]
        \item $\mb{S} \cap \De(\mb{X},\g[D]) = \emptyset$, and
        \item $\mb{S} \cup \mb{X} \setminus \{X\}$ blocks all back-door paths from $X$ to $\mb{Y}$ in $\g[D]$, for every $X \in \mb{X}$.
    \end{enumerate}
\end{definition}

\begin{definition}
\label{def:pbd}
{\normalfont (\textbf{Proper Back-Door Graph for DAGs}; cf. \citealp{perkovic2018complete})}
    Let $\mb{X}$ and $\mb{Y}$ be disjoint node sets in a DAG $\g[D]$. The proper back-door graph $\dpbd{\mb{XY}}$ is obtained from $\g[D]$ by removing all edges out of $\mb{X}$ that are on proper causal paths from $\mb{X}$ to $\mb{Y}$ in $\g[D]$.
\end{definition}

\begin{definition}
\label{def:moral}
{\normalfont (\textbf{Moral Graph for DAGs}; cf. \citealp{lauritzen1988local}; cf. \citealp{perkovic2018complete})} 
    Let $\g[D]=(\mb{V},\mb{E})$ be a DAG. The moral graph $\g[D]^m$ is formed by adding the edge $A-B$ to any structure of the form $A \to C \gets B$ for any $A,B,C \in \mb{V}$, with $A \notin \Adj(B,\g[D])$ (marrying unmarried parents) and subsequently making all edges in the resulting graph undirected.
\end{definition}

\begin{definition}
\label{def:distance}
{\normalfont (\textbf{Distance to $\mb{Z}$}; \citealp{zhang2006causal}; \citealp{perkovic2017interpreting})} 
    Let $\mb{X,Y}$ and $\mb{Z}$ be pairwise disjoint node sets in an MPDAG or PAG $\g$. Let $p$ be a path between $\mb{X}$ and $\mb{Y}$ in $\g$ such that every collider $C$ on $p$ has a possibly directed path (possibly of length $0$) to $\mb{Z}$. Define the \textit{distance to $\mb{Z}$} of $C$ to be the length of a shortest possibly directed path (possibly of length $0$) from $C$ to $\mb{Z}$, and define the \textit{distance to $\mb{Z}$} of $p$ to be the sum of the distances from $\mb{Z}$ of the colliders on $p$.
\end{definition}


\section{EXISTING RESULTS}
\label{supp:existing}

\textbf{Rules of the Do Calculus} \citep{pearl2009causality}.
Let $\mb{X,Y,Z,}$ and $\mb{W}$ be pairwise disjoint (possibly empty) node sets in a causal DAG $\g[D]$. Let $\g[D]_{\overline{\mb{X}}}$ denote the graph obtained by deleting all edges into $\mb{X}$ from $\g[D]$. Similarly, let $\g[D]_{\underline{\mb{X}}}$ denote the graph obtained by deleting all edges out of $\mb{X}$ in $\g[D]$, and let $\g[D]_{\overline{\mb{X}}\underline{\mb{Z}}}$ denote the graph obtained by deleting all edges into $\mb{X}$ and all edges out of $\mb{Z}$ in $\g[D]$. The following rules hold for all densities consistent with $\g[D]$.

\textbf{Rule 1.} If $(\mb{Y} \dsepp \mb{Z} \given \mb{X} \cup \mb{W})_{\g[D]_{\overline{\mb{X}}}}$, then
    \begin{align}
        f(\mb{y} | do(\mb{x}),\mb{z,w}) = f(\mb{y} | do(\mb{x}),\mb{w}). \label{eq:rule1-do}
    \end{align}

\textbf{Rule 2.} If $(\mb{Y} \dsepp \mb{X} \given \mb{Z} \cup \mb{W})_{\g[D]_{\underline{\mb{X}}\overline{\mb{W}}}}$, then
    \begin{align}
        f(\mb{y} | do(\mb{x}),\mb{z},do(\mb{w})) = f(\mb{y} | \mb{x},\mb{z},do(\mb{w})). \label{eq:rule2-do}
    \end{align}

\textbf{Rule 3.} If $(\mb{Y} \dsepp \mb{X} \given \mb{Z} \cup \mb{W})_{\g[D]_{\overline{\mb{X}(\mb{Z}) \cup \mb{W}}}}$, then
    \begin{align}
        \begin{split}
            f(\mb{y} | do(\mb{x}),\mb{z},do(\mb{w})) = f(\mb{y} | \mb{z},do(\mb{w})), \label{eq:rule3-do}
        \end{split}
    \end{align}
where $\mb{X(Z)} = \mb{X} \setminus \An(\mb{Z}, \g[D]_{\overline{\mb{W}}})$.

\begin{lemma}
\label{lem:wright}
{\normalfont (Wright's Rule of \citealp{wright1921correlation})}
    Let $\mb{X}=\mb{AX}+\mb{\epsilon}$, where $\mb{Q} \in \mathbb{R}^{k \times k}$, $\mb{X} = (X_1, \dots, X_k)^T$ and $\mb{\epsilon} = (\epsilon_1, \dots, \epsilon_k)^T$ is a vector of mutually independent errors with means zero. Moreover, let $Var(\mb{X}) = \mb{I}$. Let $\g[D] = (\mb{X}, \mb{E})$, be the corresponding DAG such that $X_i \to X_j$ is in $\g[D]$ if and only if $A_{ji} \neq 0$. A non-zero entry $A_{ji}$ is called the edge coefficient of $X_i \to X_j$. For two distinct nodes $X_i$, $X_j \in \mb{X}$, let $p_1, \dots, p_r$ be all paths between $X_i$ and $X_j$ in $\g[D]$ that do not contain a collider. Then $\Cov(X_i, X_j) = \sum_{s=1}^r \pi_s$, where $\pi_s$ is the product of all edge coefficients along path $p_s$, $s \in \{1, \dots, r\}$.
\end{lemma}

\begin{lemma}
\label{lem:mardia}
{\normalfont (Theorem 3.2.4 of \citealp{mardia1980multivariate})}
    Let $\mb{X} = (\mb{X_1}^T, \mb{X_2}^T)^T$ be a $p$-dimensional multivariate Gaussian random vector with mean vector $\mb{\mu} = (\mb{\mu_1}^T, \mb{\mu_2}^T)^T$ and covariance matrix $\mb{\Sigma} = \begin{bmatrix}
    \mb{\Sigma_{11}} & \mb{\Sigma}_{12} \\
    \mb{\Sigma_{21}} & \mb{\Sigma}_{22}
    \end{bmatrix}$, so that $\mb{X_1}$ is a $q$-dimensional multivariate Gaussian random vector with mean vector $\mb{\mu_1}$ and covariance matrix $\mb{\Sigma_{11}}$ and $\mb{X_2}$ is a $(p-q)$-dimensional multivariate Gaussian random vector with mean vector $\mb{\mu_2}$ and covariance matrix $\mb{\Sigma_{22}}$. Then $E[\mb{X_2}|\mb{X_1} = \mb{x_1}] = \mb{\mu_2} + \mb{\Sigma_{21}} \mb{\Sigma_{11}}^{-1}(\mb{x_1} - \mb{\mu_1})$.
\end{lemma}

\begin{lemma} 
\label{lem:markov-equiv}  
{\normalfont (cf. Theorem 1 and Proposition 3 of \citealp{lauritzen1990independence})}
    Let $\g[D] = (\mb{V},\mb{E})$ be a DAG, and let $f$ be an observational density over $\mb{V}$. Then $f$ is Markov compatible with $\g[D]$ if and only if 
    \begin{align*}
        V_i \ind \Big[ \mb{V} \setminus \big( \De(V_i, \g[D]) \cup \Pa(V_i,\g[D]) \big) \Big] | \Pa(V_i, \g[D])
    \end{align*}
    for all $V_i \in \mb{V}$, where $\ind$ indicates independence with respect to $f$.
\end{lemma}


\begin{lemma}
\label{lem:equiv-z-mpdag} 
{\normalfont (cf. Lemma 3.2 of \citealp{perkovic2017interpreting})}
    Let $\mb{X}$ and $\mb{Z}$ be disjoint node sets in an MPDAG $\g$. If $\mb{Z} \cap \PossDe(\mb{X},\g) = \emptyset$, then $\mb{Z} \cap \De(\mb{X},\g[D]) = \emptyset$ in every DAG $\g[D]$ in $[\g]$.
\end{lemma}

\begin{lemma}
\label{lem:equiv-forb}
{\normalfont (Lemma C.2 of \citealp{perkovic2017interpreting}, Lemma 9 of \citealp{perkovic2018complete})}
    Let $\mb{X}$, $\mb{Y}$, and $\mb{S}$ be pairwise disjoint node sets in an MPDAG (PAG) $\g$, where every proper possibly causal path from $\mb{X}$ to $\mb{Y}$ in $\g$ starts with a directed (visible) edge out of $\mb{X}$. Then the following statements are equivalent.
    \begin{enumerate}[label = (\roman*)]
        \item $\mb{S} \cap \fb{\g} = \emptyset$.
        \item $\mb{S} \cap \fb{\g[D]} = \emptyset$ in every DAG (MAG) $\g[D]$ in $[\g]$.
    \end{enumerate}
\end{lemma}

\begin{lemma}
\label{lem:equiv-block} 
{\normalfont (cf. Lemma C.3 of \citealp{perkovic2017interpreting}, Lemma 10 of \citealp{perkovic2018complete})}
    Let $\mb{X,Y}$ and $\mb{S}$ be pairwise disjoint node sets in an MPDAG (PAG) $\g$, where every proper possibly causal path from $\mb{X}$ to $\mb{Y}$ in $\g$ starts with a directed (visible) edge out of $\mb{X}$ and where $\mb{S} \cap \fb{\g} = \emptyset$. Then the following statements are equivalent.
    \begin{enumerate}[label = (\roman*)]
        \item $\mb{S}$ blocks all proper non-causal definite status paths from $\mb{X}$ to $\mb{Y}$ in $\g$.
        \item $\mb{S}$ blocks all proper non-causal definite status paths from $\mb{X}$ to $\mb{Y}$ in $\g[D]$ for every DAG (MAG) $\g[D]$ in $[\g]$.
    \end{enumerate}
\end{lemma}


\begin{theorem}
\label{thm:richardson}
{\normalfont (cf. Proposition 3 of \cite{lauritzen1990independence}, cf. Corollary 2 of \cite{richardson2003markov})}
    Let $\mb{X},\mb{Y}$, and $\mb{Z}$ be pairwise disjoint node sets in a DAG $\g[D]$. Further let $(\g[D]_{\An(\mb{X} \cup \mb{Y} \cup \mb{Z},\g[D])})^m$ be the moral induced subgraph of $\g[D]$ on nodes $\An(\mb{X} \cup \mb{Y} \cup \mb{Z},\g[D])$ (see Definition \ref{def:moral}). Then $\mb{Z}$ d-separates $\mb{X}$ and $\mb{Y}$ in $\g[D]$ if and only if all paths between $\mb{X}$ and $\mb{Y}$ in $(\g[D]_{\An(\mb{X} \cup \mb{Y} \cup \mb{Z},\g[D])})^m$ contain at least one node in $\mb{Z}$.
\end{theorem}

\begin{theorem}
\label{thm:ac-alt}
{\normalfont (cf. Theorem 7 of \citealp{perkovic2018complete})}
    Consider the definition of the adjustment criterion for MPDAGs (Definition \ref{def:ac}) in the specific setting of a DAG. In this setting, replacing condition \ref{def:ac-b} in Definition \ref{def:ac} with
    \begin{enumerate}[label=\emph{(\alph*)}]
        \setcounter{enumi}{1}
        \item $\mb{S}$ d-separates $\mb{X}$ and $\mb{Y}$ in $\dpbd{\mb{XY}}$ (see Definition \ref{def:pbd})
    \end{enumerate}
    results in a criterion that is equivalent to Definition \ref{def:ac} applied to a DAG.
\end{theorem}

\begin{theorem}
\label{thm:gbc-as}
{\normalfont (cf. Theorem 3.1 of \citealp{maathuis2015generalized})}
    Let $\mb{X},\mb{Y}$, and $\mb{S}$ be pairwise disjoint node sets in a causal DAG $\g[D]$. If $\mb{S}$ satisfies the generalized back-door criterion relative to $(\mb{X},\mb{Y})$ in $\g[D]$ (Definition \ref{def:gbc}), then $\mb{S}$ is an adjustment set relative to $(\mb{X},\mb{Y})$ in $\g[D]$ (Definition \ref{def:as}). 
\end{theorem}

\begin{lemma}
\label{lem:henckel-e6}
{\normalfont (cf. Lemma E.6 of \citealp{henckel2022graphical})}
    Let $\mb{X},\mb{Y}$ be disjoint node sets in an MPDAG $\g$. If there is no proper possibly causal path from $\mb{X}$ to $\mb{Y}$ that starts with an undirected edge in $\g$, then $\fb{\g} \subseteq \De(\mb{X}, \g)$. 
\end{lemma}

\begin{lemma}
\label{lem:perkovic17-35}
{\normalfont (cf. Lemma 3.5 of \citealp{perkovic2017interpreting})}
    Let $p = \langle V_1, \dots, V_k \rangle, k > 1$, be a definite status path in MPDAG $\g$. Then p is a possibly causal path in $\g$ if and only if there is no edge $V_i \leftarrow V_{i+1}$, $i \in \{1, \dots, k-1\}$ in $\g$.
\end{lemma}

\begin{lemma}
\label{lem:basic-property-pags}
{\normalfont (cf. Lemma 3.3.1 of \citealp{zhang2006causal})}
    Let $X$, $Y$, and $Z$ be distinct nodes in a PAG $\g$. If $X \bulletarrow Y \circbullet Z$, then there is an edge between $X$ and $Z$ with an arrowhead at $Z$. Furthermore, if the edge between $X$ and $Y$ is $X \rightarrow Y$, then the edge between $X$ and $Z$ is either $X \circarrow Z$ or $X \rightarrow Z$ (that is, not $X \leftrightarrow Z$).
\end{lemma}

\begin{lemma}
\label{lem:marl-cycle}
{\normalfont (cf. Lemma 7.5 of \citealp{maathuis2015generalized})}
    Let $X$ and $Y$ be two distinct nodes in a  MAG or PAG $\g$. Then $\g$ cannot have  both an edge $Y \bulletarrow X$ and a path $\langle X = V_1,\dots, V_k = Y \rangle, k>2$ where each edge $\langle V_i, V_{i+1} \rangle, i \in \{1, \dots, k-1\}$, is of one of these forms: $V_i \to V_{i+1}$ or $V_{i} \circbullet V_{i+1}$.
\end{lemma}

\begin{lemma}
\label{lem:richprime2}
{\normalfont (cf. Lemma 17 of \citealp{perkovic2018complete})}
    Let $\mb{X,Y, Z}$ and $\mb{S}$ be pairwise disjoint node sets in a MAG or PAG $\g$. Suppose  that every proper possibly causal path from $\mb{X}$ to $\mb{Y}$ in $\g$ starts with a visible edge out of $\mb{X}$ and that $\big[\mb{S} \cup \mb{Z} \big] \cap \fb{\g} = \emptyset$. Suppose furthermore that there is a path $p$ from $\mb{X}$ to $\mb{Y}$ in $\g$ such that
    \begin{enumerate}[label=(\roman*)]
        \item $p$ is a proper definite status non-causal path from $\mb{X}$ to $\mb{Y}$ in $\g$,
        \item all colliders on $p$ are in $\An(\mb{X} \cup \mb{Y} \cup \mb{Z} \cup \mb{S}, \g) \setminus \big[ \mb{X} \cup \mb{Y} \cup \fb{\g}\big]$, and
        \item no definite non-collider on $p$ is in $\mb{S} \cup \mb{Z}$.
    \end{enumerate}
    Then there is a proper definite status non-causal path from $\mb{X}$ to $\mb{Y}$ that is m-connecting given $\mb{S} \cup \mb{Z}$ in $\g$.
\end{lemma}

\begin{theorem}
\label{thm:ac-as}
{\normalfont (cf. Theorem 4.4 of \citealp{perkovic2017interpreting}, Theorems 5 and 56 of \citealp{perkovic2018complete})}
    Let $\mb{X}$, $\mb{Y}$, and $\mb{S}$ be pairwise disjoint node sets in a causal MPDAG (PAG) $\g$. Then $\mb{S}$ is an adjustment set relative to $(\mb{X,Y})$ in $\g$ (Definition \ref{def:as}) if and only if $\mb{S}$ satisfies the adjustment criterion relative to $(\mb{X,Y})$ in $\g$ (Definition \ref{def:ac}).
\end{theorem}

\begin{lemma} 
\label{lem:adding-edges-imply}
{\normalfont (cf. Lemma F.1 of \citealp{rothenhausler2018causal})}
    Let $X$ and $Y$ be nodes in an MPDAG $\g = (\mb{V,E})$ such that $X - Y$ is in $\g$. Let $\g'$ be an MPDAG constructed from $\g$ by adding $X \to Y$ and completing the orientation rules R1 - R4 of \cite{meek1995causal}. For any $Z,W \in \mb{V}$, if $Z - W$ is in $\g$ and $Z \rightarrow W$ is in $\g'$, then $W \in \De(Y,\g')$.
\end{lemma}


\begin{lemma}
\label{lem:undirected-imply}
{\normalfont (cf. Lemma F.2 of \citealp{rothenhausler2018causal})}
    Let $X$ be a node in an MPDAG $\g=(\mb{V}, \mb{E})$, and let $\mb{S}$ be a set such that for all $S \in \mb{S}$, $X - S$ is in $\g$. Then there is an MPDAG $\g' = (\mb{V}, \mb{E'})$ that is formed by taking $\g$, orienting $X \to S$ for all $S \in \mb{S}$, and completing R1-R4 of \cite{meek1995causal}.
\end{lemma}

\begin{lemma} 
\label{lem:sound-setup1-perk}
{\normalfont (cf. Lemma 59 of \citealp{perkovic2018complete})}
    Let $\mb{X},\mb{Y}$ and $\mb{S}$ be pairwise disjoint node sets in a DAG $\g[D]$ such that $\mb{S}$ satisfies the adjustment criterion relative to $(\mb{X,Y})$ in $\g[D]$ (Definition \ref{def:ac}). Let $\mb{J} \subseteq \An(\mb{X} \cup \mb{Y},\g[D]) \setminus (\De(\mb{X}, \g[D]) \cup \mb{Y})$ and $\mb{\tilde{S}} = \mb{S} \cup \mb{J}$. Then the following statements hold:
    \begin{enumerate}[label = (\roman*)]
        \item $\mb{\tilde{S}}$ satisfies the adjustment criterion relative to $(\mb{X,Y})$ in $\g[D]$, and
        \item $\int_\mb{s} f(\mb{y} \mid \mb{x,s})f(\mb{s}) d\mb{s} = \int_{\mb{\tilde{s}}} f(\mb{y} \mid \mb{x,\tilde{s}}) f(\mb{\tilde{s}}) d\mb{\tilde{s}}$, for any density $f$ consistent with $\g[D]$.
    \end{enumerate}
\end{lemma}

\begin{lemma}
\label{lem:sound-setup2-perk}
{\normalfont (Lemma 60 of \citealp{perkovic2018complete})}
    Let $\mb{X}$, $\mb{Y}$, and $\mb{S}$ be pairwise disjoint node sets in a causal DAG $\g[D]$ such that $\mb{S}$ satisfies the adjustment criterion relative to $(\mb{X,Y})$ in $\g[D]$. Let $\mb{J}=\An(\mb{X} \cup \mb{Y},\g[D]) \setminus \big(\De(\mb{X}, \g[D]) \cup \mb{Y} \big)$ and $\mb{\tilde{S}}=\mb{S} \cup \mb{J}$. Additionally, let $\mb{\tilde{S}_D} = \mb{\tilde{S}} \cap \De(\mb{X}, \g[D])$, $\mb{\tilde{S}_N} = \mb{\tilde{S}} \setminus \De(\mb{X}, \g[D])$, $\mb{Y_D} = \mb{Y} \cap \De(\mb{X},\g[D])$ and $\mb{Y_N} = \mb{Y} \setminus \De(\mb{X}, \g[D])$. Then the following statements hold:
    \begin{enumerate}[label=(\roman*)]
        \item $(\mb{X} \cup \mb{Y_N} \cup \mb{\tilde{S}}) \cap \fb{\g[D]} = \emptyset$,
        
        \item if $p = \langle H, \dots, Y_D \rangle$ is a non-causal path from $H \in \mb{X} \cup \mb{Y_N} \cup \mb{\tilde{S}}$ to $Y_D \in \mb{Y_D}$, then $p$ is blocked by $(\mb{X} \cup \mb{Y_N} \cup \mb{\tilde{S}_N}) \setminus \{H\}$ in $\g[D]$,
        
        \item $\mb{Y_D} \dsepp \mb{\tilde{S}_D} \given \mb{Y_N} \cup \mb{X} \cup \mb{\tilde{S}_N}$ in $\g[D]$, where $\mb{Y_N} = \emptyset$ is allowed,
        
        \item if $\mb{Y_N} = \emptyset$ then $\mb{\tilde{S}_N}$ satisfies the generalized back-door criterion relative to $(\mb{X,Y})$ in $\g[D]$ (Definition \ref{def:gbc}),
        
        \item the empty set satisfies the generalized back-door criterion relative to $(\mb{X} \cup \mb{Y_N} \cup \mb{\tilde{S}_N},\mb{Y_D})$ in $\g[D]$,
        
        \item $\mb{Y_D} \dsepp (\mb{Y_N} \cup \mb{\tilde{S}_N}) \given \mb{X}$ in $\g[D]_{\overline{\mb{X}}\underline{\mb{Y_N} \cup  \mb{\tilde{S}_N}}}$, and

        \item $\mb{\tilde{S}_N} \dsepp \mb{X} \given \mb{Y_N}$ in $\g[D]_{\overline{\mb{X}}}$.
    \end{enumerate}
\end{lemma}


\section{A NECESSARY CONDITION FOR IDENTIFIABILITY}
\label{supp:necessary}

This section includes the proof of Proposition \ref{prop:id-necessary}, which provides a necessary condition for the identifiability of the conditional causal effect given an MPDAG. This result is needed twice -- once for the proof of Theorem \ref{thm:cac-cas} in Section \ref{sec:cond-adj} and once for Example \ref{ex:nocas-noamen} in Section \ref{sec:exs}. Below we also provide two supporting results for the proof of Proposition \ref{prop:id-necessary} -- namely, Lemmas \ref{lem:twopaths-corrollary} and \ref{lem:markov-sub}.


\subsection{Main Result}

\begin{proposition} 
\label{prop:id-necessary}
    Let $\mb{X}$, $\mb{Y}$, and $\mb{Z}$ be pairwise disjoint node sets in a causal MPDAG $\g$. If there is a proper possibly causal path from $\mb{X}$ to $\mb{Y}$ in $\g$ that starts with an undirected edge and does not contain any element of $\mb{Z}$, then the conditional causal effect of $\mb{X}$ on $\mb{Y}$ given $\mb{Z}$ is not identifiable in $\g$.
\end{proposition}

\begin{proofof}[Proposition \ref{prop:id-necessary}]
    This lemma extends Proposition 3.2 of \cite{perkovic2020identifying} and its proof follows similar logic to that of \cite{perkovic2020identifying}.
    
    Suppose that there is a proper possibly causal path from $\mb{X}$ to $\mb{Y}$ in $\g = (\mb{V},\mb{E})$  that starts with an undirected edge and does not contain any element of $\mb{Z}$. Then by Lemma \ref{lem:twopaths-corrollary}, there is one such path -- call it $q=\langle X=V_0, \dots, V_k=Y\rangle$, $X \in \mb{X}$, $Y \in \mb{Y}, k \ge 1$ -- where the corresponding paths in two DAGs in $[\g]$ take the forms $X \to \dots \to Y$ and $X \gets V_1 \to \dots \to Y$ ($X \gets Y$ when $k=1$). Call these DAGs $\g[D]^1$ and $\g[D]^2$ with paths $q_1$ and $q_2$, respectively.

    To prove that the conditional causal effect of $\mb{X}$ on $\mb{Y}$ given $\mb{Z}$ is not identifiable in $\g$, it suffices to show that there are two families of interventional densities over $\mb{V}$ -- call them $\mb{F^*_1}$ and $\mb{F^*_2}$, where for $i \in \{1,2\}$, we define $\mb{F^*_i} = \{ f_i(\mb{v}|do(\mb{x'})) : \mb{X'} \subseteq \mb{V} \}$ -- such that the following properties hold. 
    \begin{enumerate}[label=(\roman*)]
        \item $\g[D]^1$ and $\g[D]^2$ are compatible with $\mb{F^*_1}$ and $\mb{F^*_2}$, respectively.
            \footnote{For brevity, we say a DAG is ``compatible with'' a set of interventional densities and an interventional density is ``consistent with'' a DAG as shorthand for these claims holding only were the DAG to be causal.}
            \label{prop:id-necessary-1}
        \item $f_1(\mb{v}) = f_2(\mb{v})$. \label{prop:id-necessary-2}
        \item $f_1(\mb{y} | do(\mb{x}), \mb{z}) \neq f_2(\mb{y} | do(\mb{x}), \mb{z})$. \label{prop:id-necessary-3}
    \end{enumerate}

    To define such families, we start by introducing an additional DAG and an observational density $f(\mb{v})$. That is, let $\g[D]^{1'}$ be a DAG constructed by removing every edge from $\g[D]^1$ except for the edges on $q_1$. Then let $f(\mb{v})$ be the multivariate normal distribution under the following linear structural equation model (SEM). Each random variable $A \in \mb{V}$ has mean zero and is a linear combination of its parents in $\g[D]^{1'}$ and $\epsilon_{A} \sim N(0, \sigma^2_A)$, where $\{\mb{\epsilon}_{A}: A \in \mb{V}\}$ are mutually independent. The coefficients in this linear combination are defined by the edge coefficients of $\g[D]^{1'}$. We pick these edge coefficients in conjunction with $\{\mb{\sigma}^2_{A}: A \in \mb{V}\}$ in such a way that each coefficient is in $(0,1)$ and $\Var(A) = 1$ for all $A \in \mb{V}$.

    From this, we define $\mb{F^*_1} = \{ f_1(\mb{v}|do(\mb{x'})) : \mb{X'} \subseteq \mb{V} \}$ such that $\g[D]^{1'}$ is compatible with $\mb{F^*_1}$ and such that $f_1(\mb{v}) = f(\mb{v})$. Note that $f(\mb{v})$ is Markov compatible with $\g[D]^{1'}$ by construction, and we build the interventional densities in $\mb{F^*_1}$ by replacing the intervening random variables in the SEM with their interventional values \citep{pearl2009causality}.

    To construct the second family of interventional densities, we introduce the DAG $\g[D]^{2'}$, which we form by removing every edge from $\g[D]^2$ except for the edges on $q_2$. Then note that we could have defined $f(\mb{v})$ using a linear SEM based on the parents in $\g[D]^{2'}$. In this case, the resulting observational density would again be a multivariate normal with mean vector zero and a covariance matrix with ones on the diagonal. The off-diagonal entries would be the covariances between the variables in $\g[D]^{2'}$. But note that by Lemma \ref{lem:wright}, these values will equal the product of all edge coefficients between the relevant nodes in $\g[D]^{2'}$. Since $\g[D]^{1'}$ and $\g[D]^{2'}$ contain no paths with colliders, the observational density $f(\mb{v})$ built using $\g[D]^{2'}$ will be an identical distribution to that built under $\g[D]^{1'}$. Thus, in an analogous way to $\mb{F^*_1}$, we define $\mb{F^*_2} = \{ f_1(\mb{v}|do(\mb{x'})) : \mb{X'} \subseteq \mb{V} \}$ such that $\g[D]^{2'}$ is compatible with $\mb{F^*_2}$ and such that $f_2(\mb{v}) = f(\mb{v})$.
        
    Having defined $\mb{F^*_1}$ and $\mb{F^*_2}$, we check that their desired properties hold. Note that by construction, $\g[D]^{1'}$ and $\g[D]^{2'}$ are compatible with $\mb{F^*_1}$ and $\mb{F^*_2}$, respectively. Thus \ref{prop:id-necessary-1} holds by Lemma \ref{lem:markov-sub}. Similarly by construction, \ref{prop:id-necessary-2} holds. To show that \ref{prop:id-necessary-3} holds, it suffices to show that $E[Y | do(\mb{X}=\mb{1}), \mb{Z}]$ is not the same under $f_1$ and $f_2$. 

    To calculate these expectations, we first want to apply Rules 1-3 of the do calculus (Equations \eqref{eq:rule1-do}-\eqref{eq:rule3-do}). Since $f_i(\mb{v} | do(\mb{x}))$, $i \in \{1,2\}$, is consistent with $\g[D]^{i'}$, we apply these rules using graphical relationships in $\g[D]^{i'}$. Because the path in $\g[D]^{i'}$ corresponding to $q_i$, $i \in \{1,2\}$, does not contain nodes in $\mb{Z}$ or $\mb{X} \setminus \{X\}$, then $Y \dsepp \mb{Z} | \mb{X}$ and $Y \dsepp \mb{X} \setminus \{X\} | X$ in $\g[D]^{i'}_{\overline{\mb{X}}}$. Further, $Y \dsepp X$ in $\g[D]^{1'}_{\underline{X}}$ and $Y \dsepp X$ in $\g[D]^{2'}_{\overline{X}}$. Thus by Rules 1-3 of the do calculus (Equations \eqref{eq:rule1-do}-\eqref{eq:rule3-do}), the following hold.
    \begin{align*}
        E_1[Y | do(\mb{X}=\mb{1}), \mb{Z}] &= E_1[Y | do(X=1)] = E_1[Y|X=1] := a.\\
        E_2[Y | do(\mb{X}=\mb{1}), \mb{Z}] &= E_2[Y | do(X=1)] = E_2[Y] := b,
    \end{align*}    
    where $E_i, i \in \{1,2\}$ is the expectation under $f_i$. To calculate $a$ and $b$, we rely on the observational density $f(\mb{v})$, which was constructed using $\g[D]^{1'}$. By Lemma \ref{lem:mardia}, $a$ equals the covariance of $X$ and $Y$ under $f(\mb{v})$, and by Lemma \ref{lem:wright}, $\Cov(X,Y)$ equals the product of all edge coefficients in $\g[D]^{1'}$, which were chosen to be in $(0,1)$. Therefore, $a \neq 0$. But by definition of $f(\mb{v})$, $b=0$.
\end{proofof}


\subsection{Supporting Result}

\begin{lemma}
\label{lem:twopaths-corrollary}
    Let $\mb{X}$, $\mb{Y}$, and $\mb{Z}$ be pairwise disjoint node sets in an MPDAG $\g=(\mb{V},\mb{E})$. Suppose that there is a proper possibly causal path from $\mb{X}$ to $\mb{Y}$ in $\g$ that starts with an undirected edge and does not contain nodes in $\mb{Z}$. Then there is one such path $\langle X=V_0, \dots, V_k=Y \rangle$, $X \in \mb{X}$, $Y \in \mb{Y}$, $k \ge 1$, where the corresponding paths in two DAGs in $[\g]$ take the forms $X \to \dots \to Y$ and $X \gets V_1 \to \dots \to Y$ ($X \gets Y$ when $k=1$), respectively.
\end{lemma}

\begin{proofof}[Lemma \ref{lem:twopaths-corrollary}]
    This lemma is similar to Lemma A.3 of \cite{perkovic2020identifying} and its proof borrows from the proof strategy of Lemma C.1 of \cite{perkovic2017interpreting}.
    
    Let $q^*$ be an arbitrary proper possibly causal path from $\mb{X}$ to $\mb{Y}$ in $\g$ that starts with an undirected edge and does not contain nodes in $\mb{Z}$. Then let $q=\langle X=V_0, \dots, V_k=Y\rangle$, $X \in \mb{X}$, $Y \in \mb{Y}$, $k \ge 1$, be a shortest subsequence of $q^*$ in $\g$ that also starts with an undirected edge. Note that $q$ is a proper possibly causal path from $\mb{X}$ to $\mb{Y}$ in $\g$ that starts with an undirected edge and does not contain nodes in $\mb{Z}$.

    Consider when $q$ is of definite status. Since $q$ is possibly causal, all non-endpoints of $q$ are definite non-colliders. Let $\g[D]^1$ be a DAG in $[\g]$ that contains $X \to V_1$. Then since $V_1$ is either $Y$ or a definite non-collider on $q$, the path corresponding to $q$ in $\g[D]^1$ takes the form $X \to \dots \to Y$ by induction. Let $\g[D]^2$ be a DAG in $[\g]$ with no additional edges into $V_1$ compared to $\g$ (Lemma \ref{lem:undirected-imply}). Since $\g$ contains $X - V_1$, $\g[D]^2$ contains $X \gets V_1$. When $k>1$, $\g$ contains either $V_1 - V_2$ or $V_1 \to V_2$, and so $\g[D]^2$ contains $X \gets V_1 \to V_2$. Thus by the same inductive reasoning as above, the path corresponding to $q$ in $\g[D]^2$ takes the form $X \gets V_1 \to \dots \to Y$ (or simply $X \gets Y$ when $k=1$).
    
    Consider instead when $q$ is not of definite status. Note that $k>1$. To see that $q$ contains $V_1 - V_2$, note that by the choice of $q$ and the fact that $q$ is possibly causal, $q(V_1,Y)$ is unshielded and possibly causal. Thus, $q(V_1,Y)$ is of definite status. However, $q$ is not of definite status, so $V_1$ must not be of definite status on $q$, which implies that $q$ cannot contain $V_1 \to V_2$. Since $q$ is possibly causal, it also cannot contain $V_1 \gets V_2$.

    To find two DAGs in $[\g]$ with paths corresponding to $q$ that fit our desired forms, we narrow our search to $[\g']$, where we let $\g'$ be an MPDAG constructed from $\g$ by adding $V_1 \to V_2$ and completing R1-R4 of \cite{meek1995causal}. We show below that the path corresponding to $q$ in $\g'$ takes the form $X - V_1 \to \dots \to Y$, and thus, there must be two DAGs in $[\g'] \subseteq [\g]$ with corresponding paths of the forms $X \to \dots \to Y$ and $X \gets V_1 \to \dots \to Y$.
            
    We first show that $\g'$ contains $X - V_1$ by the contraposition of Lemma \ref{lem:adding-edges-imply}. Note that we have already shown that $\g$ contains $V_1 - V_2$, that $\g'$ is formed by adding $V_1 \to V_2$ to $\g$, and that $\g$ contains $X - V_1$. It remains to show that $X,V_1 \notin \De(V_2, \g')$. To see this, note that $\g$ must contain an edge $\langle X,V_2 \rangle$, because $V_1$ is not of definite status on $q$. This edge must take the form $X \to V_2$ by the choice of $q$ and the fact that $q$ is possibly causal. Thus, $\g'$ contains $X \to V_2$ and $V_1 \to V_2$. Therefore, $X,V_1 \notin \De(V_2, \g')$. Finally, note that $\g'$ contains $V_1 \to \dots \to Y$ by R1 of \cite{meek1995causal}, since we constructed $\g'$ be adding $V_1 \to V_2$ to a path $q(V_1,Y)$ that is unshielded and possibly causal.
\end{proofof}

\begin{lemma}
\label{lem:markov-sub}
    Let $\mb{X}$, $\mb{Y}$, and $\mb{Z}$ be pairwise disjoint node sets in a causal DAG $\g[D] = (\mb{V},\mb{E})$. Then let $\g[D]^* = (\mb{V},\mb{E'})$ be a causal DAG constructed by removing edges from $\g[D]$, and let $f(\mb{v} | do(\mb{x}))$ be an interventional density over $\mb{V}$. If $f(\mb{v} | do(\mb{x}))$ is consistent with $\g[D]^*$, then it is consistent with $\g[D]$.
\end{lemma}

\begin{proofof}[Lemma \ref{lem:markov-sub}]
    Suppose that $f(\mb{v} | do(\mb{x}))$ is consistent with $\g[D]^*$. Then by definition, there exists a set of interventional densities $\mb{F^*}$ such that $\g[D]^*$ is compatible with $\mb{F^*}$. Let $f(\mb{v})$ be the density in $\mb{F^*}$ under a null intervention. Note that by the truncated factorization in Equation \eqref{eq:trunc-fact}, $f(\mb{v})$ is Markov compatible with $\g[D]^*$. Thus by Lemma \ref{lem:markov-equiv},
    \begin{align}
        V_i \ind \Big[ \mb{V} \setminus \big( \De(V_i, \g[D]^*) \cup \Pa(V_i,\g[D]^*) \big) \Big] | \Pa(V_i, \g[D]^*) \label{eq:markov-sub-1}
    \end{align}
    for all $V_i \in \mb{V}$, where $\ind$ indicates independence with respect to $f(\mb{v})$. Further, since $\De(V_i,\g[D]^*) \subseteq \De(V_i,\g[D])$, then $\De(V_i,\g[D]^*) \cap \Pa(V_i,\g[D]) = \emptyset$ and thus $\Pa(V_i,\g[D]) \subseteq \mb{V} \setminus \De(V_i, \g[D]^*)$. Therefore it follows from \eqref{eq:markov-sub-1} that
    \begin{align} 
        V_i \ind \Big[ \Pa(V_i,\g[D]) \setminus \Pa(V_i,\g[D]^*) \Big] \,\, \Big| \,\, \Pa(V_i,\g[D]^*). \label{eq:markov-sub-2}
    \end{align}
    
    Let $f(\mb{v} | do(\mb{x'}))$, $\mb{X'} \subseteq \mb{V}$, be an arbitrary density in $\mb{F^*}$. Then by definition and \eqref{eq:markov-sub-2}
    \begin{align*}
        f(\mb{v} | do(\mb{x'})) &= \prod_{V_i \in \mb{V} \setminus \mb{X'}} f(v_i|\pa(v_i,\g[D]^*)) \mathds{1}(\mb{X'} = \mb{x'})\\
                                &= \prod_{V_i \in \mb{V} \setminus \mb{X'}} f(v_i|\pa(v_i,\g[D])) \mathds{1}(\mb{X'} = \mb{x'}).
    \end{align*}
    Since $f(\mb{v} | do(\mb{x'}))$ was arbitrary, this holds for all densities in $\mb{F^*}$. Thus, $\g[D]$ is compatible with $\mb{F^*}$. Since $f(\mb{v} | do(\mb{x})) \in \mb{F^*}$, then by definition, it is consistent with $\g[D]$.
\end{proofof}


\section{PROOFS FOR SECTION \ref{sec:cond-adj}: MPDAGS - CONDITIONAL ADJUSTMENT CRITERION} 
\label{supp:adjustment}

The following results show the completeness and soundness of the conditional adjustment criterion for identifying conditional adjustment sets in DAGs. We rely on these results to show the analogous results for MPDAGs in Theorem \ref{thm:cac-cas} of Section \ref{sec:cond-adj}. Figure \ref{fig:proof-map} shows how the results in this paper fit together to prove Theorem \ref{thm:cac-cas}. Two supporting results needed for the proof of soundness in DAGs follow the main results below.

\begin{figure}
    \centering
    \begin{tikzpicture}[>=stealth',shorten >=1pt,node distance=3cm, main node/.style={minimum size=0.4cm}]
    [>=stealth',shorten >=1pt,node distance=3cm,initial/.style    ={}]
    \node[main node]                 (T3)                  {\textbf{Theorem \ref{thm:cac-cas}}};
    \node[main node,yshift=-1cm]     (T42)  [left of=T3]   {Theorem \ref{thm:completeness}};
    \node[main node,yshift=1cm]      (T43)  [left of=T3]   {Theorem \ref{thm:soundness}};
    \node[main node,yshift=-.375cm]  (L44)  [left of=T43]  {Lemma \ref{lem:sound-setup1}};
    \node[main node,yshift=.375cm]   (L45)  [left of=T43]  {Lemma \ref{lem:sound-setup2}};
    \node[main node,yshift=-.8125cm] (L6A)  [left of=L44]  {Lemma \ref{lem:comparison}\ref{lem:comparison-a}};
    \draw[->] (-2.0, .9)   to (-1,  .1);
    \draw[->] (-2.0,-.9)   to (-1, -.1);
    \draw[->] (-5.1,1.375) to (-4, 1.1);
    \draw[->] (-5.1, .625) to (-4,  .9);
    \draw[->] (-6.0, .8)  to (-6.0,1.2);
    \draw[->] (-8.0,.0875) to (-6.9, 1.175);
    \draw[->] (-8.0,-.1875) to (-6.9, .625);
    \draw[->] (-8.0,-.2875) to (-4, -1);
    \end{tikzpicture}
    \caption{Proof structure of Theorem \ref{thm:cac-cas}.}
    \label{fig:proof-map}
\end{figure}
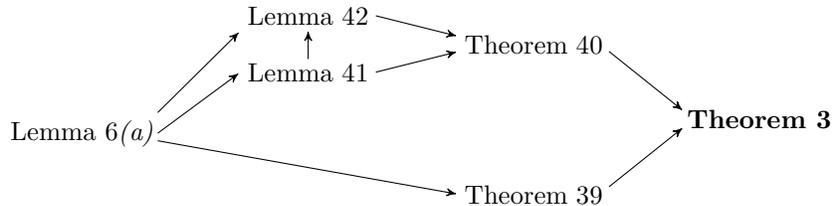


\subsection{Main Results}

\begin{proofof}[Lemma \ref{lem:comparison}] 
    \ref{lem:comparison-a} Follows from Lemma \ref{lem:henckel-e6}. \ref{lem:comparison-b} Holds by Theorem \ref{thm:cac-cas}, Lemma \ref{lem:comparison}\ref{lem:comparison-a}, and Theorem \ref{thm:ac-as}.
\end{proofof}


\begin{theorem}
\label{thm:completeness}
{\normalfont (\textbf{Completeness of the Conditional Adjustment Criterion for DAGs})}
    Let $\mb{X}$, $\mb{Y}$, $\mb{Z}$, and $\mb{S}$ be pairwise disjoint node sets in a causal DAG $\g[D]$, where $\mb{Z} \cap \De(\mb{X}, \g[D]) = \emptyset$. If $\mb{S}$ is a conditional adjustment set relative to $(\mb{X,Y,Z})$ in $\g[D]$ (Definition \ref{def:cas}), then $\mb{S}$ satisfies the conditional adjustment criterion relative to $(\mb{X,Y,Z})$ in $\g[D]$ (Definition \ref{def:cac}).
\end{theorem}

\begin{proofof}[Theorem \ref{thm:completeness}]
    Let $\mb{S}$ be a conditional adjustment set relative to $(\mb{X,Y,Z})$ in $\g[D]$, and let $f$ be a density consistent with $\g[D]$. We start by showing that $\mb{S} \cup \mb{Z}$ is an adjustment set relative to $(\mb{X,Y})$ in $\g[D]$. To do this, we calculate the following. (Justification for the numbered equations is below.)
    \begin{align}
        f(\mb{y}|do(\mb{x})) 
                &= \int_{\mb{z}} f(\mb{y,z} | do(\mb{x})) \diff \mb{z} \nonumber \\
                &= \int_{\mb{z}} f(\mb{z} | do(\mb{x})) f(\mb{y} | do(\mb{x}), \mb{z}) \diff \mb{z} \nonumber \\
                &= \int_{\mb{z}} f(\mb{z}) f(\mb{y} | do(\mb{x}), \mb{z}) \diff \mb{z} \label{eq:completeness-a} \\
                &= \int_{\mb{z}} f(\mb{z}) \int_{\mb{s}} f(\mb{y} | \mb{x,z,s}) f(\mb{s} | \mb{z}) \diff \mb{s} \diff \mb{z} \label{eq:completeness-b} \\
                &= \int_{\mb{s,z}} f(\mb{y} | \mb{x,s,z}) f(\mb{s,z}) \diff \mb{s} \diff \mb{z} \nonumber.
    \end{align}
    Equation \eqref{eq:completeness-a} follows from Rule 3 of the do calculus (Equation \eqref{eq:rule3-do}). To show that this rule holds, let $p$ be an arbitrary path from $\mb{X}$ to $\mb{Z}$ in $\g[D]_{\overline{\mb{X}}}$. Note that $p$ must begin with an edge out of $\mb{X}$. Since $\mb{Z} \cap \De(\mb{X}, \g) = \emptyset$, $p$ cannot be causal and, therefore, must have colliders. Thus, $p$ is blocked, and so $(\mb{Z} \dsepp \mb{X})_{\g[D]_{\overline{\mb{X}}}}$. Equation \eqref{eq:completeness-b} follows from the fact that $\mb{S}$ is a conditional adjustment set relative to $(\mb{X,Y,Z})$. This shows that $\mb{S} \cup \mb{Z}$ is an adjustment set relative to $(\mb{X,Y})$ in $\g[D]$. 

    By Theorem \ref{thm:ac-as}, $\mb{S} \cup \mb{Z}$ satisfies the adjustment criterion relative to $(\mb{X,Y})$ in $\g[D]$. Then by Lemma \ref{lem:comparison}\ref{lem:comparison-a}, $\mb{S}$ satisfies the conditional adjustment criterion relative to $(\mb{X,Y,Z})$ in $\g[D]$.
\end{proofof}


\begin{theorem}
\label{thm:soundness}
{\normalfont (\textbf{Soundness of the Conditional Adjustment Criterion for DAGs})}
    Let $\mb{X,Y,Z}$, and $\mb{S}$ be pairwise disjoint node sets in a causal DAG $\g[D]$, where $\mb{Z} \cap \De(\mb{X}, \g[D]) = \emptyset$. If $\mb{S}$ satisfies the conditional adjustment criterion relative to $(\mb{X,Y, Z})$ in $\g[D]$ (Definition \ref{def:cac}), then $\mb{S}$ is a conditional adjustment set relative to $(\mb{X,Y,Z})$ in $\g[D]$ (Definition \ref{def:cas}).
\end{theorem}

\begin{proofof}[Theorem \ref{thm:soundness}] 
    This theorem is analogous to Theorem 58 of \cite{perkovic2018complete} for the adjustment criterion. We use the same proof strategy and adapt the arguments to suit our needs.

    Suppose that $\mb{S}$ satisfies the conditional adjustment criterion relative to $(\mb{X,Y,Z})$ in $\g[D]$ and let $f$ be a density consistent with $\g[D]$. Our goal is to prove that
    \begin{equation}
        {f(\mb{y} | do(\mb{x}), \mb{z}) = \int_{\mb{s}} f(\mb{y} | \mb{x,z,s}) f(\mb{s} | \mb{z}) \diff \mb{s}.}
    \end{equation}

    We consider three cases below. Before this, we prove an equality that holds in all cases. Let $\mb{Y_D} = \mb{Y} \cap \De(\mb{X},\g[D])$ and $\mb{Y_N} = \mb{Y} \setminus \De(\mb{X}, \g[D])$. Then $\mb{Y_N} \dsepp \mb{X} \given \mb{Z}$ in $\g[D]_{\overline{\mb{X}}}$, since $\g[D]_{\overline{\mb{X}}}$ does not contain edges into $\mb{X}$ and since all paths from $\mb{X}$ to $\mb{Y_N}$ that start with an edge out of $\mb{X}$ in $\g[D]_{\overline{\mb{X}}}$ contain a collider -- a collider that cannot be an element of $\An(\mb{Z},\g[D])$ since $\mb{Z} \cap \De(\mb{X}, \g[D]) = \emptyset$. Rule 3 of the do calculus (Equation \eqref{eq:rule3-do}) then implies
    \begin{align}
        {f(\mb{y_N} | do(\mb{x}), \mb{z}) = f(\mb{y_N} | \mb{z}).} \label{eq:soundness-1}
    \end{align}


    \textbf{Case 1:} Assume that $\mb{Y_D} = \emptyset$ so that $\mb{Y} = \mb{Y_N}$. Then we have the following. (Justification for the numbered equations is below.)
    \begin{align}
        f(\mb{y} | do(\mb{x}) , \mb{z})
            &= f(\mb{y} | \mb{z}) \label{eq:soundness-2}\\
            &= \int_{\mb{s}} f(\mb{y} | \mb{z, s}) f(\mb{s} | \mb{z}) \diff \mb{s} \nonumber \\
            &= \int_{\mb{s}} f(\mb{y} | \mb{x, z, s}) f(\mb{s} | \mb{z}) \diff \mb{s}. \label{eq:soundness-3}
    \end{align}
    Equation \eqref{eq:soundness-2} follows from Equation \eqref{eq:soundness-1} and $\mb{Y} = \mb{Y_N}$. Equation \eqref{eq:soundness-3} follows from the following logic. Since $\mb{S}$ satisfies the conditional adjustment criterion relative to ($\mb{X,Y,Z}$) in $\g[D]$ and since $\mb{Y} = \mb{Y_N}$, it holds that $\mb{S} \cup \mb{Z}$ blocks all paths from $\mb{X}$ to $\mb{Y}$ in $\g[D]$. Thus, $\mb{X} \dsepp \mb{Y} \given \mb{S} \cup \mb{Z}$ in $\g[D]$, which implies the analogous independence statement.


    \textbf{Case 2:} Assume $\mb{Y_N} = \emptyset$ so that $\mb{Y} = \mb{Y_D}$. Define $\mb{H} = \An(\mb{X} \cup \mb{Y},\g[D]) \setminus (\De(\mb{X}, \g[D]) \cup \mb{Y} \cup \mb{Z})$, $\mb{\tilde{S}} = \mb{S} \cup \mb{H}$, $\mb{\tilde{S}_D} = \mb{\tilde{S}} \cap \De(\mb{X}, \g[D])$, and $\mb{\tilde{S}_N} = \mb{\tilde{S}} \setminus \De(\mb{X}, \g[D])$. Then we have the following. (Justification for the numbered equations is below.)
    \begin{align}
        f(\mb{y} | do(\mb{x}), \mb{z})
            &= \int_{\mb{\tilde{s}_N}} f(\mb{y} | \mb{x,z,\tilde{s}_N}) f(\mb{\tilde{s}_N} | \mb{z}) \diff \mb{\tilde{s}_N} \label{eq:soundness-4} \\
            &= \int_{\mb{\tilde{s}_N}} f(\mb{y} | \mb{x,z,\tilde{s}_N}) \int_{\mb{\tilde{s}_D}} f(\mb{\tilde{s}_D,\tilde{s}_N} | \mb{z}) \diff \mb{\tilde{s}_D} \diff \mb{\tilde{s}_N} \nonumber \\
            &= \int_{\mb{\tilde{s}_D,\tilde{s}_N}} f(\mb{y} | \mb{x,z,\tilde{s}_N}) f(\mb{\tilde{s}_D,\tilde{s}_N} | \mb{z}) \diff \mb{\tilde{s}_D} \diff \mb{\tilde{s}_N} \label{eq:soundness-5} \\
            &= \int_{\mb{\tilde{s}}} f(\mb{y} | \mb{x,z,\tilde{s}}) f(\mb{\tilde{s}} | \mb{z}) \diff \mb{\tilde{s}} \label{eq:soundness-6} \\
            &= \int_{\mb{s}} f(\mb{y} | \mb{x,z,s}) f(\mb{s} | \mb{z}) \diff \mb{s}. \label{eq:soundness-7}
    \end{align}
    Equation \eqref{eq:soundness-4} holds since by Lemma \ref{lem:sound-setup2}\ref{lem:sound-setup2-4}, $\mb{\tilde{S}_N}$ is a conditional adjustment set relative to $(\mb{X,Y,Z})$ in $\g[D]$. Equation \eqref{eq:soundness-5} holds since $\mb{\tilde{S}_D}$ is disjoint from $\mb{Y} \cup \mb{X} \cup \mb{\tilde{S}_N} \cup \mb{Z}$. Equation \eqref{eq:soundness-6} holds since by Lemma \ref{lem:sound-setup2}\ref{lem:sound-setup2-3}, we have $\mb{Y} \dsepp \mb{\tilde{S}_D} \given \mb{X} \cup \mb{\tilde{S}_N} \cup \mb{Z}$ in $\g[D]$, where the analogous independence statement follows. Finally, Equation \eqref{eq:soundness-7} results from applying Lemma \ref{lem:sound-setup1}\ref{lem:sound-setup1-b}.


    \textbf{Case 3:} Assume $\mb{Y_D} \neq \emptyset$ and $\mb{Y_N} \neq \emptyset$ and define $\mb{H}, \mb{\tilde{S}}, \mb{\tilde{S}_D}$, and $\mb{\tilde{S}_N}$ as in Case 2 above. We start by showing two equalities that rely on the do calculus. First note that by Lemma \ref{lem:sound-setup2}\ref{lem:sound-setup2-6}, $\mb{Y_D} \dsepp \mb{Y_N} \cup \mb{\tilde{S}_N} \cup \mb{Z} \given \mb{X}$ in $\g[D]_{\overline{\mb{X}}\underline{\mb{Y_N} \cup \mb{\tilde{S}_N} \cup \mb{Z}}}$. Thus by Rule 2 of the do calculus (Equation \eqref{eq:rule2-do}), we have that 
    \begin{align}
        {f(\mb{y_D} | do(\mb{x}),\mb{y_N, z,\tilde{s}_N}) = f(\mb{y_D} | do(\mb{x,y_N,z,\tilde{s}_N})).} \label{eq:soundness-8}
    \end{align}
    Second, note by Lemma \ref{lem:sound-setup2}\ref{lem:sound-setup2-7}, $\mb{\tilde{S}_N} \dsepp \mb{X} \given \mb{Y_N} \cup \mb{Z}$ in $\g[D]_{\overline{\mb{X}}}$. Thus by Rule 3 of the do calculus (Equation \eqref{eq:rule3-do}), we have that
    \begin{align}
        {f(\mb{\tilde{s}_N} | do(\mb{x}),\mb{y_N,z}) = f(\mb{\tilde{s}_N} | \mb{y_N,z}).} \label{eq:soundness-9}
    \end{align}
    Then we have the following. (Justification for the numbered equations is below.)
    \begin{align}
        f(\mb{y} | do(\mb{x}), \mb{z})
            &= \int_{\mb{\tilde{s}_N}} f(\mb{y,\tilde{s}_N} | do(\mb{x}), \mb{z}) \diff \mb{\tilde{s}_N} \nonumber \\
            &= \int_{\mb{\tilde{s}_N}} f(\mb{y_D} | \mb{\tilde{s}_N,y_N}, do(\mb{x}), \mb{z}) f(\mb{\tilde{s}_N} | \mb{y_N}, do(\mb{x}), \mb{z}) f(\mb{y_N} | do(\mb{x}), \mb{z}) \diff \mb{\tilde{s}_N} \nonumber \\
            &= \int_{\mb{\tilde{s}_N}} f(\mb{y_D} | do(\mb{x,y_N,z,\tilde{s}_N})) f(\mb{\tilde{s}_N} | \mb{y_N, z}) f(\mb{y_N} | \mb{z}) \diff \mb{\tilde{s}_N} \label{eq:soundness-10} \\
            &= \int_{\mb{\tilde{s}_N}} f(\mb{y_D} | do(\mb{x,y_N,z,\tilde{s}_N})) \int_{\mb{\tilde{s}_D}} f(\mb{\tilde{s}_N,y_N,\tilde{s}_D} | \mb{z}) \diff \mb{\tilde{s}_D} \diff \mb{\tilde{s}_N} \nonumber \\
            &= \int_{\mb{\tilde{s}_N}} f(\mb{y_D} | do(\mb{x,y_N,z,\tilde{s}_N})) \int_{\mb{\tilde{s}_D}} f(\mb{y_N} | \mb{\tilde{s},z}) f(\mb{\tilde{s}} | \mb{z}) \diff \mb{\tilde{s}_D} \diff \mb{\tilde{s}_N} \nonumber \\
            &= \int_{\mb{\tilde{s}_N}} f(\mb{y_D} | \mb{y_N,x,z,\tilde{s}_N}) \int_{\mb{\tilde{s}_D}} f(\mb{y_N} | \mb{x,z,\tilde{s}}) f(\mb{\tilde{s}} | \mb{z}) \diff \mb{\tilde{s}_D} \diff \mb{\tilde{s}_N} \label{eq:soundness-11} \\
            &= \int_{\mb{\tilde{s}}} f(\mb{y_D} | \mb{y_N,x,z,\tilde{s}}) f(\mb{y_N} | \mb{x,z,\tilde{s}}) f(\mb{\tilde{s}} | \mb{z}) \diff \mb{\tilde{s}} \label{eq:soundness-12} \\
            &= \int_{\mb{\tilde{s}}} f(\mb{y} | \mb{x,z,\tilde{s}}) f(\mb{\tilde{s}} | \mb{z}) \diff \mb{\tilde{s}} \nonumber \\
            &= \int_{\mb{s}} f(\mb{y} | \mb{x,z,}\mb{s}) f(\mb{s} | \mb{z}) \diff \mb{s}. \label{eq:soundness-13}
    \end{align}
    Equation \eqref{eq:soundness-10} holds by the applying Equations \eqref{eq:soundness-8}, \eqref{eq:soundness-9}, and \eqref{eq:soundness-1}. Equation \eqref{eq:soundness-11} holds by the following logic. By Lemma \ref{lem:sound-setup2}\ref{lem:sound-setup2-5}, the empty set is an adjustment set relative to $(\mb{X} \cup \mb{Y_N} \cup \mb{\tilde{S}_N} \cup \mb{Z}, \mb{Y_D})$ in $\g[D]$. Then by Lemma \ref{lem:sound-setup1}\ref{lem:sound-setup1-a}, $\mb{\tilde{S}}$ satisfies the conditional adjustment criterion relative to $(\mb{X,Y,Z})$ in $\g[D]$, and so $\mb{\tilde{S}} \cup \mb{Z}$ blocks all paths from $\mb{X}$ to $\mb{Y_N}$ in $\g[D]$. Thus, $\mb{Y_N} \dsepp \mb{X} \given \mb{\tilde{S}} \cup \mb{Z}$ in $\g[D]$, where the analogous independence statement follows.

    Equation \eqref{eq:soundness-12} holds since $\mb{\tilde{S}_D}$ is disjoint from $\mb{Y} \cup \mb{X} \cup \mb{\tilde{S}_N} \cup \mb{Z}$ and since by Lemma \ref{lem:sound-setup2}\ref{lem:sound-setup2-3}, we have that $\mb{Y_D} \dsepp \mb{\tilde{S}_D} \given \mb{Y_N} \cup \mb{X} \cup \mb{\tilde{S}_N} \cup \mb{Z}$ in $\g[D]$, where the analogous independence statement follows. Finally, Equation \eqref{eq:soundness-13} results from applying Lemma \ref{lem:sound-setup1}\ref{lem:sound-setup1-b}.
\end{proofof}


\subsection{Supporting Results}


\begin{lemma}
\label{lem:sound-setup1}
    Let $\mb{X}$, $\mb{Y}$, $\mb{Z}$, and $\mb{S}$ be pairwise disjoint node sets in a causal DAG $\g[D]$, where $\mb{Z} \cap \De(\mb{X}, \g[D]) = \emptyset$ and where $\mb{S}$ satisfies the conditional adjustment criterion relative to $(\mb{X,Y,Z})$ in $\g[D]$ (Definition \ref{def:cac}). Let $\mb{H} \subseteq \An(\mb{X} \cup \mb{Y},\g[D]) \setminus (\De(\mb{X}, \g[D]) \cup \mb{Y} \cup \mb{Z})$ and $\mb{\tilde{S}} = \mb{S} \cup \mb{H}$. Then:
    \begin{enumerate}[label = (\roman*)]
        \item \label{lem:sound-setup1-a} $\mb{\tilde{S}}$ satisfies the conditional adjustment criterion relative to $(\mb{X,Y,Z})$ in $\g[D]$, and
        \item \label{lem:sound-setup1-b} $\int_\mb{s} f(\mb{y} | \mb{x,z,s})f(\mb{s} | \mb{z}) \diff \mb{s} = \int_{\mb{\tilde{s}}} f(\mb{y} | \mb{x,z,\tilde{s}}) f(\mb{\tilde{s}} | \mb{z})  \diff \mb{\tilde{s}}$, for any density $f$ consistent with $\g[D]$.
    \end{enumerate}
\end{lemma}

\begin{proofof}[Lemma \ref{lem:sound-setup1}] 
    This lemma is analogous to Lemma 59 of \cite{perkovic2018complete} (Lemma \ref{lem:sound-setup1-perk}). We use the same proof strategy and adapt the arguments to suit our needs.
    
    \ref{lem:sound-setup1-a} 
    By Lemma \ref{lem:comparison}\ref{lem:comparison-a}, since $\mb{S}$ satisfies the conditional adjustment criterion relative to $(\mb{X,Y,Z})$ in $\g[D]$, then $\mb{S} \cup \mb{Z}$ satisfies the adjustment criterion relative to $(\mb{X}, \mb{Y})$ in $\g[D]$. Then by Lemma \ref{lem:sound-setup1-perk}, $\mb{\tilde{S}} \cup \mb{Z}$ satisfies the adjustment criterion relative to $(\mb{X}, \mb{Y})$. The statement follows by a second use of Lemma \ref{lem:comparison}\ref{lem:comparison-a}.

    \ref{lem:sound-setup1-b} Let $f$ be an arbitrary density consistent with $\g[D]$. We proceed with a proof by induction.


    \textbf{Base case:} Suppose $\mb{H} = \{H\}$ so that $|\mb{H}|=1$. When $H \in \mb{S}$, the claim clearly holds. Thus, we let $H \notin \mb{S}$. Note that the claim holds if either $\mb{Y} \dsepp H \given \mb{X} \cup \mb{S} \cup \mb{Z}$ or $\mb{X} \dsepp H \given \mb{S} \cup \mb{Z}$ in $\g[D]$. To see this, we calculate the following.    
    \begin{enumerate}[label=\emph{(\alph*)}]
        \item \label{pf:sound-setup1-a}
        When $(\mb{Y} \dsepp H \given \mb{X} \cup \mb{S} \cup \mb{Z})_{\g[D]}$, then
        \begin{align}
            \int_\mb{s} f(\mb{y} | \mb{x,z,s})f(\mb{s} | \mb{z}) \diff \mb{s} 
                &= \int_\mb{s} f(\mb{y} | \mb{x,z,s})\int_{h}f(\mb{s},h | \mb{z}) \diff h \diff \mb{s} \nonumber \\
                &= \int_{\mb{s},h} f(\mb{y} | \mb{x,z,s}) f(\mb{s},h | \mb{z}) \diff \mb{s} \diff h  \nonumber \\
                &= \int_{\mb{\tilde{s}}} f(\mb{y} | \mb{x,z,\tilde{s}})f(\mb{\tilde{s}} | \mb{z}) \diff \mb{\tilde{s}}, \nonumber 
        \end{align}
        where the second equality holds since $H \notin \mb{Y} \cup \mb{X} \cup \mb{S} \cup \mb{Z}$.
    
        \item \label{pf:sound-setup1-b}
        When $(\mb{X} \dsepp H \given \mb{S} \cup \mb{Z})_{\g[D]}$, then 
        \begin{align}
            \int_\mb{s} f(\mb{y} | \mb{x,z,s})f(\mb{s} | \mb{z}) \diff \mb{s} 
                &= \int_{\mb{s}}f(\mb{s} | \mb{z})\int_{h} f(\mb{y},h | \mb{x,z,s}) \diff h \diff \mb{s} \nonumber \\
                &= \int_{\mb{s},h} f(\mb{y},h | \mb{x,z,s})f(\mb{s} | \mb{z}) \diff \mb{s} \diff h \nonumber \\
                &= \int_{\mb{s},h} f(\mb{y} | \mb{x,z,s},h)f(h | \mb{x,z,s})f(\mb{s} | \mb{z}) \diff \mb{s} \diff h \nonumber \\
                &= \int_{\mb{s},h} f(\mb{y} | \mb{x,z,s},h)f(h | \mb{z,s})f(\mb{s} | \mb{z}) \diff \mb{s} \diff h \nonumber \\
                &= \int_{\mb{\tilde{s}}} f(\mb{y} | \mb{x,z,\tilde{s}}) f(\mb{\tilde{s}} | \mb{z}) \diff \mb{\tilde{s}}, \nonumber 
        \end{align}
        where the second equality holds since $H \notin \mb{S} \cup \mb{Z}$.
    \end{enumerate}

    We use the remainder of the base case to show that \ref{pf:sound-setup1-a} or \ref{pf:sound-setup1-b} must hold. For sake of contradiction, suppose that neither hold. This implies that there are two paths in $\g[D]$: one from $\mb{X}$ to $H$ that is d-connecting given $\mb{S} \cup \mb{Z}$ and one from $\mb{Y}$ to $H$ that is d-connecting given $\mb{X} \cup \mb{S} \cup \mb{Z}$. Let $p=\langle X, \ldots, H \rangle$, $X \in \mb{X}$, and $q=\langle H, \ldots, Y \rangle$, $Y \in \mb{Y}$, be such paths, respectively, where $p$ is proper. In the arguments below, we use paths related to $p$ and $q$ -- in the proper back-door graph $\dpbd{\mb{XY}}$ (see Definition \ref{def:pbd}) and in four of its moral induced subgraphs (see Definition \ref{def:moral}) -- before applying Theorems \ref{thm:richardson} and \ref{thm:ac-alt} to reach our final contradiction (that $\mb{S}$ cannot satisfy the conditional adjustment criterion relative to $(\mb{X,Y,Z})$ in $\g[D]$).

    First, we claim that both $p$ and $q$ are d-connecting given $\mb{S} \cup \mb{Z}$. This holds for $p$ by definition. For sake of contradiction, suppose that $q$ is blocked by $\mb{S} \cup \mb{Z}$. Since $q$ is d-connecting given $\mb{X} \cup \mb{S} \cup \mb{Z}$, it must contain a collider in $\An(\mb{X},\g[D]) \setminus \An(\mb{S} \cup \mb{Z},\g[D])$. Let $C$ be the closest collider to $Y$ on $q$ such that $C \in (\An(\mb{X},\g[D]) \setminus \An(\mb{S} \cup \mb{Z},\g[D])) \cup \mb{X}$, and let $r=\langle C,\ldots,X' \rangle, X' \in \mb{X}$, be a shortest causal path in $\g[D]$ from $C$ to $\mb{X}$. Then let $V$ be the node closest to $X'$ on $r$ that is also on $q(C,Y)$, and define the path $t=(-r)(X',V) \oplus q(V,Y)$. Note that $t$ is non-causal since either $(-r)(X',V)$ is of non-zero length or $X'=V=C$, so that $t$ is a path into $X'$. Further, by the definitions of $q$, $C$, and $r$, we have that $t$ is proper non-causal path from $\mb{X}$ to $\mb{Y}$ that is d-connecting given $\mb{S} \cup \mb{Z}$. But this contradicts that $\mb{S}$ satisfies the conditional adjustment criterion relative to $(\mb{X,Y,Z})$ in $\g[D]$.

    Next, we prove that the sequence of nodes in $\gpbd[D]{XY}$ corresponding to $p$ forms a path. Note that since $p$ is proper, we only need to show that $p$ does not start with an edge $X \rightarrow W$, where $W$ is a node that lies on a proper causal path  in $\g[D]$ from $X$ to $\mb{Y}$. For sake of contradiction, suppose that $p$ starts with $X \rightarrow W$ for such a $W \in \fb{\g[D]}$. Note that $p$ cannot be causal from $X$ to $H$, since $H \notin \De(\mb{X},\g[D])$ by the definition of $\mb{H}$. Thus, $p$ is non-causal and there is a collider $C'$ on $p$ such that $C' \in \De(W,\g[D])$. Since $p$ is d-connecting given $\mb{S} \cup \mb{Z}$ and $\mb{Z} \cap \De(\mb{X}, \g[D]) = \emptyset$, then $\mb{S} \cap \De(C',\g[D]) \neq \emptyset$. Further, since $\De(C',\g[D]) \subseteq \fb{\g[D]}$, this implies that $\mb{S} \cap \fb{\g[D]} \neq \emptyset$. But this contradicts that $\mb{S}$ satisfies the conditional adjustment criterion relative to $(\mb{X,Y,Z})$ in $\g[D]$.

    Similarly, we prove that the sequence of nodes in $\gpbd[D]{XY}$ corresponding to $q$ also forms a path. For this, note that all nodes in $\mb{X}$ on $q$ must be a colliders on $q$, since $q$ is d-connecting given $\mb{X} \cup \mb{S} \cup \mb{Z}$. Thus, removing edges out of $\mb{X}$ from $\g[D]$ in order to form $\gpbd[D]{XY}$ will not affect the edges on $q$.

    Let $\tilde{p}$ and $\tilde{q}$ be the paths in $\gpbd[D]{XY}$ corresponding to $p$ and $q$, respectively. Then for sake of contradiction, suppose either $\tilde{p}$ or $\tilde{q}$ is blocked given $\mb{S} \cup \mb{Z}$. Since $p$ and $q$ are d-connecting given $\mb{S} \cup \mb{Z}$, then there must be a node $C$ on $p$ or $q$ where $C$ is a collider on $p$ or $q$ and every causal path in $\g[D]$ from $C$ to $\mb{S} \cup \mb{Z}$ contains the first edge of a proper causal path from $\mb{X}$ to $\mb{Y}$ in $\g[D]$. Let $d$ be an arbitrary such causal path in $\g[D]$ from $C$ to $\mb{S} \cup \mb{Z}$. Note that $d$ is a path from $C$ to $\mb{S}$, since $d$ must contain a node in $\mb{X}$ and since $\mb{Z} \cap \De(\mb{X},\g[D]) = \emptyset$. But since $d$ contains the first edge of a proper causal path from $\mb{X}$ to $\mb{Y}$ in $\g[D]$, this implies that $\mb{S} \cap \fb{\g[D]} \neq \emptyset$, which contradicts that $\mb{S}$ satisfies the conditional adjustment criterion relative to $(\mb{X,Y,Z})$ in $\g[D]$.

    We continue the base case by reasoning with four moral induced subgraphs of $\gpbd[D]{XY}$ (see Definition \ref{def:moral}). Start by defining the following. 
    \begin{align}
        \mb{A}_{\mb{XHYSZ}} &= \An(\mb{X} \cup \mb{H} \cup \mb{Y} \cup \mb{S} \cup \mb{Z},\gpbd[D]{XY}). \nonumber \\
        \mb{A}_{\mb{XYSZ}} &= \An(\mb{X} \cup \mb{Y} \cup \mb{S} \cup \mb{Z},\gpbd[D]{XY}). \nonumber \\
        \mb{A}_{\mb{XHSZ}} &= \An(\mb{X} \cup \mb{H} \cup \mb{S} \cup \mb{Z},\gpbd[D]{XY}). \nonumber \\
        \mb{A}_{\mb{HYSZ}} &= \An(\mb{H} \cup \mb{Y} \cup \mb{S} \cup \mb{Z},\gpbd[D]{XY}). \nonumber 
    \end{align} 
    Then define $\g[D]_{\mb{XHYSZ}}$, $\g[D]_{\mb{XYSZ}}$, $\g[D]_{\mb{XHSZ}}$, and $\g[D]_{\mb{HYSZ}}$ to be the moral induced subgraphs of $\gpbd[D]{XY}$ on nodes $\mb{A}_{\mb{XHYSZ}}$, $\mb{A}_{\mb{XYSZ}}$, $\mb{A}_{\mb{XHSZ}}$, and $\mb{A}_{\mb{HYSZ}}$, respectively. In order to use Theorem \ref{thm:ac-alt}, we want to show that $\g[D]_{\mb{XYSZ}}$ contains a path from $\mb{X}$ to $\mb{Y}$ that does not contain a node in $\mb{S} \cup \mb{Z}$.

    Since $\tilde{p}$ and $\tilde{q}$ are d-connecting given $\mb{S} \cup \mb{Z}$, then by Theorem \ref{thm:richardson}, the following two paths must exist in $\g[D]_{\mb{XHSZ}}$: path $a$ from $X$ to $H$ and path $b$ from $H$ to $Y$, where neither path contains a node in $\mb{S} \cup \mb{Z}$. Note that since $\mb{A}_{\mb{XHSZ}} \subseteq \mb{A}_{\mb{XHYSZ}}$ and $\mb{A}_{\mb{HYSZ}} \subseteq \mb{A}_{\mb{XHYSZ}}$, any path in $\g[D]_{\mb{XHSZ}}$ or $\g[D]_{\mb{HYSZ}}$ will also be in $\g[D]_{\mb{XHYSZ}}$. Further, since $H \in \An(\mb{X} \cup \mb{Y},\g[D])$ by definition and since we form $\gpbd[D]{XY}$ by removing edges out of $\mb{X}$ from $\g[D]$, then $H \in \An(\mb{X} \cup \mb{Y},\gpbd[D]{XY})$. Therefore, $\mb{A}_{\mb{XHYSZ}} = \mb{A}_{\mb{XYSZ}}$ and $\g[D]_{\mb{XHYSZ}}=\g[D]_{\mb{XYSZ}}$. Thus, $a$ and $b$ are both paths in $\g[D]_{\mb{XYSZ}}$.

    We complete the base case by applying Theorems \ref{thm:richardson} and \ref{thm:ac-alt} to show our necessary contradiction. Since we can combine subpaths of $a$ and $b$ to form a path $c$ in $\g[D]_{\mb{XYSZ}}$ from $\mb{X}$ to $\mb{Y}$ that does not contain a node in $\mb{S} \cup \mb{Z}$, then by Theorem \ref{thm:richardson}, $\mb{X}$ and $\mb{Y}$ are d-connecting given $\mb{S} \cup \mb{Z}$ in $\gpbd[D]{XY}$. By Theorem \ref{thm:ac-alt}, this implies that $\mb{S} \cup \mb{Z}$ does not satisfy the adjustment criterion relative to $(\mb{X,Y})$ in $\g[D]$ (see Definition \ref{def:ac}). Therefore, by the contraposition of Lemma \ref{lem:comparison}\ref{lem:comparison-a}, $\mb{S}$ does not satisfy the conditional adjustment criterion relative to $(\mb{X,Y,Z})$ in $\g[D]$, which is a contradiction.


    \textbf{Induction step:} Assume that the result holds for $|\mb{H}| = k$, $k \in \mathbb{N}$, and let $|\mb{H}| = k+1$. Take an arbitrary $H \in \mb{H}$, and define $\mb{S'} = \mb{S} \cup \{H\}$ and $\mb{H'} = \mb{H} \setminus \{H\}$. Since the base case holds and since $\{H\} \subseteq \An(\mb{X} \cup \mb{Y},\g[D]) \setminus (\De(\mb{X}, \g[D]) \cup \mb{Y} \cup \mb{Z})$, then
    \begin{align}
        \int_\mb{s} f(\mb{y} | \mb{x,z,s})f(\mb{s} | \mb{z})  \diff \mb{s}
                &= \int_{\mb{s},h} f(\mb{y} | \mb{x,z,s},h) f(\mb{s},h | \mb{z})  \diff \mb{s} \diff h \nonumber \\
                &= \int_{\mb{s'}} f(\mb{y} | \mb{x,z,s'}) f(\mb{s'} | \mb{z})  \diff \mb{s'}. \label{eq:sound-setup1-1}
    \end{align}
    Further, by part \ref{lem:sound-setup1-a}, $\mb{S'}$ satisfies the conditional adjustment criterion relative to $(\mb{X,Y,Z})$ in $\g[D]$. Since $\mb{H'} \subseteq \An(\mb{X} \cup \mb{Y},\g[D]) \setminus (\De(\mb{X}, \g[D]) \cup \mb{Y} \cup \mb{Z})$ and $|\mb{H'}| = k$, then by the induction assumption,
    \begin{align}
        \int_\mb{s'} f(\mb{y} | \mb{x,z,s'})f(\mb{s'} | \mb{z})  \diff \mb{s'} 
            &= \int_{\mb{s',h'}} f(\mb{y} | \mb{x,z,s',h'}) f(\mb{s',h'} | \mb{z})  \diff \mb{s'} \diff \mb{h'} \nonumber \\
            &= \int_{\mb{\tilde{s}}} f(\mb{y} | \mb{x,z,\tilde{s}}) f(\mb{\tilde{s}} | \mb{z}) \diff \mb{\tilde{s}}. \label{eq:sound-setup1-2}
    \end{align}    
    Combining \eqref{eq:sound-setup1-1} and \eqref{eq:sound-setup1-2} yields the desired result.
\end{proofof}


\begin{lemma} 
\label{lem:sound-setup2}
    Let $\mb{X},\mb{Y}, \mb{Z}$, and $\mb{S}$ be pairwise disjoint node sets in a causal DAG $\g[D]$, where $\mb{Z} \cap \De(\mb{X}, \g[D]) = \emptyset$ and where $\mb{S}$ satisfies the conditional adjustment criterion relative to $(\mb{X,Y,Z})$ in $\g[D]$ (Definition \ref{def:cac}). Let $\mb{H} = \An(\mb{X} \cup \mb{Y},\g[D]) \setminus (\De(\mb{X}, \g[D]) \cup \mb{Y} \cup \mb{Z})$ and $\mb{\tilde{S}} = \mb{S} \cup \mb{H}$. Additionally, let $\mb{\tilde{S}_D} = \mb{\tilde{S}} \cap \De(\mb{X}, \g[D])$, $\mb{\tilde{S}_N} = \mb{\tilde{S}} \setminus \De(\mb{X}, \g[D])$, $\mb{Y_D} = \mb{Y} \cap \De(\mb{X},\g[D])$, and $\mb{Y_N} = \mb{Y} \setminus \De(\mb{X}, \g[D])$. Then the following statements hold:
    \begin{enumerate}[label=(\roman*)] 

        \item\label{lem:sound-setup2-1} $(\mb{X} \cup \mb{Y_N} \cup \mb{\tilde{S}} \cup \mb{Z}) \cap \fb{\g[D]} = \emptyset$,

        \item\label{lem:sound-setup2-2} if $p = \langle H, \dots, Y_D \rangle$ is a non-causal path in $\g[D]$ from $H \in \mb{X} \cup \mb{Y_N} \cup \mb{\tilde{S}} \cup \mb{Z}$ to a node $Y_D \in \mb{Y_D}$, then $p$ is blocked by $(\mb{X} \cup \mb{Y_N} \cup \mb{\tilde{S}_N} \cup \mb{Z}) \setminus \{H\}$,

        \item \label{lem:sound-setup2-3} $\mb{Y_D} \dsepp \mb{\tilde{S}_D} \given \mb{Y_N} \cup \mb{X} \cup \mb{\tilde{S}_N} \cup \mb{Z}$ in $\g[D]$,

        \item \label{lem:sound-setup2-4} if $\mb{Y_N} = \emptyset$, then $\mb{\tilde{S}_N}$ is a conditional adjustment set relative to $(\mb{X,Y,Z})$ in $\g[D]$ (Definition \ref{def:cas}),

        \item \label{lem:sound-setup2-5} the empty set is an adjustment set relative to $(\mb{X} \cup \mb{Y_N} \cup \mb{\tilde{S}_N} \cup \mb{Z}, \mb{Y_D})$ in $\g[D]$ (Definition \ref{def:as}),

        \item \label{lem:sound-setup2-6} $\mb{Y_D} \dsepp (\mb{Y_N} \cup \mb{\tilde{S}_N} \cup \mb{Z}) \given \mb{X}$ in $\g[D]_{\overline{\mb{X}}\underline{\mb{Y_N} \cup  \mb{\tilde{S}_N} \cup \mb{Z}}}$, and

        \item \label{lem:sound-setup2-7} $\mb{\tilde{S}_N} \dsepp \mb{X} \given \mb{Y_N} \cup \mb{Z}$ in $\g[D]_{\overline{\mb{X}}}$.
    \end{enumerate}
\end{lemma}

\begin{proofof}[Lemma \ref{lem:sound-setup2}]
    This lemma is analogous to Lemma 60 of \cite{perkovic2018complete} (Lemma \ref{lem:sound-setup2-perk}), which is needed for adjustment in total effect identification. We rely on this result in the proof below.

    Note that $\mb{X,Y}$, and $\mb{S} \cup \mb{Z}$ are pairwise disjoint node sets in $\g[D]$, where by Lemma \ref{lem:comparison}\ref{lem:comparison-a}, $\mb{S} \cup \mb{Z}$ satisfies the adjustment criterion relative to $(\mb{X}, \mb{Y})$ in $\g[D]$. Results \ref{lem:sound-setup2-1}-\ref{lem:sound-setup2-3} and \ref{lem:sound-setup2-6} follow directly from Lemma \ref{lem:sound-setup2-perk}. Result \ref{lem:sound-setup2-5} follows additionally from Theorem \ref{thm:gbc-as}. Result \ref{lem:sound-setup2-4} follows additionally from Theorem \ref{thm:gbc-as} and Lemma \ref{lem:comparison}\ref{lem:comparison-a}.

    \ref{lem:sound-setup2-7} Let $p$ be an arbitrary path from $X \in \mb{X}$ to $\mb{\tilde{S}_N}$ in $\g[D]_{\overline{\mb{X}}}$. By definition of $\g[D]_{\overline{\mb{X}}}$, $p$ begins with an edge out of $X$. Since, by definition, $\mb{\tilde{S}_N} \cap \De(\mb{X},\g[D]) = \emptyset$, where $\De(\mb{X},\g[D]_{\overline{\mb{X}}}) \subseteq \De(\mb{X},\g[D])$, then $p$ must contain at least one collider. Let $\mb{C}$ be the set containing the closest collider to $X$ on $p$ and its descendants in $\g[D]_{\overline{\mb{X}}}$. Note that $\mb{C} \subseteq \De(\mb{X},\g[D]_{\overline{\mb{X}}}) \subseteq \De(\mb{X},\g[D])$. By definition of $\mb{Y_N}$ and by assumption, $(\mb{Y_N} \cup \mb{Z}) \cap \De(\mb{X},\g[D]) = \emptyset$, and thus, $p$ is blocked by $\mb{Y_N} \cup \mb{Z}$.
\end{proofof}


\section{CONDITIONAL BACK-DOOR CRITERION}
\label{supp:backdoor}

This section extends Pearl's back-door criterion (\citeyear{pearl2009causality}) to the context of estimating a conditional causal effect in a DAG. Definition \ref{def:cbc} provides the extended criterion, and Lemma \ref{lem:cbc-cas} establishes that this criterion is sufficient for conditional adjustment. Lemma \ref{lem:gbc-cbc} makes a comparison between this criterion and the generalized back-door criterion of \cite{maathuis2015generalized} (Definition \ref{def:gbc}).


\begin{definition}
\label{def:cbc}
{\normalfont (\textbf{Conditional Back-door Criterion for DAGs})}
    Let $\mb{X}$, $\mb{Y}$, $\mb{Z}$, and $\mb{S}$ be pairwise disjoint node sets in a DAG $\g[D]$, where $\mb{Z} \cap \De(\mb{X},\g[D]) = \emptyset$. Then $\mb{S}$ satisfies the conditional back-door criterion relative to $(\mb{X},\mb{Y},\mb{Z})$ in $\g[D]$ if 
    \begin{enumerate}[label=\emph{(\alph*)}]
        \item $\mb{S} \cap \De(\mb{X},\g[D]) = \emptyset$, and
        \item $\mb{S} \cup \mb{Z}$ blocks all proper back-door paths from $\mb{X}$ to $\mb{Y}$.
    \end{enumerate}
\end{definition}


\begin{lemma}
\label{lem:cbc-cas}
    Let $\mb{X}$, $\mb{Y}$, $\mb{Z}$, and $\mb{S}$ be pairwise disjoint node sets in a causal DAG $\g[D]$, where  $\mb{Z} \cap \De(\mb{X},\g[D]) = \emptyset$. If $\mb{S}$ satisfies the conditional back-door criterion relative to $(\mb{X},\mb{Y},\mb{Z})$ in $\g[D]$ (Definition \ref{def:cbc}), then $\mb{S}$ is a conditional adjustment set relative to $(\mb{X},\mb{Y},\mb{Z})$ in $\g[D]$ (Definition \ref{def:cas}).
\end{lemma}

\begin{proofof}[Lemma \ref{lem:cbc-cas}]
    Let $\mb{S}$ be a set that satisfies the conditional back-door criterion relative to $(\mb{X},\mb{Y},\mb{Z})$ in $\g[D]$, and let $f$ be a density consistent with $\g[D]$. Then
    \begin{align}
        f(\mb{y}|do(\mb{x}),\mb{z}) &= \int_\mb{s} f(\mb{y},\mb{s}|do(\mb{x}),\mb{z}) \diff \mb{s} \nonumber \\
                     &= \int_\mb{s} f(\mb{y}|\mb{s},do(\mb{x}),\mb{z}) f(\mb{s}|do(\mb{x}),\mb{z}) \diff \mb{s} \nonumber \\
                     &= \int_\mb{s} f(\mb{y}|\mb{s},\mb{x},\mb{z}) f(\mb{s}|\mb{z}) \diff \mb{s} \nonumber
    \end{align}
    The first two equalities follow from the law of total probability and the chain rule. The third equality follows from Rules 2 and 3 of the do calculus (Equations \eqref{eq:rule2-do} and \eqref{eq:rule3-do}) and the d-separations shown below. 

    In order to use Rule 2 to conclude that $f(\mb{y}|\mb{s},do(\mb{x}),\mb{z})=f(\mb{y}|\mb{s},\mb{x},\mb{z})$, we show that $(\mb{Y} \dsepp \mb{X} \given \mb{S} \cup \mb{Z})_{\g[D]_{\underline{\mb{X}}}}$. Note that $\g[D]_{\underline{\mb{X}}}$ only contains back-door paths from $\mb{X}$ to $\mb{Y}$. So every path from $\mb{X}$ to $\mb{Y}$ in $\g[D]_{\underline{\mb{X}}}$ contains a proper back-door path from $\mb{X}$ to $\mb{Y}$ as a subpath. Since $\mb{S} \cup \mb{Z}$ blocks all proper back-door paths from $\mb{X}$ to $\mb{Y}$ in $\g[D]$, the d-separation holds. 

    In order to use Rule 3 to conclude that $f(\mb{s}|do(\mb{x}),\mb{z}) = f(\mb{s}|\mb{z})$, we show that $(\mb{S} \dsepp \mb{X} \given \mb{Z})_{\g[D]_{\overline{\mb{X}(\mb{Z})}}}$. This follows from the assumptions that $\mb{Z} \cap \De(\mb{X},\g[D]) = \emptyset$ and $\mb{S} \cap \De(\mb{X},\g[D]) = \emptyset$.
\end{proofof}


\begin{lemma}
\label{lem:gbc-cbc}
{\normalfont (\textbf{Comparison of Back-door Criteria for DAGs})}
    Let $\mb{X}$, $\mb{Y}$, $\mb{Z}$, and $\mb{S}$ be pairwise disjoint node sets in a DAG $\g[D]$, where $\mb{Z} \cap \De(\mb{X}, \g[D]) = \emptyset$. Then $\mb{S}$ satisfies the conditional back-door criterion relative to $(\mb{X,Y,Z})$ in $\g[D]$ (Definition \ref{def:cbc}) if and only if $\mb{S} \cup \mb{Z}$ satisfies the generalized back-door criterion relative to $(\mb{X}, \mb{Y})$ in $\g[D]$ (Definition \ref{def:gbc}).
\end{lemma}

\begin{proofof}[Lemma \ref{lem:gbc-cbc}] 
    $\Leftarrow:$ Follows immediately.
    
    $\Rightarrow:$ Since $\mb{S}$ satisfies the conditional back-door criterion relative to $(\mb{X}, \mb{Y}, \mb{Z})$ in $\g[D]$, then $\mb{S} \cap \De(\mb{X}, \g[D]) = \emptyset$. Combining this with our assumptions gives us that $(\mb{S} \cup \mb{Z}) \cap \De(\mb{X}, \g[D]) = \emptyset$. In the remainder of the proof, we show that $\mb{S} \cup \mb{Z} \cup \mb{X} \setminus \{X\}$ blocks all back-door paths from $\mb{X}$ to $\mb{Y}$. The result follows by Definition \ref{def:gbc}.

    Let $p_1$ be an arbitrary back-door path from $X \in \mb{X}$ to $Y \in \mb{Y}$ in $\g[D]$. For sake of contradiction, suppose that $p_1$ is d-connecting given $\mb{S} \cup \mb{Z} \cup \mb{X} \setminus \{X\}$. Let $X_C$ be the node in $\mb{X}$ closest to $Y$ on $p_1$, and let $p_2=p_1(X_C,Y)$. Note that $p_2$ is proper. When $X_C=X$, then $p_2=p_1$ is a back-door path. When $X_C \neq X$, then because $p_1$ is d-connecting given $\mb{S} \cup \mb{Z} \cup \mb{X} \setminus \{X\}$, we have that $X_C$ is a collider on $p_1$, and therefore, $p_2$ is again a back-door path. Thus, $p_2$ is a proper back-door path from $\mb{X}$ to $\mb{Y}$ that, by assumption, must be blocked by $\mb{S} \cup \mb{Z}$. 

    Let $A$ be the node on $p_2$ immediately following $X_C$. That is, $p_2$ contains $X_C \gets A$. Note that since $p_2$ is blocked given $\mb{S} \cup \mb{Z}$, then $A \neq Y$. Thus, we consider the path $p_3 = p_2(A,Y)$. Since $p_1$ is d-connecting given $\mb{S} \cup \mb{Z} \cup \mb{X} \setminus \{X\}$, where $A$ is a non-collider on $p_1$, then $A \notin \mb{S} \cup \mb{Z} \cup \mb{X} \setminus \{X\}$ and thus, $p_3$ is also d-connecting given $\mb{S} \cup \mb{Z} \cup \mb{X} \setminus \{X\}$. Similarly, since $p_2$ is blocked given $\mb{S} \cup \mb{Z}$, where $A$ is not a collider on $p_2$ and $A \notin \mb{S} \cup \mb{Z} \cup \mb{X} \setminus \{X\}$, then $p_3$ is also blocked by $\mb{S} \cup \mb{Z}$. 

    Since $p_3$ is d-connecting given $\mb{S} \cup \mb{Z} \cup \mb{X} \setminus \{X\}$ and blocked given $\mb{S} \cup \mb{Z}$, then $p_3$ must contain at least one collider in $\An(\mb{X} \setminus \{X\},\g[D]) \setminus \An(\mb{S} \cup \mb{Z},\g[D])$. Let $C$ be the closest such collider to $Y$ on $p_3$ and let $r=\langle C,\ldots,X' \rangle, X' \in \mb{X}$, be a shortest causal path from $C$ to $\mb{X}$ in $\g[D]$. While there must be a causal path from $C$ to $\mb{X} \setminus \{X\}$ in $\g[D]$, note that $r$ need not be one, and thus, we allow for the possibility that $X'=X$.

    Let $B$ be the node closest to $X'$ on $r$ that is also on $p_3(C,Y)$, and define the path $t=(-r)(X',B) \oplus p_3(B,Y)$. Note that since $p_2$ is proper, $(-r)(X',B)$ is at least of length one, and therefore, $t$ is a back-door path. Further, since $p_3$ is d-connecting given $\mb{S} \cup \mb{Z} \cup \mb{X} \setminus \{X\}$ and by the definition of $C$ and $r$, we have that $t$ is proper back-door path from $\mb{X}$ to $\mb{Y}$ that is d-connecting given $\mb{S} \cup \mb{Z}$. But this contradicts that $\mb{S}$ satisfies the conditional back-door criterion relative to $(\mb{X,Y,Z})$ in $\g[D]$.
\end{proofof}


\section{PROOFS FOR SECTION \ref{sec:constructing}: MPDAGS - CONSTRUCTING CONDITIONAL ADJUSTMENT SETS}
\label{supp:constructing}

This section includes the proofs of two results from Section \ref{sec:constructing}: Lemma \ref{lem:parent-set} and Theorem \ref{thm:adjust-and-o}. We also provide three supporting results needed for these proofs.


\subsection{Main Results}

\begin{proofof}[Lemma \ref{lem:parent-set}]
    By Lemma \ref{lem:henckel-e6}, $\Pa(X, \g)$ must satisfy condition \ref{def:cac-a} of Definition \ref{def:cac}, so it suffices to show that $\Pa(X, \g) \cup \mb{Z}$ blocks all non-causal definite status paths from $X$ to $\mb{Y}$ in $\g$. Note that since $\mb{Y} \cap \Pa(X, \g) = \emptyset$, any definite status path from $X$ to $\mb{Y}$ in $\g$ that starts with an edge into $X$ is blocked by $\Pa(X, \g) \cup \mb{Z}$. 
    
    Further, any non-causal definite status path from $X$ to $\mb{Y}$ in $\g$ that starts with an edge out of $X$ or an undirected edge must contain a collider. Additionally, the closest collider to $X$ on any such path and all of its descendants in $\g$ must be in $\PossDe(X, \g)$  by Lemma \ref{lem:concat}. Then since $\big[ \Pa(X, \g) \cup \mb{Z} \big] \cap \PossDe(X, \g) = \emptyset$, these paths are also blocked by $\Pa(X, \g) \cup \mb{Z}$.
\end{proofof}

\begin{proofof}[Theorem \ref{thm:adjust-and-o}]
    By Theorem \ref{thm:cac-cas}, it suffices to show that $\adjustb{\g}$ and $\optb{\g}$ separately satisfy the conditional adjustment criterion relative to $(\mb{X,Y,Z})$ in $\g$ (Definition \ref{def:cac}). We start by noting that $\adjustb{\g}$ and $\optb{\g}$ are both disjoint from $\fb{\g} \cup \mb{X} \cup \mb{Y} \cup \mb{Z}$, so it suffices to prove that \ref{pf:adjust-and-o-casea} $\adjustb{\g} \cup \mb{Z}$ and \ref{pf:adjust-and-o-caseb} $\optb{\g} \cup \mb{Z}$ block all proper non-causal definite status paths from $\mb{X}$ to $\mb{Y}$ in $\g$. We prove \ref{pf:adjust-and-o-casea} and \ref{pf:adjust-and-o-caseb} below. For these proofs, note that $\mb{Z} \cap \fb{\g} = \emptyset$ by the assumption that $\mb{Z} \cap \PossDe(\mb{X}, \g) = \emptyset$ and by Lemma \ref{lem:henckel-e6}. 

        \begin{enumerate*}[label=(\alph*)]
        \item\label{pf:adjust-and-o-casea}
        $\textbf{Adjust(}\mb{X,Y,Z}, \pmb{\g} \textbf{)} \pmb{\cup} \mb{Z} \textbf{:}$ Suppose for sake of contradiction that there is a proper non-causal definite status path from $\mb{X}$ to $\mb{Y}$ in $\g$ that is d-connecting given $\adjustb{\g} \cup \mb{Z}$. Let $p = \langle X, \dots, Y \rangle$ be a shortest such path.
        \end{enumerate*}

        Since $p$ is proper, no non-endpoint on $p$ is in $\mb{X}$. Suppose for sake of contradiction that there exists $Y' \in \mb{Y}$ that is a non-endpoint on $p$. By choice of $p$, this implies that $p(X,Y')$ is possibly causal. Then by Lemma \ref{lem:perkovic17-35}, since $p$ is non-causal, $p(Y',Y)$ must contain a collider on $p$. Let $C$ be the closest such collider to $Y'$ (possibly $C=Y'$). Note that by Lemma \ref{lem:perkovic17-35}, $C \in \PossDe(Y',\g)$, so by Lemma \ref{lem:concat}, $\De(C,\g) \subseteq \PossDe(Y',\g)$, where $Y' \in \PossDe(X,\g)$. Thus, $\De(C,\g) \subseteq \fb{\g}$. However, this contradicts that $p$ is d-connecting given $\adjustb{\g} \cup \mb{Z}$. Therefore, no non-endpoint on $p$ is in $\mb{X} \cup \mb{Y}$.

        We now consider cases \ref{pf:adjust-and-o-casea-1} and \ref{pf:adjust-and-o-casea-2} below.
        \begin{enumerate}[label = (\arabic*)]
            \item\label{pf:adjust-and-o-casea-1} 
            Consider when there is no collider on $p$. Since $p$ is d-connecting given $\adjustb{\g} \cup \mb{Z}$, no node on $p$ is in $\adjustb{\g} \cup \mb{Z}$. Then by Equation \eqref{eq:adjust-mpdag}, no node on $p$ is in $\PossAn(\mb{X \cup Y},\g) \setminus [\fb{\g} \cup \mb{X} \cup \mb{Y} \cup \mb{Z}]$. However, note that by Lemma \ref{lem:perkovic17-35}, every non-endpoint on $p$ is a possible ancestor of an endpoint on $p$ and thus is in $\PossAn(\mb{X \cup Y}, \g) \setminus (\mb{X} \cup \mb{Y} \cup \mb{Z})$. Combining these, we have that all non-endpoints on $p$ are in $\fb{\g}$. But this implies that there is no set that is both disjoint from $\fb{\g}$ and can block $p$. By Theorem \ref{thm:cac-cas}, this contradicts our assumption that there is a conditional adjustment set relative to $(\mb{X,Y,Z})$ in $\g$.
            
            \item\label{pf:adjust-and-o-casea-2} 
            Consider when there is at least one collider $C$ on $p$. For sake of contradiction, suppose that there are more than three nodes on $p$. Then there is a non-collider $B \notin \mb{X} \cup \mb{Y}$ such that $C \gets B$ or $B \to C$ is on $p$. Since $p$ is d-connecting given $\adjustb{\g} \cup \mb{Z}$, then $B \notin \mb{Z}$ and $B \in \An(\adjustb{\g} \cup \mb{Z}, \g)$. By Equation \eqref{eq:adjust-mpdag} and Lemma \ref{lem:concat}, $B \in [\PossAn(\mb{X \cup Y},\g) \cup \An(\mb{Z}, \g)] \setminus (\mb{X} \cup \mb{Y} \cup \mb{Z})$. Additionally, since $p$ is d-connecting given $\adjustb{\g} \cup \mb{Z}$, then $B \notin \adjustb{\g} \equiv [\PossAn(\mb{X \cup Y},\g) \cup \An(\mb{Z}, \g)] \setminus (\fb{\g} \cup \mb{X} \cup \mb{Y} \cup \mb{Z})$. Combining these, we have that $B \in \fb{\g}$. Since there is a causal path in $\g$ from $B$ to every node in $\De(C,\g)$, by Lemma \ref{lem:concat}, $\De(C,\g) \subseteq \fb{\g}$. However, this would contradict that $p$ is d-connecting given $\adjustb{\g} \cup \mb{Z}$.

            Hence, $p$ must be of the form $X \to C \leftarrow Y$, where $C \in \An(\adjustb{\g} \cup \mb{Z},\g)$ and thus by Equation \eqref{eq:adjust-mpdag} and Lemma \ref{lem:concat}, $C \in \PossAn(\mb{X \cup Y},\g) \cup \An(\mb{Z}, \g)$. Note that $C \notin \An(\mb{Z}, \g)$, since otherwise, $\mb{Z} \cap \PossDe(\mb{X}, \g) \neq \emptyset$. Further, $C \notin \PossAn(\mb{Y}, \g)$, because otherwise by Lemma \ref{lem:concat}, $C \in \possmediatb{\g}$, which would imply $\De(C,\g) \subseteq \fb{\g}$ which we have shown is a contradiction. Therefore, $C \in \PossAn(\mb{X}, \g)$.

            Let $q = \langle C = Q_1, \dots, Q_m= X'\rangle, m \ge 2$, be a shortest possibly causal path in $\g$ from $C$ to $\mb{X}$. Further, define the node $Q_j, j \in \{1, \dots, m\}$, as follows. When $q$ has no directed edges, let $Q_j = Q_m$. When $q$ has at least one directed edge, let $Q_j$ be the node on $q$ closest to $Q_1$ such that $Q_j \to Q_{j+1}$ is on $q$. Note that by Lemma \ref{lem:shortest-subseq}, $q$ is unshielded. Thus by R1 of \cite{meek1995causal}, $q$ takes the form $Q_1 - \dots - Q_j \to \dots \to Q_m$.

            Pause to consider the path $X \to Q_1 \gets Y$. Note that $X \gets Y$ cannot be in $\g$, because no set can block this proper non-causal definite status path from $\mb{X}$ to $\mb{Y}$ in $\g$. By Theorem \ref{thm:cac-cas}, this would contradict our assumption that there is a conditional adjustment set relative to $(\mb{X,Y,Z})$ in $\g$. Similarly, $X \to Y$ and $X - Y$ are not in $\g$, because this would imply $\De(C,\g) \subseteq \fb{\g}$, which we have shown is a contradiction. Thus, $X \to Q_1 \gets Y$ is an unshielded collider in $\g$.

            We complete this case by showing that $\g$ contains $X \to Q_j \gets Y$. If $j=1$, we are done. If instead $j>1$, then consider the node $Q_2$. Since $X \to Q_1 - Q_2$ and $Y \to Q_1 - Q_2$ are in $\g$, so is a path $\langle X,Q_2, Y \rangle$ by R1 of \cite{meek1995causal}. The unshielded paths $X \to Q_2 - Y$ and $X - Q_2 \leftarrow Y$ contradict that R1 of \cite{meek1995causal} is completed in $\g$. Further, the path $Q_2 \to Y \to Q_1 - Q_2$ or $Q_2 \to X \to Q_1 - Q_2$ contradicts that R2 of \cite{meek1995causal} is completed in $\g$, and the path $X - Q_2 - Y$ contradicts that R3 of \cite{meek1995causal} is completed in $\g$. This leaves only one option for $\langle X,Q_2, Y \rangle$, and that is $X \to Q_2 \leftarrow Y$.

            If $j=2$, we are done. If instead $j>2$, then we consider the node $Q_3$. By identical logic to that above, we can show that $\g$ contains $X \to Q_3 \gets Y$. Continuing in this way, we have that $\g$ contains $X \to Q_j \gets Y$.

            With this shown, we derive our final contradictions. When $j=m$, then $\g$ contains $X' \gets Y$. But this is a proper non-causal definite status path from $\mb{X}$ to $\mb{Y}$ that no set can block, which we have shown is a contradiction. When $j<m$, then $\g$ contains the following two paths: $X' \gets \dots \gets Q_j \gets Y$ and $X \to Q_j \leftarrow Y$. These paths are proper non-causal definite status paths from $\mb{X}$ to $\mb{Y}$ that cannot both be blocked by the same set, which again is a contradiction.
        \end{enumerate}

        \begin{enumerate*}[label=(\alph*)]\setcounter{enumi}{1}
        \item\label{pf:adjust-and-o-caseb}
        $\textbf{O(}\mb{X,Y}, \pmb{\g} \textbf{)} \pmb{\cup} \mb{Z} \textbf{:}$ Let $p'$ be an arbitrary proper non-causal definite status path from $X \in \mb{X}$ to $\mb{Y}$ in $\g$, and let $Y$ be the node in $\mb{Y}$ closest to $X$ on $p'$ such that $p'(X,Y)$ is still a proper non-causal definite status path from $\mb{X}$ to $\mb{Y}$ in $\g$. Then let $p = p'(X,Y)$, where $p=\langle X=V_1, \dots, V_k = Y \rangle, k \ge 2$. Additionally, note that by assumption, $Y \in \possmediatb{\g}$.
        \end{enumerate*}
        
        We now consider cases \ref{pf:adjust-and-o-caseb-1} and \ref{pf:adjust-and-o-caseb-2} below. In both cases, we show that $p$ -- and therefore $p'$ -- is blocked by $\optb{\g} \cup \mb{Z}$.
        \begin{enumerate}[label = (\arabic*)]
        
            \item\label{pf:adjust-and-o-caseb-1} Suppose that $p$ ends with $V_{k-1} \gets Y$ or $V_{k-1}-Y$. If $p$ has no colliders, then by Lemma \ref{lem:perkovic17-35}, $(-p)$ is a possibly causal path from $Y$ to $X$. Since $Y \in \possmediatb{\g}$, this implies that $V_2, \dots, V_{k-1} \in \fb{\g}$. But then there is no set that is both disjoint from $\fb{\g}$ and can block $p$. By Theorem \ref{thm:cac-cas}, this contradicts our assumption that there is a conditional adjustment set relative to $(\mb{X,Y,Z})$ in $\g$. Hence, there must be a collider on $p$.
            
            Let $C$ be the closest collider to $Y$ on $p$. By Lemma \ref{lem:perkovic17-35}, $C \in \PossDe(Y, \g)$. Thus by Lemma \ref{lem:concat}, $\De(C, \g) \subseteq \PossDe(Y, \g)$. By assumption, $Y \in \possmediatb{\g}$, which implies that $\De(C, \g) \subseteq \fb{\g}$. Since $\big[ \optb{\g} \cup \mb{Z} \big] \cap \fb{\g} = \emptyset$, $p$ is blocked by $\optb{\g} \cup \mb{Z}$. 
           
            \item\label{pf:adjust-and-o-caseb-2} Suppose that $p$ ends with $V_{k-1} \to Y$. Note that $p$ is not a possibly causal path from $X$ to $Y$, so by Lemma \ref{lem:perkovic17-35}, there must be an edge $V_{i-1} \gets V_{i}$, $i \in \{2,\dots,k-1\}$, on $p$. In particular, let $V_i$ be the closest node to $Y$ on $p$ such that $V_{i-1} \gets V_{i}$ is on $p$. 
            
            In order to complete this proof, we want to show that either $\{V_i, \dots, V_{k-1}\} \cap [\optb{\g} \cup \mb{Z}] \neq \emptyset$ or $\{V_i, \dots, V_{k-1}\} \subset \possmediatb{\g}$. In both cases, we will show that $p$ is blocked by $\optb{\g} \cup \mb{Z}$. To do this, we briefly note that by the choice of $V_i$, the path $p(V_i, Y)$ is possibly causal and every node in $\{V_i, \dots, V_{k-1}\}$ is a non-collider on $p$. Further by the choice of $p$, no node in $\{V_i, \dots, V_{k-1}\}$ is in $\mb{X} \cup \mb{Y}$. We turn to consider each node in $\{V_i, \dots, V_{k-1}\}$, working backward through the set.

            Consider the node $V_{k-1}$. If $V_{k-1} \in \optb{\g} \cup \mb{Z}$, then since $V_{k-1}$ is a non-collider on $p$, $p$ is blocked by $\optb{\g} \cup \mb{Z}$, and we are done. Consider when $V_{k-1} \notin \optb{\g} \cup \mb{Z}$. Since $Y \in \possmediatb{\g}$ and since $V_{k-1} \to Y$ is in $\g$, then either $V_{k-1} \in \possmediatb{\g}$ or $V_{k-1} \in \Pa(\possmediatb{\g},\g)$. We show the latter is impossible. If $V_{k-1} \in \Pa(\possmediatb{\g},\g)$ and $V_{k-1} \notin \optb{\g} \cup \mb{Z}$, then by Equation \eqref{eq:o-set}, we have that $V_{k-1} \in \fb{\g}$. But by Lemma \ref{lem:henckel-e6}, this implies that $V_{k-1} \in \De(\mb{X},\g)$. Since $\g$ contains $V_{k-1} \to Y$, then $V_{k-1} \in \possmediatb{\g}$. But this contradicts that $V_{k-1} \in \Pa(\possmediatb{\g},\g)$ by the definition of a parent set. Therefore, either $V_{k-1} \in \optb{\g} \cup \mb{Z}$ and we are done, or $V_{k-1} \in \possmediatb{\g}$.
            
            In the latter case, we turn to consider $V_{k-2}$ if such a node exists. If $p$ contains $V_{k-2} \to V_{k-1}$, then since $V_{k-1} \in \possmediatb{\g}$, we can use the same logic as above to show that either $V_{k-2} \in \optb{\g} \cup \mb{Z}$ and we are done, or $V_{k-2} \in \possmediatb{\g}$. If $p$ contains $V_{k-2} - V_{k-1}$, then since $V_{k-1} \in \possmediatb{\g}$, we have that $V_{k-2} \in \fb{\g} \subseteq \De(\mb{X},\g)$. Because $p(V_{k-2},Y)$ is possibly causal, then by Lemma \ref{lem:concat}, $V_{k-2} \in \possmediatb{\g}$.
            
            Working backward in this way, either a node on $p(V_i, Y)$ is in $\optb{\g} \cup \mb{Z}$ and we are done, or $V_j \in \possmediatb{\g}$ for all $j \in \{i, \dots, k-1\}$. In the latter case, we have that $V_i \in \possmediatb{\g}$ and that every node in $\{V_i, \dots, V_{k-1}\} \subseteq \possmediatb{\g} \subseteq \fb{\g}$ is a non-collider on $p$. We can now apply the same argument as in \ref{pf:adjust-and-o-caseb-1} above to show that $p(X,V_i)$ -- and therefore $p$ -- is blocked given $\optb{\g} \cup \mb{Z}$. 
       \end{enumerate}
\end{proofof}


\subsection{Supporting Results}

\begin{lemma}
\label{lem:shortest-subseq}
    Let $X$ and $Y$ be distinct nodes in an MPDAG $\g=(\mb{V},\mb{E})$ and let $p$ be a possibly causal path from $X$ to $Y$ in $\g$. Then any shortest subsequence of $p$ forms an unshielded, possibly causal path from $X$ to $Y$. 
\end{lemma}

\begin{proofof}[Lemma \ref{lem:shortest-subseq}]
    This result is similar to Lemma 3.6 of \cite{perkovic2017interpreting}, but we derive a slightly more general statement.
    
    Let $k$ be the number of nodes on $p$. Pick an arbitrary shortest subsequence of $p$ and call it $p^*$, where $p^*=\langle X=V_0, \dots, V_\ell=Y \rangle$, $0 < \ell \le k$. Note that there is no edge $V_i \gets V_j, 0 \le i < j \le k$ in $\g$, since this would contradict that $p$ is possibly causal. Thus, $p^*$ is also possibly causal by definition. Further note that $p^*$ is unshielded, since if any triple on the path is shielded, it either contradicts that $p^*$ is possibly causal (i.e. $V_i \gets V_{i+2}$ cannot be in $p^*$) or that $p^*$ is a shortest subsequence of $p$ (i.e. $V_i \to V_{i+2}$ and $V_i - V_{i+2}$ cannot be in $p^*$).
\end{proofof}


\begin{lemma} 
\label{lem:nobackpaths}
    Let $p=\langle P_0, \dots, P_k \rangle$ be a path in an MPDAG $\g$. Then $p$ is possibly causal if and only if $\g$ does not contain any path $P_i \gets \dots \gets P_j$, $0 \le i < j \le k$.
\end{lemma}

\begin{proofof}[Lemma \ref{lem:nobackpaths}]
    Suppose that $\g$ does not contain any path $P_i \gets \dots \gets P_j$, $0 \le i < j \le k$. Then $\g$ does not contain any edge $P_i \gets P_j$, $0 \le i < j \le k$. Therefore, by definition, $p$ is possibly causal in $\g$.

    Now suppose $p$ is possibly causal in $\g$. For sake of contradiction, suppose $\g$ contains a path $q$ from $P_i$ to $P_j$, $0 \le i < j \le k$, of the form $P_i=Q_0 \gets Q_1 \gets \dots \gets Q_{\ell-1} \gets Q_{\ell} =P_j$.

    Consider the subpath of $p$ from $P_i$ to $P_j$. Note that this subpath is a possibly causal path. Let $r=\langle P_i = R_0, R_1, \dots, R_m= P_j\rangle$ be a shortest subsequence of this subpath. By Lemma \ref{lem:shortest-subseq}, $r$ is an unshielded, possibly causal path.

    Consider the edge $r(R_0, R_1)$. $R_0 \gets R_1$ cannot be in $r$, since $r$ is possibly causal. Neither is $R_0 \to R_1$ in $r$ since $r$ being unshielded would imply, by R1 of \cite{meek1995causal}, that $\g$ contains the cycle $P_i= R_0 \to R_1 \to \dots \to R_m=P_j=Q_{\ell} \to Q_{\ell-1} \to \dots \to Q_0=P_i$. Thus $r$ contains $R_0 - R_1$. 

    However, note that no DAG in $[\g]$ can contain the edge $R_0 \to R_1$, since $r$ being unshielded would imply, by R1 of \cite{meek1995causal}, that the DAG contains the cycle $P_i= R_0 \to R_1 \to \dots \to R_m=P_j=Q_{\ell} \to Q_{\ell-1} \to \dots \to Q_0=P_i$. This contradicts that $r$ contains $R_0 - R_1$. Thus we conclude that $\g$ does not contain any path $P_i \gets \dots \gets P_j$, $0 \le i < j \le k$.
\end{proofof}


\begin{lemma}
\label{lem:concat}
    Let $X$, $Y$, and $Z$ be distinct nodes in an MPDAG $\g$.
    \begin{enumerate}[label=(\roman*)]
        \item \label{lem:concat-pc-p} 
        If $p$ is a possibly causal path from $X$ to $Y$ and $q$ is a causal path from $Y$ to $Z$, then $p \oplus q$ is a possibly causal path from $X$ to $Z$.
    
        \item \label{lem:concat-c-pc} 
        If $p$ is a causal path from $X$ to $Y$ and $q$ is a possibly causal path from $Y$ to $Z$, then $p \oplus q$ is a possibly causal path from $X$ to $Z$.
    \end{enumerate}
\end{lemma}

\begin{proofof}[Lemma \ref{lem:concat}]
    Let $p= \langle X=P_0, P_1, \dots, P_k=Y \rangle$ and let $q= \langle Y=Q_0, Q_1, \dots, Q_r=Z \rangle$. Before beginning the main arguments, we note that $p$ and $q$ cannot share any nodes other than $Y$, and thus, we can define a path $p \oplus q$. To see this, for sake of contradiction, suppose $p$ and $q$ share at least one node other than $Y$. Let $\mb{S}$ denote the collection of such nodes, and consider the node in $\mb{S}$ with the lowest index on $q$. That is, consider $Q_j \in \mb{S}$ such that $j \le \ell$ for all $Q_{\ell} \in \mb{S}$. Let $Q_j=P_i$ for some $P_i \neq Y$ on $p$. Note that since $q$ or $p$ is causal, $\g$ contains either $P_k=Q_0 \to Q_1 \to \dots \to Q_j=P_i$ or $Q_j=P_i \to P_{i+1} \to \dots \to Y = Q_0$. By Lemma \ref{lem:nobackpaths}, the first option contradicts that $p$ is possibly causal and the second contradicts that $q$ is possibly causal. Thus we conclude that $p$ and $q$ cannot share any nodes other than $Y$.
    
    For $p \oplus q$ to be possibly causal in $\g$ we only need to show that there is no backward edge between any two nodes on $p \oplus q$. Note that there is no edge $P_{i_1} \gets P_{j_1}$ for $0 \le i_1 < j_1 \le k$, or $Q_{i_2} \gets Q_{j_2}$ for $0 \le i_2 < j_2 \le r$ in $\g$, by choice of $p$ and $q$. 
    
    \ref{lem:concat-pc-p}
    Assume for sake of contradiction that there exists an edge $P_i \gets Q_j$ in $\g$ for $i \in \{0, \dots, k-1\}$ and $j \in \{1, \dots, r\}$. Note that $P_i$ is on $p$ and not $q$, and analogously, $Q_j$ is on $q$ and not $p$, since we have shown $p$ and $q$ cannot share nodes other than $Y$. Also note that since $q$ is causal, it contains $Y \to Q_1 \to \dots \to Q_j$.
    
    Consider the subpath $p(P_i, Y)$. Since $p$ is possibly causal, so is this subpath. Pick an arbitrary shortest subsequence of $p( P_i, Y)$ and call it $t$, where $t=\langle P_i=T_0, \dots, T_m=Y \rangle$, $m \ge 1$. By Lemma \ref{lem:shortest-subseq}, $t$ forms an unshielded, possibly causal path from $P_i$ to $Y$.
    
    Consider the edge $t(P_i, T_1)$. Edge $P_i \gets T_1$ cannot be on $t$, since $t$ is possibly causal. Then $P_i \to T_1$ or $P_i - T_1$ must be in $\g$. However, note that no DAG in $[\g]$ can contain the edge $P_i \to T_1$, since $t$ being unshielded would imply, by R1 of \cite{meek1995causal}, that the DAG contains the cycle $P_i \to T_1 \to \dots \to Y \to \dots \to Q_j \to P_i$. This contradicts that $t$ contains $P_i - T_1$ or $P_i \to T_1$. Thus, there does not exist an edge $P_i \gets Q_j$ in $\g$. 
    
    \ref{lem:concat-c-pc}
    Assume for sake of contradiction that there exists an edge $P_i \gets Q_j$ in $\g$ for $i \in \{0, \dots, k-1\}$ and $j \in \{1, \dots, r\}$. Note that $P_i$ is on $p$ and not $q$, and analogously, $Q_j$ is on $q$ and not $p$, since we have shown $p$ and $q$ cannot share nodes other than $Y$. Also note that since $p$ is causal, it contains $P_i \to P_{i+1} \to \dots \to Y$.
    
    Consider the subpath $q( Y, Q_j)$. Since $q$ is possibly causal, so is this subpath. Pick an arbitrary shortest subsequence of $q( Y, Q_j)$ and call it $t$, where $t=\langle Y=T_0, \dots, T_m=Q_j \rangle$, $m \ge 1$. By Lemma \ref{lem:shortest-subseq}, $t$ forms an unshielded, possibly causal path from $Y$ to $Q_j$.
    
    Consider the edge $t(Y, T_1)$. Edge $Y \gets T_1$ cannot be on $t$, since $t$ is possibly causal. Then $Y \to T_1$ or $Y - T_1$ must be in $\g$. However, note that no DAG in $[\g]$ can contain the edge $Y \to T_1$, since $t$ being unshielded would imply, by R1 of \cite{meek1995causal}, that the DAG contains the cycle $Y \to T_1 \to \dots \to Q_j \to P_i \to P_{i+1} \to \dots \to Y$. This contradicts that $t$ contains $Y - T_1$ or $Y \to T_1$. Thus, there does not exist an edge $P_i \gets Q_j$ in $\g$.
\end{proofof}


\section{PROOF FOR SECTION \ref{sec:cond-adj-pag}: PAGS - CONDITIONAL ADJUSTMENT CRITERION}
\label{supp:comparison-pag}

This section includes the proof of Theorem \ref{thm:cac-cas-pag} and one result (Lemma \ref{lem:equiv-z-pag}) needed for the proof of Lemma \ref{lem:comparison-pag}. The statements of Theorem \ref{thm:cac-cas-pag} and Lemma \ref{lem:comparison-pag} can be found in Section \ref{sec:cond-adj-pag}.

Figure \ref{fig:proof-map-pag} shows how the results in this paper fit together to prove Theorem \ref{thm:cac-cas-pag}. Note that Theorem \ref{thm:cac-cas-pag} is an analogous result to Theorem \ref{thm:cac-cas} (Section \ref{sec:cond-adj}), where the former applies to PAGs and the latter to MPDAGs. However, while the proof of Theorem \ref{thm:cac-cas} relies directly on completeness and soundness proofs for DAGs (see Figure \ref{fig:proof-map} in Supplement \ref{supp:adjustment}), the proof of Theorem \ref{thm:cac-cas-pag} relies on them indirectly through Theorem \ref{thm:cac-cas}.

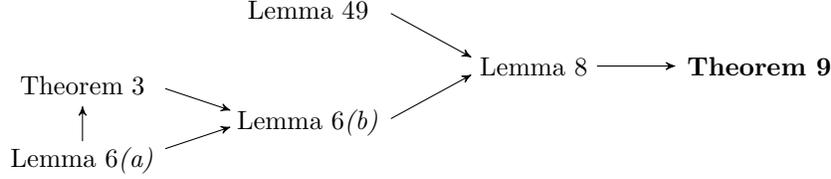
\begin{figure}
    \centering
    \begin{tikzpicture}[>=stealth',shorten >=1pt,node distance=3cm, main node/.style={minimum size=0.4cm}]
    [>=stealth',shorten >=1pt,node distance=3cm,initial/.style    ={}]
    \node[main node]               (T9)                  {\textbf{Theorem \ref{thm:cac-cas-pag}}};
    \node[main node]               (L8)   [left of=T3]   {Lemma \ref{lem:comparison-pag}};
    \node[main node,yshift=-.75cm] (L6B)  [left of=L8]   {Lemma \ref{lem:comparison}\ref{lem:comparison-b}};
    \node[main node,yshift=.75cm]  (L52)  [left of=L8]   {Lemma \ref{lem:equiv-z-pag}};
    \node[main node,yshift=.5cm]   (T3)   [left of=L6B]  {Theorem \ref{thm:cac-cas}};
    \node[main node,yshift=-.5cm]  (L6A)  [left of=L6B]  {Lemma \ref{lem:comparison}\ref{lem:comparison-a}};
    \draw[->] (-4.9, .7)   to (-3.8,  .1);
    \draw[->] (-4.9,-.7)   to (-3.8, -.1);
    \draw[->] (-7.9,-.3)   to (-7,-.6);
    \draw[->] (-7.9,-1.1)  to (-7,-.8);
    \draw[->] (-9,-1.0)    to (-9,-.5);
    \draw[->] (L8) to (T9);
    \end{tikzpicture}
    \caption{Proof structure of Theorem \ref{thm:cac-cas-pag}.}
    \label{fig:proof-map-pag}
\end{figure}

\begin{proofof}[Theorem \ref{thm:cac-cas-pag}]
    Follows from Lemma \ref{lem:comparison-pag} and Theorem \ref{thm:ac-as}.
\end{proofof}

\begin{lemma}
\label{lem:equiv-z-pag} 
    Let $\mb{X}$ and $\mb{Z}$ be disjoint node sets in a PAG $\g$. Then the following statements are equivalent.
    \begin{enumerate}[label = (\roman*)]
        \item $\mb{Z} \cap \PossDe(\mb{X},\g) = \emptyset$. \label{lem:equiv-z-pag-a}
        \item $\mb{Z} \cap \De(\mb{X},\g[D]) = \emptyset$ in every DAG $\g[D]$ represented by $\g$. \label{lem:equiv-z-pag-b}
    \end{enumerate}
\end{lemma}

\begin{proofof}[Lemma \ref{lem:equiv-z-pag}]
    $\neg \ref{lem:equiv-z-pag-a} \Rightarrow \neg \ref{lem:equiv-z-pag-b}$ Let $p$ be a possibly causal path from $\mb{X}$ to $\mb{Z}$ in $\g=(\mb{V},\mb{E})$ and let $p^*=\langle X=V_0, \dots, V_k=Z \rangle$, $k \ge 1$, $X \in \mb{X}$, $Z \in \mb{Z}$, be an unshielded possibly causal subsequence of $p$ in $\g$. 

    Since $p^*$ contains $X \circcirc V_1$, $X \circarrow V_1$ or $X \to V_1$, there must be some MAG $\g[M]$ in $[\g]$ with the edge $X \to V_1$. Let $p^{**}$ be the path in $\g[M]$ corresponding to $p^*$ in $\g$. Then since $p^*$ is unshielded, so is $p^{**}$, and so $p^{**}$ takes the form $X \to V_1 \to \dots \to V_k$. Let $\g[D]$ be a DAG created from $\g[M]$, by retaining all the nodes in $\g[M]$ and all the directed edges in $\g[M]$ and by adding a node $L_{AB}$ and edges $L_{AB} \to B$ and $L_{AB} \to A$ for each bidirected edge $A \leftrightarrow B$ in $\g[M]$ (this DAG is titled the canonical DAG by \citealp{richardson2002ancestral}). Now, DAG $\g[D]$ contains a causal path from $X$ to $Z$.

    $\neg \ref{lem:equiv-z-pag-b} \Rightarrow \neg \ref{lem:equiv-z-pag-a}$ If there is a DAG $\g[D]$ represented by $\g$ with a causal path from $X \in \mb{X}$ to $Z \in \mb{Z}$, then any MAG $\g[M]$ of $\g[D]$ that contains $X$ and $Z$ will contain a causal path from $X$ to $Z$. This is due to the fact that a MAG of a DAG will preserve ancestral relationships  between observed variables. Then the path in $\g$ that corresponds to $q$ in $\g[M]$ cannot have any arrowheads pointing in the direction of $X$, and so it must be possibly causal.
\end{proofof}


\section{PROOFS FOR SECTION \ref{sec:constructing-pag}: PAGS - CONSTRUCTING CONDITIONAL ADJUSTMENT SETS}
\label{supp:constructing-pag}

This section includes the proof of Theorem \ref{thm:adjust-and-o-pag}, which can be found in Section \ref{sec:constructing-pag}. We provide one supporting result needed for the proof of this theorem.

We make an important remark here on R software. Note that by Lemmas \ref{lem:comparison} and \ref{lem:comparison-pag}, any algorithms developed for checking the existence of an unconditional adjustment set (Definition \ref{def:as}) also apply to conditional adjustment sets -- provided that $\mb{Z} \cap \PossDe(\mb{X}, \g) = \emptyset$. First consider the R package \texttt{dagitty} \citep{textor2016robust}. Suppose the condition on $\mb{Z}$ is satisfied and let $\mb{S}$ be a set such that $\mb{S} \cap (\mb{X} \cup \mb{Y} \cup \mb{Z}) = \emptyset$. Then, one can apply the function $\texttt{isAdjustmentSet}$ of the package \texttt{dagitty} to a PAG $\g$, set $\mb{S} \cup \mb{Z}$, exposure $\mb{X}$, and outcome $\mb{Y}$ to learn whether $\mb{S}$ is a conditional adjustment set relative to $(\mb{X},\mb{Y},\mb{Z})$ in $\g$. Next consider the R package \texttt{pcalg} \citep{kalisch2012pcalg}. Suppose the condition on $\mb{Z}$ is satisfied and let $\mb{S}$ be a set such that $\mb{S} \cap (\mb{X} \cup \mb{Y} \cup \mb{Z}) = \emptyset$. Then, one could apply the function \texttt{gac} of the package \texttt{pcalg} to the MPDAG or PAG $\g$ and to the node sets $\mb{X}$, $\mb{Y}$, and $\mb{S} \cup \mb{Z}$. These functions will return \texttt{TRUE} if and only if $\mb{S}$ is a conditional adjustment set relative to $(\mb{X},\mb{Y},\mb{Z})$ in $\g$, and \texttt{FALSE} otherwise.


\subsection{Main Result}

\begin{proofof}[Theorem \ref{thm:adjust-and-o-pag}]
    Suppose that $\adjustb{\g}$ does not satisfy the conditional adjustment criterion relative to $(\mb{X,Y,Z})$ in $\g$. Since $\adjustb{\g} \cap \fb{\g} = \emptyset$ by construction, it must be that there is a proper definite status non-causal path from $\mb{X}$ to $\mb{Y}$ that is m-connecting given $\adjustb{\g} {\cup \mb{Z}}$. By Lemma \ref{lem:adjust-and-o-pag-helper}, there is then a proper definite status non-causal path $p$ from $\mb{X}$ to $\mb{Y}$ in $\g$ such that all {definite non-colliders on $p$ are in $\fb{\g}$ (case \ref{opt-pag-prop2} of Lemma \ref{lem:adjust-and-o-pag-helper}) and all colliders on $p$ are in $\An(\mb{X} \cup \mb{Y} \cup \mb{Z}, \g)$ (cases \ref{opt-pag-prop3} and \ref{opt-pag-prop6} of Lemma \ref{lem:adjust-and-o-pag-helper}). Since $\An(\mb{X} \cup \mb{Y} \cup \mb{Z}, \g) \subseteq \An(\mb{X} \cup \mb{Y} \cup \mb{Z} \cup \mb{S}, \g)$, for any set $\mb{S}$ that satisfies $[\mb{S} \cup \mb{Z}] \cap [\mb{X} \cup \mb{Y} \cup \fb{\g}] = \emptyset$, Lemma \ref{lem:richprime2} implies that there is also a proper definite status non-causal path from $\mb{X}$ to $\mb{Y}$ in $\g$ that is open given $\mb{S}$. Since this is true for an arbitrary set $\mb{S}$ that satisfies condition \ref{def:cac-pag-a} of Definition \ref{def:cac-pag},} it follows that there cannot be any set that satisfies the conditional adjustment criterion relative to to $(\mb{X,Y,Z})$ in $\g$.
\end{proofof}


\subsection{Supporting Result}

\begin{lemma}
\label{lem:adjust-and-o-pag-helper}
    Let $\mb{X}$, $\mb{Y}$, and $\mb{Z}$, be pairwise disjoint node sets in a PAG $\g$, where $\mb{Z} \cap \PossDe(\mb{X}, \g) = \emptyset$ and where every proper possibly causal path from $\mb{X}$ to $\mb{Y}$ in $\g$ starts with a visible edge out of $\mb{X}$. Suppose furthermore, that there exists a set $\mb{S}$ that satisfies the conditional adjustment criterion for $(\mb{X,Y,Z})$ in $\g$. If there is a proper definite status non-causal path from $\mb{X}$ to $\mb{Y}$ in $\g$ that is m-connecting given $\adjustb{\g} {\cup \mb{Z}}$ (see definition in Theorem \ref{thm:adjust-and-o-pag}), then there is a path $p$ from $\mb{X}$ to $\mb{Y}$ in $\g$ such that the following hold.
    \begin{enumerate}[label =(\roman*)]
        \item\label{opt-pag-prop1} Path $p$ is a proper definite status non-causal path from $\mb{X}$ to $\mb{Y}$ in $\g$.
        
        \item\label{opt-pag-prop2} All definite non-colliders on $p$ are in $\fb{\g}$.
        
        \item\label{opt-pag-prop3} There is at least one collider on $p$, and all colliders on $p$ are in $\mb{C_1} \cup \mb{C_2}$, where $\mb{C_1}$ and $\mb{C_2}$ are disjoint sets such that
        \begin{align*}
                & \mb{C_1} \subseteq \PossAn(\mb{X} \cup \mb{Y} {\cup \mb{Z}}, \g) \setminus \big[ \An(\mb{X} \cup \mb{Y} \cup \mb{Z}, \g) \cup {\mb{X} \cup \mb{Y} \cup \fb{\g}} \big]\ and \\
                &\mb{C_2} \subseteq \An(\mb{X} \cup \mb{Y} \cup \mb{Z}, \g) \setminus \big[ \mb{X} \cup \mb{Y} \cup \fb{\g}\big].
        \end{align*}
        
        \item\label{opt-pag-prop4} None of the colliders on $p$ can be possible descendants of a non-collider on $p$.
        
        \item\label{opt-pag-prop5} For any collider $C \in \mb{C_1}$ on $p$ there is an unshielded possibly directed path from $C$ to $\mb{X} \cup \mb{Y} \cup \mb{Z}$ that does not start with $\circcirc$.
        
        \item\label{opt-pag-prop6} $\mb{C_1} = \emptyset$, that is for any collider $C \in \mb{C_1}$ on $p$ there is an unshielded directed path from $C$ to $\mb{X} \cup \mb{Y} \cup \mb{Z}$.
    \end{enumerate}
\end{lemma}

\begin{proofof}[Lemma \ref{lem:adjust-and-o-pag-helper}]
    Consider the sets of all proper definite status non-causal paths from $\mb{X}$ to $\mb{Y}$ in $\g$ that are m-connecting given $\adjustb{\g} {\cup \mb{Z}}$ and choose among them a shortest path with a shortest distance to $\mb{X} \cup \mb{Y} \cup \mb{Z}$ (Definition \ref{def:distance}). Let this path be called $p$, where $p= \langle X = V_1, V_2, \dots, V_k =Y \rangle$, $X \in \mb{X}, Y \in \mb{Y}$, $k \ge 2$. By choice of $p$, \ref{opt-pag-prop1} is satisfied. We will now show that $p$ also satisfies properties \ref{opt-pag-prop2}-\ref{opt-pag-prop6} above. 

    First, consider properties \ref{opt-pag-prop2} and \ref{opt-pag-prop3}. Since $p$ is m-connecting given $\adjustb{\g} {\cup \mb{Z}}$, any collider on $p$ is in $\An(\adjustb{\g} \cup \mb{Z}, \g)$. Furthermore, since $\adjustb{\g} \cup \mb{Z} = \PossAn(\mb{X} \cup \mb{Y} \cup \mb{Z}, \g) \setminus \big[ \mb{X} \cup \mb{Y}  \cup \fb{\g} \big]$, and since in a PAG $\g$ for any set $\mb{W}$, $\An(\PossAn(\mb{W}, \g)) = \PossAn(\mb{W}, \g)$, we have that any collider on $p$ is in $\PossAn(\mb{X} \cup \mb{Y} \cup \mb{Z}, \g)$. Furthermore, since by definition, $\De(\fb{\g}, \g) = \fb{\g}$, we have that no collider on $p$ can be in $\fb{\g}$. Hence, all colliders on $p$ are in $\PossAn(\mb{X} \cup \mb{Y} \cup \mb{Z}, \g) \setminus \fb{\g}.$

    Also, since $p$ is proper, a node in $\mb{X}$ cannot be a non-endpoint node on $p$. Now, since $p$ is additionally chosen as a shortest proper non-causal definite status path from $\mb{X}$ to $\mb{Y}$ that is m-connecting given $\adjustb{\g}$, it holds that either a node in $\mb{Y}$ is not a non-endpoint node on $p$, or there is a node  $Y' \in \mb{Y} \setminus \{Y\}$ on $p$ such that   $p(X, Y')$ is a possibly causal path from $X$ to $Y'$. Moreover, in this case $p(X,Y')$ must be a causal path in $\g$ (because $p$ must start with a visible edge and because $A \bulletarrow B \circbullet C$ cannot be a subpath of a definite status path). Since $p$ itself is a non-causal path in $\g$, there is a collider on $p$ that is a descendant of $Y'$. But since $Y' \in \fb{\g}$, this collider would then also have to be in $\fb{\g}$, which we have ruled out as an option in the previous paragraph. Hence, a node on $\mb{Y}$ is also not a non-endpoint node on $p.$
    
    Then all colliders on $p$ are in $\PossAn(\mb{X} \cup \mb{Y} \cup \mb{Z}, \g) \setminus [\fb{\g} \cup \mb{X} \cup \mb{Y}]$. Also, any definite non-collider on $p$ is a possible ancestor of a collider on $p$ or of an endpoint on $p$. Hence, every definite non-collider on $p$ is in $\PossAn(\mb{X} \cup \mb{Y} \cup \mb{Z}, \g) \setminus [\mb{X} \cup \mb{Y}].$ But, since $p$ is m-connecting given $\adjustb{\g} {\cup \mb{Z}}$, none of the definite non-colliders on $p$ are in $\PossAn(\mb{X} \cup \mb{Y} \cup \mb{Z} , \g)  \setminus \big[ \mb{X} \cup \mb{Y} \cup \fb{\g} \big]$. Therefore, any definite non-collider on $p$ is in $\fb{\g}$. This proves property \ref{opt-pag-prop2}.

    Next, consider property \ref{opt-pag-prop3}. We have already shown that any collider on $p$ is in $\PossAn(\mb{X} \cup \mb{Y} \cup \mb{Z}, \g) \setminus [\fb{\g} \cup \mb{X} \cup \mb{Y}]$. So it is only left to show that at least one collider is on $p$. Since we know that $p$ must be blocked by  $\mb{S} \cup \mb{Z}$ for some set $\mb{S}$, where $\mb{S} \cap {\big[ \mb{X} \cup \mb{Y} \cup \mb{Z} \cup \fb{\g} \big]} = \emptyset$, and since all definite non-colliders on $p$ are in  $\fb{\g}$, there is at least one collider $C$ on $p$.

    Property \ref{opt-pag-prop4} follows almost directly now, since by \ref{opt-pag-prop2}, all definite non-colliders on $p$ are in $\fb{\g}$ and by \ref{opt-pag-prop3}, none of the colliders can be in $\fb{\g}$. The claim then holds since by definition of the $\fb{\g}$ in a PAG, $\PossDe(\fb{\g}, \g) = \fb{\g}$.

    Next, we show properties \ref{opt-pag-prop5} and \ref{opt-pag-prop6}. Let $C \in \mb{C_1}$ be a collider on $p$. Then $C \notin \big[\mb{X} \cup \mb{Y} \cup \mb{Z} \big]$ and that there is an unshielded possibly directed path $r = \langle C, Q, \dots, V\rangle$ from $C$ to a node $V \in \mb{X} \cup \mb{Y} {\cup \mb{Z}}$.
   
    \ref{opt-pag-prop5} Suppose for a contradiction that edge $\langle C,Q \rangle$ on $r$ is of type $C \circcirc Q$ (possibly $Q =V$). We derive a contradiction by constructing a proper definite status non-causal path from $\mb{X}$ to $\mb{Y}$ that is m-connecting given $\adjustb{\g} {\cup \mb{Z}}$ and shorter than $p$, or of the same length as $p$ but with a shorter distance to $\mb{X} \cup \mb{Y} \cup \mb{Z}$ (Definition \ref{def:distance}).

    Let $A$ and $B$ be nodes on $p$ such that $A \bulletarrow C \arrowbullet B$ is a subpath of $p$ (possibly $A=X$, $B=Y$). Then paths $A \bulletarrow C \circcirc Q$ and $B \bulletarrow C \circcirc Q$ together with Lemma \ref{lem:basic-property-pags} imply that $A \bulletarrow Q \arrowbullet B$ is in $\g$.

    Suppose first that $A \neq X$, and $B \neq Y$. Note that by property \ref{opt-pag-prop4} above, if $A \neq X$, then $A \leftrightarrow C$ is in $\g$. Moreover, if $A \leftrightarrow C$ is in $\g$, then $A \leftrightarrow Q$ is in $\g$, otherwise path $\langle A,Q,C \rangle$ and edge $A \leftrightarrow C$  contradict Lemma \ref{lem:marl-cycle}. Hence, if $A \neq X$, the collider/definite non-collider status of $A$ is the same on $p$ and on $p(X,A) \oplus \langle A,Q\rangle$. Analogous reasoning  can be employed in the case when $B \neq Y$, to show that $B \leftrightarrow Q$, that is, the collider/definite non-collider status of $B$ is the same on $p$ and on $\langle Q, B \rangle \oplus p(B,Y)$. 

    Now, we return to the general case where we allow $A = X$ and $B = Y$. In each of the cases below we will derive the contradiction by finding a path $s$ from $\mb{X}$ to $\mb{Y}$ in $\g$ that is a proper non-causal definite status path in $\g$ and m-connecting given $\adjustb{\g} {\cup \mb{Z}}$. Additionally, the path $s$ will either be shorter than $p$ or of the same length as $p$, but with a shorter distance to $\mb{X} \cup \mb{Y} {\cup \mb{Z}}$ (Definition \ref{def:distance}) which implies a contradiction with our choice of $p$.

     Suppose first that $Q$ is not a node on $p$. 
    \begin{itemize}
        \item If $Q \notin \mb{X} \cup \mb{Y}$, then
        \begin{itemize}
            \item if $A \neq X$ and $B \neq Y$, then  let $s= p(X,A) \oplus \langle A,Q,B \rangle \oplus p(B,Y)$. By the reasoning above, this path transformation amounts to replacing $A \leftrightarrow C \leftrightarrow B$ on $p$ with $A \leftrightarrow Q \leftrightarrow B$ on $s$ thereby creating a path with the same properties as $p$ but with a shorter distance to $\mb{X} \cup \mb{Y} {\cup \mb{Z}}$ (Definition \ref{def:distance}).
            \item If $A = X$, and $B \neq Y$, then  let $s = \langle A,Q,B \rangle \oplus p(B,Y)$. This path transformation amounts to replacing $X \bulletarrow C \leftrightarrow B$ on $p$, with $X \bulletarrow Q \leftrightarrow B$ on $s$, thereby creating a path  with the same properties as $p$ and of the same length as $p$ but with a shorter distance  to $\mb{X} \cup \mb{Y} {\cup \mb{Z}}$ (Definition \ref{def:distance}).
            \item If $A \neq X$, and $B = Y$, then let $s = p(X,A) \oplus \langle A,Q,B \rangle $. This path transformation amounts to replacing $A \leftrightarrow C  \arrowbullet Y$ on $p$, with $A \leftrightarrow Q \arrowbullet Y$ on $s$, thereby creating a path  with the same properties as $p$, that is  of the same length, but with a shorter distance  to $\mb{X} \cup \mb{Y} {\cup \mb{Z}}$ (Definition \ref{def:distance}).
            \item If $A = X$, and $B = Y$, then let $s \langle A,Q,B \rangle$. Now $s$ is of the form $X \bulletarrow Q \arrowbullet Y$ and clearly satisfies all the same properties as $p$ while being of the same length, but with a shorter distance  to $\mb{X} \cup \mb{Y} {\cup \mb{Z}}$ (Definition \ref{def:distance}).
        \end{itemize}

        \item If $Q \equiv X'$, $X' \in \mb{X}$, then:
          \begin{itemize}
            \item if $B \neq Y$, let $s = \langle Q,B \rangle \oplus p(B,Y)$. This path transformation amounts to replacing $X \dots C \leftrightarrow B$ on $p$, with $X' \leftrightarrow B$ on $s$, thereby creating a shorter path  with the same properties as $p$.
            \item If $B =Y$, then let $s = \langle Q, B\rangle$. Due to the discussion above, $s$ is of the form $X' \arrowbullet Y$ in $\g$. 
        \end{itemize}
 
        \item Otherwise, $Q \equiv Y', Y' \in \mb{Y}$. If $Q \in \mb{Y} \cap \fb{\g}$, this would imply that $C \in \fb{\g}$, which contradicts \ref{opt-pag-prop3}. So $Q$ must be in $\mb{Y} \setminus \fb{\g}$. Then:
        \begin{itemize}
            \item if $A \neq X$, then let $s = p(X,A) \oplus \langle A,Q \rangle$. This path transformation amounts to replacing  $A \leftrightarrow C \dots Y$ on $p$, with $A \leftrightarrow Y'$ on $s$, thereby creating a shorter path  with the same properties as $p$.
            \item If $A = X$, then let $s = \langle A, Q \rangle$. Due to the discussion above, $s$ is of the form $X \arrowbullet Y'$ in $\g$. 
        \end{itemize}
    \end{itemize}

    Otherwise, $Q$ is on $p$. Therefore, $Q \notin \mb{X} \cup \mb{Y}$. Also, $Q$ is a collider on $p$, otherwise $Q \in \fb{\g}$ and $C \in \fb{\g}$, because of $C \circcirc Q$. 
    \begin{itemize}
        \item Suppose first that $Q$ is on $p(C,Y)$. Then:
        \begin{itemize}
            \item  if $A \neq X$, then let $s = p(X,A) \oplus \langle A,Q \rangle \oplus p(Q,Y)$. This path transformation amounts to replacing  $A \leftrightarrow C \leftrightarrow \dots \leftrightarrow Q $ on $p$, with $A \leftrightarrow Q$ on $s$, thereby creating a shorter path  with the same properties as $p$. 
            \item If $A = X$, then let $s = \langle A,Q \rangle \oplus p(Q,Y)$. This path transformation amounts to replacing  $X  \leftrightarrow C \leftrightarrow \dots \leftrightarrow Q$ on $p$, with $X \leftrightarrow Q$ on $s$, thereby creating a shorter path  with the same properties as $p$. 
        \end{itemize}
        
        \item Next, suppose that $Q$ is on $p(X,C)$. Then depending  on whether  $B = Y$, we can choose one of the following paths as the path $s$:
        \begin{itemize}
            \item  if $B \neq Y$, then let $s = p(X,Q) \oplus \langle Q,B \rangle \oplus p(B,Y) $. This path transformation amounts to replacing  $Q \leftrightarrow \dots \leftrightarrow C \leftrightarrow B$ on $p$, with $Q \leftrightarrow B$ on $s$, thereby creating a shorter path  with the same properties as $p$. 
            \item If $B = Y$, then let $s= p(X,Q) \oplus \langle Q,B \rangle$.  Similarly to above, this path transformation amounts to replacing  $Q \leftrightarrow \dots \leftrightarrow C \leftrightarrow Y$ on $p$, with $Q \leftrightarrow Y$ on $s$, thereby creating a shorter path  with the same properties as $p$. 
        \end{itemize}
    \end{itemize}

    \ref{opt-pag-prop6} Since we showed above that the starting edge $\langle C, Q \rangle$ on $r = \langle C,Q, \dots, V \rangle$ is not of the form $C \circcirc Q$, and since $r$ is an unshielded possibly directed path from $C$ to $V \in \mb{X} \cup \mb{Y} {\cup \mb{Z}}$, in order to prove property \ref{opt-pag-prop6} it is enough to show that $\langle C, Q \rangle$ is also not of the form $C \circarrow Q$ (since $P_1 \bulletarrow P_2 \circbullet P_3$ cannot be a subpath of any unshielded possibly directed path in $\g$, \citealp{zhang2008completeness}). Suppose for a contradiction that $\langle C,Q\rangle$ is exactly of that form. Since $A \bulletarrow C \arrowbullet B$ and $C \circarrow Q$ are in $\g$, by Lemma \ref{lem:basic-property-pags}, $A \bulletarrow Q \arrowbullet B$ is in $\g$. 
    
    Now, our goal is to identify a nodes $A'$ and $B'$ on $p$ that satisfy the following. Node $A'$ is on $p(X,A)$, and edge $A' \bulletarrow Q$ is in $\g$. Additionally, $A' = X$ or $A'$ is a non-endpoint node on $p$ that has the same definite non-collider/collider status on $p$ and on $p(X,A') \oplus \langle A', Q \rangle$. Similarly, $B'$ is on $p(B,Y)$, and edge $B' \bulletarrow Q$ is in $\g$. Additionally, $B' = Y$ or $B'$ is a non-endpoint node on $p$ that has the same definite non-collider/collider status on $p$ and on $\langle Q, B' \rangle \oplus p(B',Y)$. We only show how to find node $A'$ on $p(X,A)$, since the argument for finding $B'$ on $p(B,Y)$ is exactly symmetric.
    
    \begin{itemize}
        \item Consider the path $p(X,C) = \langle X = V_1, V_2, \dots, V_{i-1} = A, V_i = C \rangle $. Note that by \ref{opt-pag-prop4} and the properties of unshielded paths, $p(X,C)$ is of the form $X \bulletarrow V_2 \leftrightarrow \dots \leftrightarrow A \leftrightarrow C$ or $X \leftarrow V_2 \leftarrow \dots \leftarrow V_j \leftrightarrow \dots \leftrightarrow C$, for some $V_j$, $j \in \{2, \dots, i-1\}$. 
        
        Hence, if there is any non-endpoint node $W$ on $p(X,A)$ such that $W \leftrightarrow Q$, this node has the same definite collider / non-collider status on both $p$ and on $p(X,W) \oplus \langle W, Q \rangle$. Then we choose $A' \equiv W$. Otherwise, if there is a non-endpoint node $W$ on $p(X,A)$ such that $-p(W, X)$ is of the form $W \to \dots \to X$, and an edge $W \to Q$ or $W \circarrow Q$ is in $\g$, then $W$ is a definite non-collider on both $p$ and $p(X,W) \oplus \langle W, Q \rangle$ and we choose $A' \equiv W$.

        We will now show that if neither of the above choices for $A'$ are possible in $\g$, then $p(X,C)$ is of the form $X \leftrightarrow V_2 \leftrightarrow \dots \leftrightarrow C$, and for every node $V_j$, $j \in \{1, \dots, i\}$ on $p(X,C)$, the edge $V_j \to Q$ or $V_j \circarrow Q$ is in $\g$. In this case, we choose $A' \equiv X$.
        
        Hence, consider first node $V_{i-1} = A$ on $p$. By above $V_{i-1} \bulletarrow Q$ is in $\g$. Also, by our assumption  $V_{i-1} \leftrightarrow Q$ is not in $\g$, so we must have either $V_{i-1} \to Q$ or $V_{i-1} \circarrow Q$ is in $\g$. Similarly, by the assumption above we now know that edge $\langle V_{i-2}, V_{i-1} \rangle$ is not of the form $V_{i-2} \leftarrow V_{i-1}$, so we can conclude that $V_{i-2} \leftrightarrow V_{i-1}$ is in $\g$. 
        
        Now, $V_{i-2} \leftrightarrow V_{i-1} \leftrightarrow C \circarrow Q$ and either $V_{i-1} \to Q$ or $V_{i-1} \circarrow Q$ is in $\g$. If $V_{i-1} \to Q$ is in $\g$, then $R4$ of \cite{zhang2008completeness} would imply that $V_{i-2} \in \Adj(Q, \g)$. Moreover, since $V_{i-2} \leftrightarrow V_{i-1} \to Q$ is in $\g$, $R2$ of \cite{zhang2008completeness} would imply that $V_{i-2} \bulletarrow Q$ is in $\g$, and our assumption further lets us conclude that $V_{i-2} \to Q$, or $V_{i-2} \circarrow Q$ is in $\g.$
        
        If $V_{i-1} \circarrow Q$ is in $\g$, then $V_{i-2} \leftrightarrow V_{i-1} \circarrow Q$ and Lemma \ref{lem:basic-property-pags} imply that, $V_{i-2} \bulletarrow Q$ is in $\g$. Hence, as above either $V_{i-2} \to Q$, or $V_{i-2} \circarrow Q$ is in $\g.$
        
        If $V_{i-2} = X$ we are done. Otherwise, we can repeat the same argument as in the preceding three paragraphs to conclude that $V_{i-3} \leftrightarrow V_{i-2} \leftrightarrow V_{i-1} \leftrightarrow C$ is in $\g$, and either $V_{i-3} \to Q$ or $V_{i-3} \circarrow Q$ are in $\g$. If $X \neq V_{i-3}$, we can keep applying the same argument, until we reach $X$.
    \end{itemize}

    Now that we have chosen the appropriate $A'$ and $B'$ the remaining argument is very similar to case \ref{opt-pag-prop5}.  In each of the cases below we will derive the contradiction by finding a path $s$ from $\mb{X}$ to $\mb{Y}$ in $\g$ that is a proper non-causal definite status path in $\g$ and m-connecting given $\adjustb{\g} {\cup \mb{Z}}$. Additionally, the path $s$ will either be shorter than $p$ or of the same length as $p$, but with a shorter distance to $\mb{X} \cup \mb{Y} {\cup \mb{Z}}$ (Definition \ref{def:distance}) which implies a contradiction with our choice of $p$.

Suppose first that $Q$ is not on $p$:
    \begin{itemize}
        \item If $Q \notin \mb{X} \cup \mb{Y}$, then
        \begin{itemize}
            \item if $A' \neq X$ and $B' \neq Y$, then let $s = p(X,A') \oplus \langle A',Q,B' \rangle \oplus p(B',Y)$. By the reasoning above, this path transformation amounts to replacing $p(A',B')$ on $p$ with  $\langle A',Q,B' \rangle$ on $s$ such that the collider / definite non-collider status of $A'$ and $B'$ is the same on both paths. Therefore, $s$ is a path with the same properties as $p$, but either shorter than $p$ or of the same length but with a shorter distance to $\mb{X} \cup \mb{Y} {\cup \mb{Z}}$ (Definition \ref{def:distance}).
            \item If $A' = X$, and $B' \neq Y$, then let $s= \langle A',Q,B' \rangle \oplus p(B',Y)$. By the reasoning above, this path transformation amounts to replacing $p(X,B')$ on $p$ with  $\langle X,Q,B' \rangle$ on $s$ such that the collider / definite non-collider status of $B'$ is the same on both paths, and $s$ is a non-causal path because of $Q \arrowbullet B'$ edge. Therefore, $s$ is a path with the same properties as $p$ but either shorter than $p$ or of the same length but with a shorter distance to $\mb{X} \cup \mb{Y} {\cup \mb{Z}}$ (Definition \ref{def:distance}).
            \item If $A' \neq X$, and $B' = Y$, then let $s = p(X,A') \oplus \langle A',Q,B' \rangle$. This path transformation amounts to replacing $p(A',Y)$ on $p$ with  $\langle A',Q,Y \rangle$ on $s$ such that the collider / definite non-collider status of $B'$ is the same on both paths, and $s$ is a non-causal path because of $Q \arrowbullet Y$ edge. Therefore, $s$ is a path with the same properties as $p$ but either shorter than $p$ or of the same length but with a shorter distance to $\mb{X} \cup \mb{Y} {\cup \mb{Z}}$ (Definition \ref{def:distance}).
            \item If $A' = X$ and $B' = Y$, $\langle A',Q,B' \rangle$. Then $s$ is of the form $X \bulletarrow Q \arrowbullet Y$ and $Q \in \An(\adjustb{\g} \cup \mb{Z},\g)$ and $Q$ has a shorter distance to $\mb{X} \cup \mb{Y} \cup \mb{Z}$ than $C$.    
        \end{itemize}

        \item If $Q \equiv X'$, $X' \in \mb{X}$, then:
          \begin{itemize}
            \item if $B' \neq Y$, then let $s = \langle Q,B' \rangle \oplus p(B',Y)$. This path transformation amounts to replacing $p(X,B')$ on $p$ with  $\langle X', B' \rangle$ on $s$ such that the collider / definite non-collider status of $B'$ is the same on both paths, and $s$ is a non-causal path because of $X' \arrowbullet B'$ edge. Therefore, $s$ is a path with the same properties as $p$ shorter than $p$.
            \item If $B' = Y$, then let $s = \langle Q, B'\rangle$, where based on the reasoning above, $s$ is of the form $X' \arrowbullet Y$.
        \end{itemize}
 
        \item Otherwise, $Q \equiv Y'$, $Y' \in \mb{Y}$. Then
        \begin{itemize}
            \item if $A' \neq X$, then $s = p(X,A') \oplus \langle A',Q \rangle$. Note that in this case $s$ is of the form $X \leftrightarrow \dots \leftrightarrow A' \leftrightarrow Y'$, or $X \leftarrow \dots \leftarrow A' \circarrow Y'$, or $X \leftarrow \dots \leftarrow A' \to Y'$. In all cases, $s$ is a proper non-causal definite status path from $\mb{X}$ to $\mb{Y}$ that is m-connecting given $\adjustb{\g} \cup \mb{Z}.$
            \item If $A' = X$, then let $s =\langle A', Q \rangle$. We now discuss why $s$ is of the form $X \leftrightarrow Y'$ in $\g$. 
           
            Note that $X \circarrow Y'$ cannot be in $\g$, since there exists a set $\mb{S}$ that can satisfy the conditional adjustment criterion relative to $\mb{X,Y,Z}$ in $\g$.     
        If instead $X \to Y'$ is a visible edge in $\g'$, then there is either a node $D \notin \Adj(Y', \g)$ such that $D \bulletarrow X$ is in $\g$ or there is a collection of nodes $D_1, \dots, D_k$, such that $D_1 \notin \Adj(Y', \g)$, $D_2, \dots, D_k \in \Pa(Y', \g)$, and $D_1 \bulletarrow D_2 \leftrightarrow \dots \leftrightarrow D_k \leftrightarrow X$ is in $\g$. Without loss of generality we will assume that we are in the fist case, that is $D \bulletarrow X$ is in $\g$ and $D \notin \Adj(Y', \g)$, since the latter case has an analogous proof to what follows. 

        By above, the only way way that $A' \equiv X$ is if $X \leftrightarrow V_2 \leftrightarrow \dots \leftrightarrow C$ is in $\g$ and if for all nodes $V_j\in \{V_2, \dots, V_{i-2}, V_{i-1}, V_i\}, V_j \to Y'$, or $V_j \circarrow Y'$ is in $\g$. Now since, $D \bulletarrow X \leftrightarrow V_2 \leftrightarrow \dots \leftrightarrow V_{i-1}  \leftrightarrow C$ is also in $\g$, and $D \notin \Adj(Y', \g)$, we can use $R4$ of \cite{zhang2008completeness} iteratively to conclude that $V_j \to Y'$ is in $\g$ for all $j \in \{1,\dots, i\}$. However, as $V_i \equiv C$, this contradicts our assumption that $C \circarrow Y'$ is in $\g$, for $Y' = Q$.
            
        \end{itemize} 
    \end{itemize}

    Otherwise, $Q$ is on $p$. Therefore, $Q \notin \mb{X} \cup \mb{Y}$. 
    \begin{itemize}
        \item Suppose first that $Q$ is on $p(C,Y)$. By \ref{opt-pag-prop3}, \ref{opt-pag-prop4}, and the definition of $\fb{\g}$, we have that $p(C, Y)$ is of one of the following forms:
        \begin{itemize}
            \item $C \leftrightarrow \dots \leftrightarrow Q \leftrightarrow \dots \leftrightarrow V_k \arrowbullet Y$ for $k > i$, or
            \item $C \leftrightarrow \dots \leftrightarrow Q \leftrightarrow \dots \leftrightarrow T_1 \to \dots \to Y$, for some $T_1$ on $p(C,Y)$, or
            \item $C \leftrightarrow \dots \leftrightarrow T_2 \to \dots \to Q \to \dots \to Y$, for some $T_2$ on $p(C,Y)$, or
             \item $C \leftrightarrow \dots \leftrightarrow Q \to \dots  \dots \to Y$.
        \end{itemize}
        
        Then
        \begin{itemize}
            \item If $A' \neq X$, then $s= p(X,A') \oplus \langle A',Q \rangle \oplus p(Q,Y)$. Note that by above forms of $p(C,Y)$ $s$ is always a is a proper non-causal path from $\mb{X}$ to $\mb{Y}$. Additionally, by above listed options for $p(C,Y)$ we know that $Q$ has the same collider / definite non-collider status on both $p$ and $s$. Hence, $s$ is also an m-connecting path given $\adjustb{\g} \cup \mb{Z}$. Since $s$ is also shorter than $p$ we obtain our contradiction. 
            \item If $A' \equiv X$, we let $s= p(X,Q) \oplus p(Q,Y)$. 
             Path $s$ is proper, since $p$ itself is proper and $Q \notin \mb{X \cup Y}$. Furthermore, by the above listed options for $p(C,Y)$ we know that $Q$ has the same collider / definite non-collider status on both $p$ and $s$ and that $s$ is a definite status path. Hence, $s$ is also an m-connecting path given $\adjustb{\g} \cup \mb{Z}$. If $s$ is a non-causal path in $\g$, we obtain a contradiction with the choice of $p$. 

             Hence, suppose for a contradiction that $s$ is a possibly causal path from $\mb{X}$ to $\mb{Y}$ in $\g$. By assumption, it must be that $X \to Q$ is a visible edge in $\g$. Now, similarly to the previous case, since $X \to Q$ is a visible edge in $\g$, there is either a node $D \notin \Adj(Q, \g)$ such that $D \bulletarrow X$ is in $\g$ or there is a collection of nodes $D_1, \dots, D_k$ such that $D_1 \notin \Adj(Q, \g)$, $D_2, \dots, D_k \in \Pa(Q, \g)$, and $D_1 \bulletarrow D_2 \leftrightarrow \dots \leftrightarrow D_k \leftrightarrow X$ is in $\g$. We again assume without loss of generality that we are in the former case, that is $D \bulletarrow X$ is in $\g$ and $D \notin \Adj(Q, \g)$. 
        
        Since $A' \equiv X$, by the same reasoning as in the previous case above we know that $X \leftrightarrow V_2 \leftrightarrow \dots \leftrightarrow C$ is in $\g$ and that for all nodes $V_j\in \{V_1, \dots, V_{i-1}, V_i\}, V_j \to Q$, or $V_j \circarrow Q$ is in $\g$. Now since, $D \bulletarrow X \leftrightarrow  V_2 \leftrightarrow \dots \leftrightarrow C$ is in $\g$, and since $D \notin \Adj(Q, \g)$, we can use $R4$ of \cite{zhang2008completeness} iteratively to conclude that $V_j \to Q$ is in $\g$ for all $j \in \{1,\dots, i\}$. However, as $V_i \equiv C$, this contradicts our assumption that $C \circarrow Q$ is in $\g$.        
        \end{itemize}
    
        \item Lastly, suppose that $Q$ is on $p(X,C)$. Analogously to above, by \ref{opt-pag-prop3}, \ref{opt-pag-prop4}, and the definition of $\fb{\g}$, we have that $p(X,C)$ is of one of the following forms:
        \begin{itemize}
            \item $X \bulletarrow V_2 \leftrightarrow \dots \leftrightarrow Q \leftrightarrow \dots \leftrightarrow C$, or
            \item  $X \leftarrow \dots \leftarrow T_1 \leftrightarrow \dots \leftrightarrow Q \leftrightarrow \dots \leftrightarrow C$, for some $T_1$ on $p(X,C)$,  or
            \item $X \leftarrow \dots \leftarrow Q \leftarrow \dots \leftarrow T_2 \leftrightarrow \dots \leftrightarrow C$, for some  $T_2$ on $p(X,C)$, or 
            \item $X \leftarrow \dots  \leftarrow Q \leftrightarrow \dots \leftrightarrow C$. 
        \end{itemize}

        Then
        \begin{itemize}
            \item If $B' \neq Y$, we have that $s = p(X,Q) \oplus \langle Q,B' \rangle \oplus p(B,Y)$ is  a proper non-causal path from $\mb{X}$ to $\mb{Y}$ that is shorter than $p$. Additionally, $Q$ is of the same collider / definite non-collider status on both $p$ and $s$ and therefore, $s$ is not only of definite status, but also m-connecting given $\adjustb{\g} {\cup \mb{Z}}$ in $\g$ which leads to a contradiction.
            \item If $B' \equiv Y$, then $s= p(X,Q) \oplus \langle Q,B' \rangle$ is a proper definite status non-causal path that is m-connecting given $\adjustb{\g} {\cup \mb{Z}}$ in $\g$ and shorter than $p$.
        \end{itemize}
    \end{itemize}
\end{proofof}


\vfill
\end{document}